\documentclass[preprint]{ptephy_om}

\preprintnumber{KYUSHU-HET-337, RIKEN-iTHEMS-Report-25}

\usepackage{mathrsfs}

\usepackage{graphicx,xcolor,tikz}
\usepackage{pgfplots}
\usetikzlibrary{3d, decorations.pathreplacing}

\usepackage{braket,bm}

\numberwithin{equation}{section}

\usepackage{hyperref}

\usepackage{comment}

\usepackage{orcidlink}

\newcommand{\no}{\notag}

\newcommand{\cE}{\mathcal E}
\newcommand{\cH}{\mathcal H}
\newcommand{\cI}{\mathcal I}

\newcommand{\cM}{\mathcal M}
\newcommand{\cO}{\mathcal O}
\newcommand{\cR}{\mathcal R}

\newcommand{\cZ}{\mathcal Z}
\newcommand{\cW}{\mathcal W}

\newcommand{\scrO}{\mathscr O}
\newcommand{\scrH}{\mathscr H}
\newcommand{\fO}{\mathfrak{O}}

\newcommand{\bbZ}{\mathbb{Z}}

\newcommand{\rmi}{\mathrm{i}}\newcommand{\rmd}{\mathrm{d}}

\newcommand{\cyan}[1]{\textcolor{cyan}{#1}}

\newcommand{\magenta}[1]{#1}

\newcommand{\order}[1]{\mathop{O}(#1)}
\newcommand{\littleorder}[1]{\mathop{o}(#1)}
\newcommand{\lorder}[1]{\littleorder{#1}}

\newcommand{\rhoge}{\rho_{\mathrm{Gibbs}}}
\newcommand{\rhogge}{\rho_{\mathrm{GGE}}}

\DeclareMathOperator{\Eop}{\mathbb{E}}
\DeclareMathOperator{\Vop}{\mathbb{V}}
\DeclareMathOperator{\Pop}{\mathbb{P}}

\DeclareMathOperator{\lcm}{\mathrm{lcm}}

\newcommand{\psiin}{\psi_{\text{in}}}
\newcommand{\kket}[1]{\ket{\ket{#1}\!}}
\newcommand{\bbra}[1]{\bra{\!\bra{#1}}}

\title{Eigenstate Thermalization Hypothesis\\
with projective representation}

\author[1]{Soma Onoda \orcidlink{0009-0008-9886-9279}
\thanks{onoda.soma@phys.kyushu-u.ac.jp}}
\affil[1]{Department of Physics, Kyushu University, 744 Motooka, Nishi-ku,
Fukuoka 819-0395, Japan}

\author[2]{Osamu Fukushima \orcidlink{0000-0001-7205-5324}
\thanks{osamu.fukushima@riken.jp}}
\affil[2]{Center for Interdisciplinary Theoretical and Mathematical Sciences (iTHEMS),
RIKEN, Wako 351-0198, Japan}

\author[3,2]{Ryusuke Hamazaki \orcidlink{0000-0003-3793-6016}
\thanks{ryusuke.hamazaki@riken.jp}}
\affil[3]{Nonequilibrium Quantum Statistical Mechanics RIKEN Hakubi Research Team, Pioneering Research Institute (PRI), RIKEN, 2-1 Hirosawa, Wako, Saitama 351-0198, Japan
}

\author[2]{Okuto Morikawa \orcidlink{0000-0002-0044-4491}
\thanks{okuto.morikawa@riken.jp}}

\begin{document}
\begin{abstract}
The Eigenstate Thermalization Hypothesis (ETH) provides a sufficient condition for thermalization of isolated quantum systems. While the standard ETH is formulated in the absence of degeneracy, physical systems often possess symmetries that induce degenerate energy eigenstates. In this paper, we investigate ETH in the presence of nontrivial projective representations of Abelian symmetries, which arise naturally from 't~Hooft anomalies. We argue that such projective structures can lead to degenerate excited states, and how the ETH can be formulated under such degeneracies.
In the presence of projective charges supplied by symmetry operators, our projective-representation ETH indicates that the stationary values of the operators are described by the generalized Gibbs ensemble instead of the standard Gibbs ensemble. Our findings elucidate the role of symmetry and degeneracy in quantum thermalization and pave the way for further exploration of the ETH in anomalous symmetry settings.
\end{abstract}
\maketitle

\tableofcontents

\section{Introduction}

Understanding thermalization---the process by which isolated quantum systems reach thermal equilibrium---remains one of the central challenges in modern theoretical physics. A widely accepted mechanism is the \textit{Eigenstate Thermalization Hypothesis} (ETH).
It states that
the energy eigenstates themselves are thermal for certain observables and offers a sufficient condition for thermalization in isolated systems, thereby providing a concrete framework in which this longstanding problem can be formulated.
The ETH has been numerically verified in a broad variety of non-integrable systems, for which matrix elements of observables in the excited energy eigenstates take a universal form that ensures thermal expectation values~\cite{Srednicki:1999bhx,Deutsch:1991msp,Srednicki:1994mfb,Dymarsky:2016ntg,rigol2008thermalization,santos2010onset,ikeda2011eigenstate,steinigeweg2013eigenstate, kim2014testing, beugeling2014finite, steinigeweg2014pushing, alba2015eigenstate, beugeling2015off, mondaini2017eigenstate, nation2018off, hamazaki2019random, khaymovich2019eigenstate,yoshizawa2018numerical, jansen2019eigenstate,PhysRevLett.126.120602,PhysRevLett.129.030602,sugimoto2023rigorous}.
 Implications of the ETH extend beyond condensed matter physics, influencing research in high-energy theory, including holography~\cite{Liska:2022vrd,deBoer:2016bov,Basu:2017kzo,Datta:2019jeo,Fitzpatrick:2015zha,Besken:2019bsu,Dymarsky:2019etq,Lashkari:2016vgj}, gauge theories~\cite{Mueller:2021gxd, Yao:2023pht, Ebner:2023ixq}, and also higher-form symmetries~\cite{Fukushima:2023svf}.

\subsection{Notion of thermalization}
To make the discussion precise, we begin by defining what we mean by ``thermalization.'' Consider an observable $\cO$. We say that $\cO$ thermalizes in the thermodynamic limit~$V\to\infty$ (where $V$ denotes the system size) if the following two conditions are satisfied:
\begin{itemize}
    \item The expectation value of $\cO$ approaches a stationary value at late times.
    \item This stationary value agrees with the prediction of the thermal ensemble.
\end{itemize}
We especially require these two conditions for any initial state~$\ket{\psiin}$, whose energy fluctuation is not too large (as detailed later).

These conditions can be formalized using the long-time average
\begin{align}
    \overline{\bra{\psiin}e^{\rmi Ht}\cO e^{-\rmi Ht}\ket{\psiin}}
    := \lim_{T \to \infty} \frac{1}{T} \int_0^T \rmd t \bra{\psiin} e^{\rmi Ht} \cO e^{-\rmi Ht} \ket{\psiin},
\end{align}
and the fluctuation
\begin{align}
    \sigma(\cO) := \left( \overline{   \left| \bra{\psiin}e^{\rmi Ht}\cO e^{-\rmi Ht}\ket{\psiin}-\overline{\bra{\psiin} e^{\rmi Ht} \cO e^{-\rmi Ht} \ket{\psiin}} \right|^2} \right)^{1/2}.
\end{align}
Then, assuming that $\|\cO\|=O(V^0)$, we define thermalization from the conditions\footnote{For a given function~$f(V)$, the symbols $\order{f(V)}$ and $\lorder{f(V)}$ are defined by the following relations:
\begin{align}
\lim_{V\to\infty}\left|\frac{\order{f(V)}}{f(V)}\right|<\infty,\qquad\lim_{V\to\infty}\left|\frac{\lorder{f(V)}}{f(V)}\right|=0.
\end{align}
}
\begin{align}
    \sigma(\cO) = \lorder{V^0}, \label{stationary}
\end{align}
and
\begin{align}
    \overline{\bra{\psiin}e^{\rmi Ht}\cO e^{-\rmi Ht}\ket{\psiin}} = \Tr[\rho_{\text{thermal}}\cO] + \lorder{V^0}. \label{stationary=thermal}
\end{align}
Here, $\rho_{\text{thermal}}$ is usually chosen as the Gibbs ensemble.

For Hamiltonians with non-degenerate and non-resonant eigenvalues, the ETH is known to justify both conditions above. 
The ETH in this case is formulated as
\begin{align}\label{ETH}
\braket{E_i|\mathcal{O}|E_j}
\simeq \scrO(E_i/V)\delta_{ij},
\end{align}
where $\ket{E_i}$ and $E_i$ are the eigenstates and eigenvalues of the Hamiltonian $H$, $\scrO$ is the smooth function of energy density, and $\simeq$ means that it is exact in the thermodynamic limit.
For sufficiently complicated systems with continuous Abelian symmetry, local conserved charges arise, but ETH is still expected to hold \textit{within} each symmetry (charge) sector.\footnote{For discrete symmetries, no associated Noether charges exist, and thus they do not usually affect the ETH for local observables~\cite{Hamazaki:2015ied}. In the case of higher-form symmetries, we can consider ETH-violating operators that are non-local but have a support much smaller than the total system size~\cite{Fukushima:2023svf}. Even in such scenarios, ETH is still expected to hold within each symmetry sector when the system is sufficiently complicated.}
We also note that the finite-size generalization  of the ETH, proposed in Ref.~\cite{Srednicki:1999bhx}, is given as
\begin{align}\label{Srednikiconj}
\braket{E_i|\mathcal{O}|E_j}
= \scrO(E_i/V)\delta_{ij}+e^{-S(\mathcal{E})/2}f(\mathcal{E}/V,\omega)R_{ij},
\end{align}
where $\cE := (E_i + E_j)/2$ is the averaged energy, $\omega := E_i - E_j$ is the  energy difference, and  $S(\cE)=\order{V}$ is the thermodynamic entropy. The function $f(\cE/V,\omega)$ is supposed to be smooth, and 
each element of the matrix $R_{ij}$ is quasi-random and of order $\order{1}$. In the thermodynamic limit, Eq.~\eqref{Srednikiconj} reduces to Eq.~\eqref{ETH}.

\subsection{Symmetry-induced structure in thermalization}
If the system possesses a symmetry and the energy eigenstates are non-degenerate, then it is shown that the diagonal matrix elements of a charged operator in the energy eigenbasis vanish due to symmetry selection rules. This implies that the long-time average of the expectation value of such charged operators vanishes for any initial state.
%
For ordinary Abelian symmetries, one expects (non-degenerate) block-diagonal structures and symmetry-resolved thermal ensembles. 

In contrast, a qualitatively new situation arises if the system possesses non-Abelian symmetries, which inevitably lead to degeneracies in the energy spectrum. In such cases, the arguments applicable to Abelian symmetries no longer hold. 
For instance, in systems with $SU(2)$ symmetry, matrix elements of observables decompose into Clebsch--Gordan coefficients and reduced matrix elements via the Wigner--Eckart theorem.  
It has been proposed~\cite{Murthy:2022dao} and later numerically investigated~\cite{Lasek:2024ess,Patil:2025ump,Noh:2025osn} that the ETH is then applied to the reduced matrix elements.

Here, as another route to degeneracies different from non-Abelian symmetries, we consider a situation where an Abelian symmetry acts \emph{projectively} on the Hilbert space. In such a case, there are degeneracies because the symmetry generators commute only up to a $U(1)$ phase as
\begin{align}
  U_{g_1}U_{g_2} = e^{\rmi\phi(g_1,g_2)}\, U_{g_1g_2},
\end{align}
with a nontrivial factor $\phi$.\footnote{This is discussed in more detail in Section~\ref{Projective representation-Hilbert-space}.} Nontrivial \textit{projective representations} of global symmetries in canonical quantization are a 
hallmark of systems exhibiting 't~Hooft anomalies.\footnote{In $(0+1)$-dimensional quantum mechanics, for instance, the presence of an 't~Hooft anomaly always implies a nontrivial projective representation. While this correspondence between an 't~Hooft anomaly and a projective representation does not strictly hold in higher dimensions, many quantum field theories do exhibit projective symmetry operators as a signature of 't~Hooft anomalies.}
The 't~Hooft anomaly is an obstruction to gauging global symmetries and has the important property of being invariant under the renormalization group. As a result, it imposes insightful constraints on the infrared dynamics of quantum field theories. For example, systems with 't~Hooft anomalies cannot have a trivially gapped ground state.

Furthermore, when the anomaly manifests, not only the ground state but also excited states exhibit degeneracy. Since the ETH concerns the structure of highly excited states, such projective representations can potentially alter the thermalization properties.

\subsection{Projective-representation ETH}
In this paper, we propose a generalization of ETH in the context of projective representations arising from Abelian symmetries, rather than from non-Abelian groups.
We formulate an ETH ansatz with projective representations (projective-representation ETH; prETH) of a $\bbZ_{N_1}\times\bbZ_{N_2}$ symmetry for matrix elements concerning symmetry-charged operators $\cO^{q_1,q_2}$ (the charges $q_1$ and~$q_2$ are associated to $\bbZ_{N_1}$ and~$\bbZ_{N_2}$). The prETH ansatz reduces to the conventional ETH for neutral operators and retains the standard structure conjectured by Srednicki~\cite{Srednicki:1999bhx,Dymarsky:2016ntg}.

We naturally classify charged operators  into three classes: \textit{neutral}, \textit{Type~I} (whose diagonal matrix elements vanish in the thermodynamic limit), and \textit{Type~II} (whose diagonals remain $\order{1}$).
Our prETH predicts the bahaviors of the stationary values for these three types of operators.
For neutral and Type~I operators, the stationary value equals the thermal prediction, whereas Type~II operators generically retain memory of projective charges carried by the initial state, exhibiting non-thermal behavior.
Examples of such Type II operator include the case where
the charge of a charged operator is supplied by the projective phase of a symmetry operator.

We demonstrate that the appropriate stationary ensemble for the charged sector is a \textit{non-commutative generalized Gibbs ensemble} (GGE) built from a complete set of Hermitian charges constructed out of symmetry generators (see Sec.~\ref{sec:gibbs_ensemble}). This GGE correctly captures stationary values of Type~II operators, extending the ideas for symmetry-resolved and non-Abelian cases~\cite{Murthy:2022dao}.
We also discuss the possibility of \textit{anomalous finite-size scaling},
reminiscent of the scenario discussed in Ref.~\cite{Murthy:2022dao}. We will extend and clarify this phenomenon in this paper.
In particular, while neutral observables exhibit the conventional $\order{V^{-1}}$ corrections, we discuss a possible scenario for which Type~II observables show enhanced $\order{V^{-1/2}}$ corrections.


We also test these predictions in two complementary settings: a $(1+1)$‑dimensional $\bbZ_N \times \bbZ_N$ spin model realizing nontrivial projective sectors and a $(2+1)$‑dimensional $\bbZ_2$ lattice gauge theory. Exact diagonalization of highly excited states confirms (i) the prETH structure of matrix elements, (ii) the classification into neutral/Type~I/Type~II observables, and (iii) the adequacy of the GGE description for charged observables.


\subsection{Organization}
This paper is organized as follows. In Section~\ref{Projective representation-Hilbert-space}, we review the conventional projective representation, especially focusing on the $\bbZ_N\times\bbZ_N$ projective symmetry. In Section~\ref{Eigenstate thermalization hypothesis and projective representation}, we extend the ETH framework to systems exhibiting nontrivial projective representations of Abelian symmetries, motivated by the operator formalism of 't~Hooft anomalies. We also give a classification of charged operators. Section~\ref{sec:gibbs_ensemble} is devoted to a generalized Gibbs ensemble to give a valid description of the stationary states, where the standard Gibbs ensemble breaks down.
Section~\ref{sec:examples} introduces a physical origin and structure of projective symmetry in Hamiltonians with concrete examples: numerical tests in the $\bbZ_N \times \bbZ_N$ spin model and $\bbZ_2$ lattice gauge theory. In Section~\ref{Anomalous scaling}, we also analyze how anomalous finite-size scaling may emerge due to the effects of the projective charge. Finally, Section~\ref{sec:conclusion} summarizes our results and discusses future directions.

\section{Projective representation}
\label{Projective representation-Hilbert-space}
\subsection{Projective representation}

First of all, we provide a brief review of projective representations.
Let us consider a system with a Hamiltonian $H$ that possesses a symmetry group $G$, assumed to be an Abelian group. The symmetry operators $U_g$, which act on the Hilbert space as a unitary representation of $G$, satisfy
\begin{align}
   \forall g \in G\qquad [U_g, H] = 0.
\end{align}
A projective representation is characterized by a phase factor such that, for all group elements $g_1$, $g_2 \in G$, the following relation holds:
\begin{align}
    \text{$\forall g_1$, $g_2 \in G$}\qquad U_{g_1} U_{g_2} &= e^{\rmi\phi(g_1, g_2)} U_{g_1 g_2}, & e^{\rmi\phi(g_1, g_2)} &\in U(1),\\
    U_{g_1} U_{g_2} &= e^{\rmi\phi(g_1, g_2)-\rmi\phi(g_2, g_1)}U_{g_2} U_{g_1},\label{noncommute projective}
\end{align}
where $e^{\rmi\phi(g_1, g_2)}$ is referred to as the projective phase.

It is important to note that each $U_g$ is defined only up to a $U(1)$ phase
\begin{align}
    U_g \to e^{\rmi\tilde\phi(g)} U_g.
\end{align}
A projective representation is said to be \textit{trivial} (i.e., a linear representation) if there exists such a redefinition that eliminates the projective phase entirely
\begin{align}
    \exists e^{\rmi\Tilde{\phi}(g)} \in U(1),\ \text{$\forall g_1$, $g_2 \in G$} \qquad e^{\rmi\phi(g_1, g_2) - \rmi\Tilde{\phi}(g_1) - \rmi\Tilde{\phi}(g_2)+\rmi\Tilde{\phi}(g_1g_2)} = 1.
\end{align}
Conversely, a \textit{nontrivial} projective representation is one for which no such a redefinition can trivialize the projective phase.

Let us now consider the case where a nontrivial projective representation exists. It is known that there necessarily exist elements $g_1$, $g_2 \in G$ such that
\begin{align}
    e^{\rmi\phi(g_1, g_2)} \neq e^{\rmi\phi(g_2, g_1)}.
\end{align}
In this situation, we introduce simultaneous eigenstates of the Hamiltonian $H$ and a symmetry generator $U_{g_1}$
\begin{align}
    H \ket{E_i, \alpha} &= E_i \ket{E_i, \alpha}, & U_{g_1} \ket{E_i, \alpha} &= e^{\rmi\alpha} \ket{E_i, \alpha}.
\end{align}
Then, it follows that the energy eigenstates must be degenerate. This is demonstrated using Eq.~\eqref{noncommute projective} and the fact that $U_{g_1}$ is a unitary operator, namely
\begin{align}
    \bra{E_i, \alpha} U_{g_2} \ket{E_i, \alpha}
    &= \bra{E_i, \alpha} (U_{g_1})^\dagger U_{g_2} U_{g_1} \ket{E_i, \alpha} \no \\
    &= e^{\rmi\phi(g_1, g_2) - \rmi\phi(g_2, g_1)} \bra{E_i, \alpha} U_{g_2} \ket{E_i, \alpha},
\end{align}
and therefore,
\begin{align}
    \bra{E_i, \alpha} U_{g_2} \ket{E_i, \alpha} = 0.
\end{align}
This implies that $U_{g_2} \ket{E_i, \alpha}$ is orthogonal to $\ket{E_i, \alpha}$.
Together with the fact that $U_2\ket{E_i,\alpha}$ is also the eigenstate of $H$ with eigenvalue $E_i$, we conclude that the energy eigenstate is necessarily degenerate.

\subsection{$\bbZ_{N}\times\bbZ_{N}$ symmetry}

We briefly review projective representations of discrete groups in this section.
For simplicity, we consider a Hamiltonian system with a $\bbZ_{N_1}\times\bbZ_{N_2}$ symmetry, where we set~$N_1=N_2=N$ in the following sections. For a more general case that $N_1\neq N_2$, see Appendix~\ref{N_1, N_2}. In the operator formalism, we introduce two symmetry operators $U_1$ and $U_2$ that correspond to the generators of the $\bbZ_{N_1}$ and $\bbZ_{N_2}$ subgroups, respectively, and then they satisfy the following properties:
\begin{align}
    &(U_1)^{N_1}=1,& &(U_2)^{N_2}=1,\label{rep-of-Z_N}\\
    &[(U_1)^\alpha, (U_1)^\beta]=0,& &[(U_2)^\alpha, (U_2)^\beta]=0,\label{anomaly-free}
\end{align}
for any $\alpha$, $\beta\in\bbZ$.
The first line indicates that $U_1$ and $U_2$ represent the~$\bbZ_{N_1}$ and $\bbZ_{N_2}$ symmetries, respectively.  
The second line expresses that the elements of the~$\bbZ_N$ subgroups commute with each other.

We suppose that the generators $U_1$ and $U_2$ also satisfy 
\begin{align}
    U_2U_1=e^{-\frac{2\pi \rmi}{N}}U_1U_2,\label{Projective-representation}
\end{align}
which represents a minimal nontrivial projective phase between the~$\bbZ_{N_1}$ and $\bbZ_{N_2}$. We note that the minimal nontrivial projective phase is consistent with the property~\eqref{rep-of-Z_N} and corresponds to the mixed 't~Hooft anomaly between the two $\bbZ_{N}$'s.

Next, we construct a basis of the Hilbert space using the eigenstates of the Hamiltonian and the symmetry operators. Since the symmetry operators commute with the Hamiltonian~$H$, they can be simultaneously diagonalized. Let $\kket{E_i}$ be an eigenstate of the Hamiltonian corresponding to the $i$th eigenvalue that is neutral under the action of $U_1$. That is,
\begin{align}
    H\kket{E_i}&=E_i\kket{E_i},& U_1\kket{E_i}&=\kket{E_i}.\label{neutral-state}
\end{align}
Then, using Eq.~\eqref{Projective-representation}, we find that,
for $\alpha=0$, $1$, \dots, $N-1$,
\begin{align}
    U_1(U_2)^\alpha\kket{E_i}&=e^{\frac{2\pi \rmi}{N}\alpha}(U_2)^\alpha\kket{E_i},& H(U_2)^\alpha\kket{E_i}&=E_i(U_2)^\alpha\kket{E_i}.
\end{align}
Therefore, without changing the energy, one can shift the eigenvalue of $U_1$ and generate a degenerate subspace arising from the projective representation.\footnote{Under the assumption~\eqref{Projective-representation}, the action of $U_2$ can shift the eigenvalue of $U_1$ arbitrarily. As a result, energy eigenstates are not classified into distinct symmetry sectors as they would be in a case with trivial projective phases---an anomaly-free case.} In this paper, we consider only degeneracy caused by projective representations. We can thus conclude that the total Hilbert space $\cH$ without any other symmetry is spanned by the eigenstates $(U_2)^\alpha\kket{E_i}$, and that a general state can be expressed as
\begin{align}
    \ket{\psi}=\sum_{i}\sum_{\alpha=0}^{N-1}z_{i,\alpha}(U_2)^\alpha\kket{E_i}\quad \text{with $\sum_{i}\sum_{\alpha=0}^{N-1}|z_{i,\alpha}|^2=1$, $z_{i,\alpha}\in\mathbb{C}$}.
\end{align}

\magenta{To summarize, the double-ket $\kket{E_i}$ denotes a
\textit{representative} of an $N$-fold degenerate energy multiplet induced by the
projective symmetry. Concretely, we choose $\kket{E_i}$ to be an energy eigenstate satisfying Eq.~\eqref{neutral-state} and generate the full degenerate multiplet by acting with $U_2$.
Due to the projective relation $U_2U_1=e^{-2\pi i/N}U_1U_2$, these states carry distinct $U_1$-eigenvalues while sharing the same energy $E_i$.
In the following, the symbol $\kket{\cdot}$ is \emph{only} used to indicate the chosen representative within each degenerate multiplet.}

\section{Eigenstate thermalization hypothesis and projective representation}
\label{Eigenstate thermalization hypothesis and projective representation}

In this section, we formulate a modified version of the Eigenstate Thermalization Hypothesis (ETH) in the presence of projective representations, i.e., prETH. 
We begin by considering how the long-time average of operator expectation values is represented through matrix elements of operators. In particular, we focus on charged operators~$\cO^{q_1,q_2}$ that satisfy
\begin{align}
    U_1^\dagger\cO^{q_1,q_2}U_1 &= e^{\frac{2\pi \rmi}{N}q_1} \cO^{q_1,q_2},&
    U_2^\dagger\cO^{q_1,q_2}U_2 &= e^{\frac{2\pi \rmi}{N}q_2} \cO^{q_1,q_2},\label{Charged-operator}
\end{align}
and their time-averaged expectation values given in Section~\ref{subsection:Time-averaged expectation value}.
In general, any operator acting on the Hilbert space $\cH$ can be expressed as a linear combination of such charged operators.\footnote{Let $\Tilde{O}$ be a generic operator including charged operators. We define
\begin{align}
    \Tilde{\cO}^{q_1,q_2} := \frac{1}{N^2} \sum_{\alpha=0}^{N-1} \sum_{\beta=0}^{N-1} e^{-\frac{2\pi \rmi}{N}(q_1\alpha + q_2\beta)} (U_2)^{-\beta}(U_1)^{-\alpha} \Tilde{O} (U_1)^{\alpha}(U_2)^{\beta},\label{Fourier-trans-for-Z_N}
\end{align}
and then $\Tilde{\cO}^{q_1,q_2}$ satisfies Eq.~\eqref{Charged-operator}, and is thus a charged operator. Inverting this relation, we have
\begin{align}
    \Tilde{O} = \sum_{q_1=0}^{N-1} \sum_{q_2=0}^{N-1} \Tilde{\cO}^{q_1,q_2}.
\end{align}
This relation demonstrates that any operator can be decomposed into a linear combination of charged operators.} Therefore, we lose no generality by restricting our attention to these charged operators.
In Section~\ref{Details of eigenstate thermalization hypothesis}, we then formulate the ETH in our setting, evaluate the time-averaged expectation values for charged operators, and show that the values become independent of the details of the initial state. 


\subsection{Time-averaged expectation value}\label{subsection:Time-averaged expectation value}

We consider the initial state
\begin{align}
    \ket{\psiin} := \sum_{i}\sum_{\alpha=0}^{N-1}c_{i,\alpha}(U_2)^\alpha\kket{E_i}\label{def-initial-state}
\end{align}
that has the following properties:
\begin{align}
    &\bra{\psiin}H\ket{\psiin} =: \Bar{E},\label{Energy}\\
    &(\Delta E)^2:=\bra{\psiin}(H - \Bar{E})^2\ket{\psiin} =  \order{V},\label{Var-energy}
\end{align}
where $V$ denotes the volume of the system. For convenience in later discussions, we rewrite these conditions using the energy density $\varepsilon := H/V$ as follows:
\begin{align}
    &\bra{\psiin}\varepsilon\ket{\psiin} =\Bar{\varepsilon}:=\Bar{E}/V,\label{Energy-density}\\
    &\bra{\psiin}\left(\varepsilon-\Bar{\varepsilon}\right)^2\ket{\psiin}= \order{V^{-1}}.\label{Var-energy-density}
\end{align}
These conditions imply that the fluctuation in energy density is macroscopically negligible.

We now consider the long-time average of the expectation value of a charged operator:
\begin{align}
    \overline{\bra{\psiin}e^{\rmi Ht}\cO^{q_1, q_2}e^{-\rmi Ht}\ket{\psiin}}
    :=\lim_{T\to\infty}\frac{1}{T}\int_0^{T}\rmd t\bra{\psiin} e^{\rmi Ht}\cO^{q_1,q_2}e^{-\rmi Ht}\ket{\psiin}.\label{Time-averaged-expectation-value}
\end{align}
Substituting Eq.~\eqref{def-initial-state} into the long-time average of the expectation value and computing explicitly, we obtain
\begin{align}
    &\lim_{T\to\infty}\frac{1}{T}\int_0^{T}\rmd t\,\bra{\psiin} e^{\rmi Ht}\cO^{q_1,q_2}e^{-\rmi Ht}\ket{\psiin}\no\\
    &=\lim_{T\to\infty}\frac{1}{T}\int_0^{T}\rmd t \sum_{i,j}\sum_{\alpha,\beta}c_{i,\alpha}^*c_{j,\beta} e^{\rmi(E_i-E_j)t}\bbra{E_i}(U_2)^{-\alpha} \cO^{q_1,q_2}(U_2)^\beta\kket{E_j}\no\\    &=\sum_{i}\sum_{\alpha,\beta}c_{i,\alpha}^* c_{i,\beta}\bbra{E_i}(U_2)^{-\alpha}\cO^{q_1, q_2}(U_2)^\beta\kket{E_i},
\end{align}
where we have used the identity $\lim_{T \to \infty}\frac{1}{T}\int_0^{T}\rmd t\, e^{\rmi(E_i - E_j)t} = \delta_{ij}$ obtained from the assumption that there is no degeneracies aside from the ones due to the projective representation.
Since $U_1\kket{E_i} = \kket{E_i}$ and the relations~\eqref{Projective-representation} and \eqref{Charged-operator} hold, we obtain
\begin{align}
    \bbra{E_i}(U_2)^{-\alpha} \cO^{q_1, q_2}(U_2)^\beta \kket{E_i}
    = e^{\frac{2\pi \rmi}{N}(\alpha + q_1 - \beta)} \bbra{E_i}(U_2)^{-\alpha} \cO^{q_1, q_2}(U_2)^\beta \kket{E_i}.\label{selection via projevtive phase}
\end{align}
This leads to a selection rule: the above matrix element vanishes unless $\beta=\alpha+q_1\bmod N$. Therefore,
\begin{align}
    &\sum_{i}\sum_{\alpha,\beta} c_{i,\alpha}^* c_{i,\beta} \bbra{E_i}(U_2)^{-\alpha} \cO^{q_1, q_2} (U_2)^\beta \kket{E_i} \no\\
   &= \sum_{i}\left(\sum_{\alpha=0}^{N-q_1-1}c_{i,\alpha}^* c_{i,\alpha+q_1} e^{\frac{2\pi \rmi}{N}q_2 \alpha}+\sum_{\alpha=N-q_1}^{N-1}c_{i,\alpha}^* c_{i,\alpha+q_1-N} e^{\frac{2\pi \rmi}{N}q_2 \alpha}\right)\no\\
   &\qquad\qquad
   \times\bbra{E_i} \cO^{q_1, q_2}(U_2)^{q_1}(U_1)^{-q_2} \kket{E_i}.\label{Time-averaged-expectation-value after selection}
\end{align}
Here, we have explicitly included the factor $(U_1)^{-q_2}$ using the identity $\kket{E_i} = (U_1)^{-q_2} \kket{E_i}$, in order to emphasize that the $\cO^{q_1, q_2}(U_2)^{q_1}(U_1)^{-q_2}$ is neutral under the action of both $U_1$ and $U_2$. We note that the form~$\bbra{E_i} \cO^{q_1, q_2}(U_2)^{q_1}(U_1)^{-q_2} \kket{E_i}$ is independent of the basis within the degenerate subspace.\footnote{\magenta{
Let us consider a general state in the degenerate subspace,
\begin{align*}
\kket{\widetilde{E_i}}
:=\sum_{\alpha=0}^{N-1} b_{i,\alpha}(U_2)^{\alpha}\kket{E_i},
\qquad \text{with}\qquad
\sum_{\alpha=0}^{N-1}\bigl|b_{i,\alpha}\bigr|^2=1.
\end{align*}
By an explicit calculation, one finds that the expectation value
$\bbra{E_i}\cO^{q_1,q_2}(U_2)^{q_1}(U_1)^{-q_2}\kket{E_i}$
is independent of the choice of basis within the degenerate subspace:
\begin{align}
\bbra{\widetilde{E_i}}\cO^{q_1,q_2}(U_2)^{q_1}(U_1)^{-q_2}\kket{\widetilde{E_i}}
&=\left(\sum_{\alpha=0}^{N-1}|b_{i,\alpha}|^2\right)
\bbra{E_i}\cO^{q_1,q_2}(U_2)^{q_1}(U_1)^{-q_2}\kket{E_i}\notag\\
&=\bbra{E_i}\cO^{q_1,q_2}(U_2)^{q_1}(U_1)^{-q_2}\kket{E_i}.
\end{align}
}}

To rewrite the $\alpha$-dependent coefficient in Eq.~\eqref{Time-averaged-expectation-value after selection}, we introduce the projection operator $P_i$ onto the degenerate eigenspace corresponding to the energy eigenvalue $E_i$. We can substitute $(U_1)^{q_2}(U_2)^{-q_1}P_i$ for $\cO^{q_1,q_2}$, so that we obtain
\begin{align}
&\lim_{T\to\infty}\frac{1}{T}\int_0^{T}\rmd t\,\bra{\psiin} e^{\rmi Ht}(U_1)^{q_2}(U_2)^{-q_1}P_ie^{-\rmi Ht}\ket{\psiin}\no\\
    &=\bra{\psiin} (U_1)^{q_2} (U_2)^{-q_1} P_i \ket{\psiin}\no\\
    &=\sum_{\alpha=0}^{N-q_1-1}c_{i,\alpha}^* c_{i,\alpha+q_1} e^{\frac{2\pi \rmi}{N}q_2 \alpha}+\sum_{\alpha=N-q_1}^{N-1}c_{i,\alpha}^* c_{i,\alpha+q_1-N} e^{\frac{2\pi \rmi}{N}q_2 \alpha}.
\end{align}
Hence, we finally arrive at
\begin{align}
    &\overline{\bra{\psiin} e^{\rmi Ht}\cO^{q_1, q_2}e^{-\rmi Ht}\ket{\psiin}} \no\\
    &=\sum_i \bra{\psiin} (U_1)^{q_2} (U_2)^{-q_1} P_i \ket{\psiin} 
    \bbra{E_i} \cO^{q_1, q_2}(U_2)^{q_1}(U_1)^{-q_2} \kket{E_i}.\label{Time-average}
\end{align}


\subsection{Eigenstate Thermalization Hypothesis for projective representations}\label{Details of eigenstate thermalization hypothesis}

We now propose the (finite-size generalization of) ETH for general charged operators $\cO^{q_1,q_2}$
as a sufficient condition of thermalization.
Motivated by Eq.~\eqref{Srednikiconj}, the explicit form of the projective-representation ETH (prETH) is given by 
\begin{align}
    \bbra{E_i} \cO^{q_1,q_2}(U_2)^{q_1} (U_1)^{-q_2}\kket{E_j}
    =
    \scrO^{(q_1,q_2)}(\cE/V)\delta_{ij}+
    e^{-S(\cE)/2}f^{(q_1,q_2)}(\cE/V,\omega)R_{ij},
    \label{projective-ETH}
\end{align}
where we have introduced the averaged energy $\cE := (E_i + E_j)/2$, energy difference $\omega := E_i - E_j$, and the thermodynamic entropy $S(\cE)=\order{V}$. The functions $\scrO^{(q_1,q_2)}(\cE/V)$ and $f^{(q_1,q_2)}(\cE/V,\omega)$ are supposed to be smooth.
Each element of the matrix $R_{ij}$ is quasi-random and of order $\order{1}$. The suppression of the off-diagonal term 
\begin{align}
e^{-S(\cE)/2} f^{(q_1,q_2)}(\cE/V,\omega) R_{ij} 
= e^{-\order{V}/2}
\end{align}
ensures the small temporal fluctuations of the expectation value of 
$\cO^{q_1,q_2}$ under time evolution around the stationary state~(see appendix~\ref{Justification of Stationarization}). In other words, the prETH~\eqref{projective-ETH} provides a justification for the existence of the stationary state~\eqref{stationary} in the definition of thermalization.

\magenta{{We stress that Eq.~\eqref{projective-ETH} is \textit{not} a conventional ETH ansatz for the diagonal matrix element
$\langle E_i,\alpha|\cO^{q_1,q_2}|E_i,\alpha\rangle$ of the charged operator itself.}
Rather, it is an ansatz for the \textit{symmetry-neutral ``dressed'' operator} $\cO^{q_1,q_2}\,(U_2)^{q_1}(U_1)^{-q_2}$, whose diagonal expectation value within each degenerate multiplet is basis-independent and well-defined.
In particular, the ``smooth dependence on energy'' refers to this multiplet-invariant diagonal quantity, which naturally appears in Eq.~\eqref{Time-average}.
}

We now examine the prETH~\eqref{projective-ETH} in concrete settings, beginning with introducing different classes of operators. Depending on the structure of the charge and the thermodynamic limit, charged operators $\cO^{q_1,q_2}$ can be classified into the following categories:
\begin{itemize}
    \item \textbf{Neutral operators}: these are operators with $(q_1,q_2)=(0,0)$ and are invariant under both $U_1$ and $U_2$. In this case, $\cO^{0,0}$ directly coincides with the standard ETH formulation. The matrix elements $\bbra{E_i} \cO^{0,0} \kket{E_j}$ take the familiar ETH form with a thermal expectation value on the diagonal and random fluctuations on the off-diagonal parts, i.e., Eq.~\eqref{Srednikiconj}.

    \item \textbf{Type I charged operators}: these are charged operators whose diagonal terms  vanish in the thermodynamic limit, namely,
    \begin{align}
    \bbra{E_i} \cO^{q_1, q_2}(U_2)^{q_1}(U_1)^{-q_2}\kket{E_i}=\lorder{V^0}.\label{Conjecture-matrix-element-zero}
\end{align}
In terms of the prETH ansatz~\eqref{projective-ETH}, this statement corresponds to the condition
\begin{align}
    \scrO^{(q_1,q_2)}(\cE/V)=\lorder{V^0}.
\end{align}

    We conjecture that local operators $\cO^{q_1,q_2}$ whose supports are negligible compared to that of the symmetry operator $(U_1)^{q_2}(U_2)^{-q_1}$ in the thermodynamic limit belong to this class.  Let us state a naive explanation for this conjecture. In the thermodynamic limit, the operator~$(U_2)^{q_1}(U_1)^{-q_2}$ generally induces a change in a large fraction of quantum states (in terms of the support where the operator acts), whereas the operator~$\cO^{q_1,q_2}$ induces only a local change. Therefore, one may conjecture that the state~$\cO^{q_1,q_2}(U_2)^{q_1}(U_1)^{-q_2}\kket{E_i}$  significantly differs from the state~$\kket{E_i}$, making their overlap small. Later, we numerically verify this conjecture for specific settings.

\item \textbf{Type II charged operators}: these are charged operators whose diagonal terms do not vanish in the thermodynamic limit, i.e.,
    \begin{align}
    \bbra{E_i} \cO^{q_1, q_2}(U_2)^{q_1}(U_1)^{-q_2}\kket{E_i}=\order{1}>\lorder{V^0}.
\end{align}
If we assume that they satisfy the prETH~\eqref{projective-ETH}, the smooth function~$\scrO^{(q_1,q_2)}(\cE/V)$ takes a non-vanishing value. 
Later, we discuss that the generalized Gibbs ensemble, instead of the standard Gibbs ensemble, correctly describes the stationary state for this class of operators under the assumption of the prETH. 
In addition, these operators may exhibit either conventional or anomalous finite-size scaling; we will discuss it in Section~\ref{Anomalous scaling}.

Given the conjecture stated in Type I charged operator, we argue that such operators are generally non-local. Important examples for Type II charged operators are given as follows:
   \begin{align}
        \cO^{q_1,q_2} &= \cO^{0,0}\, (U_1)^{q_2}(U_2)^{-q_1}.\label{Type II neutral times symmetry}
        \end{align}
    Here, $\cO^{0,0}$ is a local neutral operator whose expectation values with respect to the eigenstates do not vanish, which means
    \begin{align}
    \scrO^{(q_1,q_2)}(\cE/V) &=\scrO^{(0,0)}(\cE/V)\neq 0.
    \end{align}
     The charge of $\cO^{0,0}\, (U_1)^{q_2}(U_2)^{-q_1}$ is entirely supplied by the symmetry operators $U_1$ and $U_2$.


\end{itemize}
In what follows, provided that $||\cO^{q_1,q_2}||=\order{V^0}$, we analyze each case in turn and clarify the role of symmetry, projective representations, and degeneracies in determining the stationary state.

\paragraph{Neutral operators}
In the case of neutral operators, the situation reduces to that of the conventional ETH without degeneracy~\cite{d2016quantum}, and we can proceed in the same manner with it. From the prETH~\eqref{projective-ETH}, the long-time average of the expectation value of the neutral operator~$\cO^{0,0}$ can be evaluated as follows:
\begin{align}
    \overline{\bra{\psiin} e^{\rmi Ht}\cO^{0,0} e^{-\rmi Ht}\ket{\psiin}} 
    &:= \sum_i \bra{\psiin}P_i\ket{\psiin} \bbra{E_i} \cO^{0,0}  \kket{E_i} \no \\
    &= \sum_i \bra{\psiin}P_i\ket{\psiin} \scrO^{0,0}(\varepsilon_i) + e^{-\order{V}/2} \no \\
    &= \scrO^{0,0}(\Bar{\varepsilon}) + \sum_{n\geq 2} (\scrO^{0,0})^{(n)}(\Bar{\varepsilon}) \sum_i \bra{\psiin}P_i\ket{\psiin} (\varepsilon_i - \Bar{\varepsilon})^n + e^{-\order{V}/2}, \label{Time-averaged-exp-ETH}
\end{align}
where in the second line, we have used Eq.~\eqref{projective-ETH}, and in the third line, the smoothness of the function $\scrO^{0,0}(\varepsilon_i)$ is assumed to use a Taylor expansion around the expectation value~\eqref{Var-energy-density}. Additionally, we have used $\sum_i \bra{\psiin}P_i\ket{\psiin}(\varepsilon_i - \Bar{\varepsilon}) = 0$ to obtain the third line.

For $n \geq 2$, we note that
\begin{align}
&\left|\sum_i \bra{\psiin}P_i\ket{\psiin}(\varepsilon_i - \Bar{\varepsilon})^n\right|\no\\
&\leq \sum_i \bra{\psiin}P_i\ket{\psiin} |\varepsilon_i - \Bar{\varepsilon}|^n\no\\
&=\sum_{i:|\varepsilon_i-\Bar{\varepsilon}|\leq\Delta_n} \bra{\psiin}P_i\ket{\psiin} |\varepsilon_i - \Bar{\varepsilon}|^n+\sum_{i:|\varepsilon_i-\Bar{\varepsilon}|>\Delta_n} \bra{\psiin}P_i\ket{\psiin} |\varepsilon_i - \Bar{\varepsilon}|^n.\label{decomposition taylor expansion}
\end{align}
Here, let $\Delta_n$ be defined as the smallest width of energy density that satisfies
\begin{align}
    \sum_{i:|\varepsilon_i-\Bar{\varepsilon}|\leq\Delta_n} \bra{\psiin}P_i\ket{\psiin} |\varepsilon_i - \Bar{\varepsilon}|^n
    > \sum_{i:|\varepsilon_i-\Bar{\varepsilon}|>\Delta_n} \bra{\psiin}P_i\ket{\psiin} |\varepsilon_i - \Bar{\varepsilon}|^n.
    \label{def: Delta n}
\end{align}
Assuming that at least $\Delta_n=\order{1}$,\footnote{This assumption is plausible if the distribution 
$\bra{\psiin}P_i\ket{\psiin}$ is sharply localized around the expectation value 
$\Bar{\varepsilon}$. At least, it is completely justified when the initial state belongs to the microcanonical shell
\begin{align}
  \cH_{E,\Delta E}&=\text{span}\{(U_2)^\alpha\kket{E_i}: \, |E_i-E|\leq\Delta E,\;\alpha=0,\dots,N-1\}\subset\cH,\label{mc subspace}\\
  E&=\order{V},\qquad \Delta E=\lorder{V^0}.\label{bound energy shell}
\end{align}}
we obtain the following bound:
\begin{align}
\sum_{i:|\varepsilon_i-\Bar{\varepsilon}|\leq\Delta_n} \bra{\psiin}P_i\ket{\psiin} |\varepsilon_i - \Bar{\varepsilon}|^n
&\leq (\Delta_n)^{n-2}\sum_i \bra{\psiin}P_i\ket{\psiin} |\varepsilon_i-\Bar{\varepsilon}|^2 \no \\
&\leq \order{V^{-1}},\label{bound using Delta n}
\end{align}
where, in deriving the last inequality, we have used Eq.~\eqref{Var-energy-density}.
Therefore, combining Eqs.~\eqref{decomposition taylor expansion}, \eqref{def: Delta n}, and \eqref{bound using Delta n}, we get
\begin{align}
\left|\sum_i \bra{\psiin}P_i\ket{\psiin}(\varepsilon_i - \Bar{\varepsilon})^n\right|
\leq \order{V^{-1}}.
\label{n-geq-2-inequality}
\end{align}
Therefore, we finally obtain
\begin{align}
    \overline{\bra{\psiin} e^{\rmi Ht}\cO^{0,0} e^{-\rmi Ht}\ket{\psiin}} 
    = \scrO^{0,0}(\Bar{\varepsilon}) + \order{V^{-1}}, \label{time-average-neutral-op-scaling}
\end{align}
which is independent of initial states.

\paragraph{Type I charged operators}
Next, we discuss the thermalization of Type I charged operators.
In this case, as already stated above, we conjecture that the matrix elements of $\cO^{q_1, q_2}$ satisfy the following relation:
\begin{align}
    \bbra{E_i} \cO^{q_1, q_2}(U_2)^{q_1}(U_1)^{-q_2}\kket{E_i}=\lorder{V^0}.
\end{align}
That is, we conjecture $\bbra{E_i} \cO^{q_1, q_2}(U_2)^{q_1}(U_1)^{-q_2}\kket{E_i} \to 0$ in the thermodynamic limit $V\to\infty$. Thus, we obtain
\begin{align}
    &\overline{\bra{\psiin} e^{\rmi Ht}\cO^{q_1, q_2}e^{-\rmi Ht}\ket{\psiin}} \no\\
    &:=\sum_i\bra{\psiin}(U_1)^{q_2}(U_2)^{-q_1}P_i\ket{\psiin}\bbra{E_i} \cO^{q_1, q_2}(U_2)^{q_1}(U_1)^{-q_2}\kket{E_i}
    \to0\label{type I becomes 0}
\end{align}
in the thermodynamic limit.

\paragraph{Type II charged operators}
Finally, let us consider Type II charged operators that have a non-zero diagonal term in the thermodynamic limit.

In subsection~\eqref{subsection:Time-averaged expectation value}, we obtained the time-averaged expectation value as
\begin{align}
    \overline{\bra{\psiin} e^{\rmi Ht}\cO^{q_1, q_2}e^{-\rmi Ht}\ket{\psiin}}
    := \sum_i \bra{\psiin}(U_1)^{q_2}(U_2)^{-q_1} P_i\ket{\psiin}
    \bbra{E_i}\cO^{q_1, q_2}\kket{E_i}.
    \label{Time-average-charged-op-with-symmetry-op}
\end{align}
Using the prETH~\eqref{projective-ETH} and a Taylor expansion, this becomes
\begin{align}
    &\sum_i \bra{\psiin}(U_1)^{q_2}(U_2)^{-q_1} P_i\ket{\psiin}
    (\scrO^{q_1,q_2}(\varepsilon_i) + e^{-\order{V}/2}) \no\\
    &= \bra{\psiin}(U_1)^{q_2}(U_2)^{-q_1}\ket{\psiin}\scrO^{q_1,q_2}(\Bar{\varepsilon})
    + \sum_i \bra{\psiin}(U_1)^{q_2}(U_2)^{-q_1}P_i\ket{\psiin}
    (\scrO^{q_1,q_2})'(\Bar{\varepsilon})(\varepsilon_i-\Bar{\varepsilon}) \no\\
    &\qquad
    + \sum_{n \geq 2} (\scrO^{q_1,q_2})^{(n)}(\Bar{\varepsilon})
    \sum_i \bra{\psiin}(U_1)^{q_2}(U_2)^{-q_1}P_i\ket{\psiin}
    (\varepsilon_i-\Bar{\varepsilon})^n
    + e^{-\order{V}/2}.
    \label{Time-averaged-charged-op-with-symmetry-op-and-ETH}
\end{align}
The second term on the right-hand side can be estimated under the Jensen inequality and the bound 
$|\bra{\psiin}(U_1)^{q_2}(U_2)^{-q_1}P_i\ket{\psiin}| \leq \bra{\psiin}P_i\ket{\psiin}$:
\begin{align}
    \left| \sum_i \bra{\psiin}(U_1)^{q_2}(U_2)^{-q_1}P_i\ket{\psiin}(\varepsilon_i - \Bar{\varepsilon}) \right|
    &\leq \left( \sum_i \bra{\psiin}P_i\ket{\psiin}(\varepsilon_i - \Bar{\varepsilon})^2 \right)^{1/2} \no\\
    &\leq \order{V^{-1/2}}.
    \label{Bound-used-jansen}
\end{align}
The third term on the right-hand side in Eq.~\eqref{Time-averaged-charged-op-with-symmetry-op-and-ETH} can be estimated by using Eq.~\eqref{n-geq-2-inequality} as
\begin{align}
    \left| \sum_i \bra{\psiin}(U_1)^{q_2}(U_2)^{-q_1}P_i\ket{\psiin}
    (\varepsilon_i - \Bar{\varepsilon})^n \right|
    \leq\sum_i \bra{\psiin}P_i\ket{\psiin}
    \left|\varepsilon_i - \Bar{\varepsilon}\right|^n 
    \leq \order{V^{-1}}.
    \label{n-geq-2-inequality-with-charged-op}
\end{align}
Hence, using Eqs.~\eqref{Time-averaged-exp-ETH}, \eqref{Bound-used-jansen}, and~\eqref{n-geq-2-inequality-with-charged-op}, we obtain
\begin{align}
    \overline{\bra{\psiin} e^{\rmi Ht}\cO^{q_1,q_2}e^{-\rmi Ht}\ket{\psiin}}
    = \bra{\psiin}(U_1)^{q_2}(U_2)^{-q_1}\ket{\psiin}\scrO^{q_1,q_2}(\Bar{\varepsilon})+\order{V^{-1/2}}.
    \label{time-average-charged-op-scaling}
\end{align}

\subsection{\magenta{Comparison with the commuting (anomaly-free) case}}

\magenta{It is instructive to contrast our setting with the conventional situation where all symmetry generators mutually commute.
If $[U_1,U_2]=[H,U_1]=[H,U_2]=0$, one can choose a common eigenbasis of $H$, $U_1$, and $U_2$, and classify energy eigenstates by fixed symmetry eigenvalues.
In that case, the standard expectation is that ETH applies \textit{within each symmetry sector}, i.e., to matrix elements between states sharing the same set of symmetry eigenvalues.}

\magenta{In our case, although the symmetry group is Abelian, the generators obey a \textit{projective} relation $U_2U_1=e^{-2\pi i/N}U_1U_2$, so $U_1$ and $U_2$ cannot be simultaneously diagonalized.
This invalidates the sector-wise ETH logic based on simultaneous eigenvalues, and enforces an $N$-fold degeneracy associated with each energy $E_i$.
The prETH ansatz~\eqref{projective-ETH} is designed to capture the universal structure of matrix elements \textit{within such symmetry-enforced degenerate multiplets} by focusing on the multiplet-invariant dressed operator $\cO^{q_1,q_2}(U_2)^{q_1}(U_1)^{-q_2}$. We again stress that this dressed operator is physically relevant to the long-time dynamics of $\mathcal{O}^{q_1,q_2}$, as seen from Eq.~\eqref{Time-average}.
}

\section{Generalized Gibbs ensemble for projective representation}\label{sec:gibbs_ensemble}
In this section, we discuss how the stationary values of operators explained in the previous section are effectively described by the statistical ensemble.
We first show that the standard Gibbs ensemble describes the stationary values for neutral and Type I charged operators, under the assumption of the prETH in Eq.~\eqref{projective-ETH}.
However, we also discuss that because of the prETH, the stationary values of Type II charged operators are described by the generalized Gibbs ensemble, instead of the standard Gibbs ensemble.

\subsection{Standard Gibbs ensemble and its breakdown}
Let us first define 
the standard Gibbs ensemble by
\begin{align}
\rhoge := \frac{e^{-\beta H}}{\tr{e^{-\beta H}}}.
\end{align}
Here, the inverse temperature $\beta$ is determined such that
\begin{align}
\tr{\rhoge H} = \Bar{E}.
\end{align}

\paragraph{Neutral operators}
We first evaluate the expectation value of the neutral operator $\cO^{0,0}$ in the Gibbs ensemble as
\begin{align}
\tr{\rhoge\cO^{0,0}} 
&= \sum_{i}\sum_{\alpha=0}^{N-1} \frac{e^{-\beta E_i}}{\tr{e^{-\beta H}}} \bbra{E_i}(U_2)^{-\alpha} \cO^{0,0}(U_2)^\alpha \kket{E_i} \no \\
&= \sum_{i} \frac{e^{-\beta E_i}}{\tr{e^{-\beta H}}/N} \bbra{E_i} \cO^{0,0} \kket{E_i} .
\end{align}
Using the prETH~\eqref{projective-ETH} and a Taylor expansion of $\scrO^{0,0}(\varepsilon_i)$ around $\Bar{\varepsilon}$, we have
\begin{align}
\tr{\rhoge\cO^{0,0}} 
&= \sum_{i} \frac{e^{-\beta E_i}}{\tr{e^{-\beta H}}/N} \left( \scrO^{0,0}(\varepsilon_i) + e^{-\order{V}/2} \right) \no \\
&= \scrO^{0,0}(\Bar{\varepsilon}) + \order{V^{-1}}.
\end{align}
Thus, the prETH~\eqref{projective-ETH} concludes that
\begin{align}
    \overline{\bra{\psiin} e^{\rmi Ht}\cO^{0,0} e^{-\rmi Ht}\ket{\psiin}} 
    = \tr{\rhoge\cO^{0,0}} + \order{V^{-1}}. \label{Thermalization-for-neutral-op}
\end{align}
In the thermodynamic limit $V \to \infty$, thermalization~\eqref{stationary=thermal} of the neutral operator~$\cO^{0,0}$ is therefore justified as
\begin{align}
    \lim_{V\rightarrow\infty}\overline{\bra{\psiin} e^{\rmi Ht}\cO^{0,0} e^{-\rmi Ht}\ket{\psiin}} = \lim_{V\rightarrow\infty}\tr{\rhoge\cO^{0,0}}.
\end{align}

\paragraph{Type I charged operators}
Next, we consider the behavior of charged operators~$\cO^{q_1, q_2}$ in this ensemble.
Noting that $\forall \gamma \in \bbZ$, $(U_1)^\gamma \kket{E_i} = \kket{E_i}$ and its selection rule, we see
\begin{align}
    \tr{\rhoge\cO^{q_1, q_2}} 
    &= \sum_{i,\alpha} \frac{e^{-\beta E_i}}{\tr{e^{-\beta H}}}
    \bbra{E_i} (U_2)^{-\alpha}\cO^{q_1, q_2}(U_2)^\alpha \kket{E_i} \no\\
    &= \sum_{i,\alpha} \frac{1}{N} \sum_{\gamma=0}^{N-1} \frac{e^{-\beta E_i}}{\tr{e^{-\beta H}}}
    e^{\frac{2\pi \rmi}{N}(q_1 \gamma + q_2 \alpha)}
    \bbra{E_i}\cO^{q_1, q_2}\kket{E_i} \no\\
    &\propto \delta_{q_1, 0}\delta_{q_2, 0}.
    \label{Gibbsed-Charged-op-zero}
\end{align}
If $q_1 \neq 0$ or $q_2 \neq 0$, then $\tr{\rhoge\cO^{q_1, q_2}} = 0$. Note that this selection rule holds for general charged operators.

For Type I charged operators, since Eq.~\eqref{type I becomes 0} holds, together with the selection rule~\eqref{Gibbsed-Charged-op-zero}, the following relation can be justified in the thermodynamic limit~$V\to\infty$:
\begin{align}
    \overline{\bra{\psiin} e^{\rmi Ht}\cO^{q_1, q_2}e^{-\rmi Ht}\ket{\psiin}}
    =\tr{\rhoge\cO^{q_1, q_2}}=0.
\end{align}
That is, the standard Gibbs ensemble suffices in this case as well.

\paragraph{Type II charged operators}
Finally, let us consider the behavior of Type II charged operators in this ensemble.
In the case of Type II charged operators, 
$\bra{\psiin}(U_1)^{q_2}(U_2)^{-q_1}\ket{\psiin}$ and 
$\scrO^{q_1,q_2}(\Bar{\varepsilon})$ in Eq.~\eqref{time-average-charged-op-scaling} 
do not vanish in the thermodynamic limit.
In other words, in the thermodynamic limit, the selection rule~\eqref{Gibbsed-Charged-op-zero} 
and Eq.~\eqref{time-average-charged-op-scaling} are not compatible.
Consequently, for non-vanishing $\bra{\psiin}(U_1)^{q_2}(U_2)^{-q_1}\ket{\psiin}$ 
and $\scrO^{q_1,q_2}(\Bar{\varepsilon})$, one finds
\begin{align}
    &\overline{\bra{\psiin} e^{\rmi Ht}\cO^{q_1, q_2}e^{-\rmi Ht}\ket{\psiin}}\neq0,\,\text{and}\, \tr{\rhoge\cO^{q_1, q_2}}=0,\no\\
    &\Longrightarrow\overline{\bra{\psiin} e^{\rmi Ht}\cO^{q_1, q_2}e^{-\rmi Ht}\ket{\psiin}}\neq\tr{\rhoge\cO^{q_1, q_2}}
\end{align}
and hence the Gibbs ensemble fails to describe the stationary value of $\cO^{q_1, q_2}$.
We argue that this observation signals a breakdown of thermalization to the standard Gibbs ensemble~\eqref{stationary=thermal}.

We comment on this observation by focusing on the class of Type II charged operators that can be written as $\cO^{q_1,q_2} = \cO^{0,0}\,(U_1)^{q_2}(U_2)^{-q_1}$.
Note that, as argued in Ref.~\cite{Fukushima:2023svf}, if $U_1$ and $U_2$ correspond to $0$-form symmetries, their supports span the entire system, and this thermalization breakdown may be attributed to the fact that there is only a small ``bath,'' i.e., the complement of the support of $\cO^{0,0}(U_1)^{q_2}(U_2)^{-q_1}$. However, if $U_1$ and $U_2$ correspond to higher-form symmetries, then their supports can be made sufficiently small compared to the ``bath''~(system) in the infinite-volume limit, even though they are nonlocal. In such cases, the thermalization breakdown cannot be attributed to the small bath, but rather reflects a fundamental feature of the symmetry of the Hamiltonian.

In the next subsection, we will demonstrate that stationary values of Type II operators are described by a generalized Gibbs ensemble that incorporates information about non-local and non-commutative conserved quantities, instead of the Gibbs ensemble.

\subsection{Generalized Gibbs Ensemble}\label{Generalized Gibbs Ensemble}
We show that the stationary states of all classes of the operators discussed in Section~\ref{Details of eigenstate thermalization hypothesis}, including Type II operators, can be described by the Generalized Gibbs Ensemble (GGE).
 We propose the following ``non-commutative'' GGE~\cite{Yunger_Halpern_2016}:
\begin{align}
    \rhogge &:= \frac{1}{\cZ} \exp\left(-\beta H - \sum_{r=1}^{N^2-1} \mu_{r} Q^r\right),
    \label{def-GGE} 
    \\
    \cZ &:= \tr\left[\exp\left(-\beta H - \sum_{r=1}^{N^2-1} \mu_{r} Q^r\right)\right],
    \label{Partition-function-GGE} 
\end{align}
where $Q^r$ are linearly independent Hermitian operators constructed as linear combinations of the conserved quantities $(U_1)^{r_1}(U_2)^{r_2}$ for $r_1$, $r_2=0$, $1$, \dots, $N-1$.\footnote{$Q^r$ are not a unique choice; any set of $N^2$ independent and Hermitian conserved quantities would suffice.}

More specifically, one can always choose $N^2-1$ independent operators from the set of the real parts and the imaginary parts, i.e.
\begin{align}\begin{split}
    \Big\{&
    \cR^{r_1,r_2}:=\big[(U_1)^{r_1}(U_2)^{r_2} + \big((U_1)^{r_1}(U_2)^{r_2}\big)^\dagger\big]/2 ,
    \\
    &\cI^{r_1,r_2}:=\big[(U_1)^{r_1}(U_2)^{r_2} - \big((U_1)^{r_1}(U_2)^{r_2}\big)^\dagger\big]/(2i)
     \Big\}_{r_1,r_2=0,\dots,N-1}.
\end{split}\end{align}
Because the operators $\cR^{r_1,r_2}$ and $\cI^{r_1,r_2}$ are subject to the equivalence relation $(r_1,r_2)\sim (N-r_1,N-r_2)$, it suffices to choose one representative from each equivalence class. 
The explicit choice of representatives depends on whether $N$ is odd or even.  
\begin{itemize}
  \item \textbf{Odd $N$:}
  Except for $(0,0)$, each label $(r_1,r_2)$ is paired with $(N-r_1,N-r_2)$. Hence, there are $(N^2-1)/2$ independent labels. As representatives, we take the lexicographically smaller elements in each pair. Since both $\cR$ and $\cI$ are non-vanishing for these labels, the total number of independent operators is 
  $2 \times (N^2-1)/2 = N^2-1$.  

  \item \textbf{Even $N$:}
  In addition to $(0,0)$, the labels $(0,N/2)$, $(N/2,0)$, and $(N/2,N/2)$ are also self-conjugate under the pairing. Excluding these, the treatment is identical to the odd case, where lexicographic representatives are chosen.  
  For the special cases:
  \begin{itemize}
    \item For $(0,N/2)$ and $(N/2,0)$, the operator $\cI$ vanishes, so only $\cR$ remains.
    \item For $(N/2,N/2)$, the operator $\cR$ vanishes if $N/2$ is odd, while $\cI$ vanishes if $N/2$ is even. Thus, we retain the non-vanishing operator in each case.
  \end{itemize}
  Summing up the contributions, the total number of independent operators is $2 \times (N^2-4)/2 + 2 + 1 = N^2-1$.  
\end{itemize}
The Hermitian operators $Q^r$ commute with the Hamiltonian of the system by construction.
The $N^2$ parameters, $\beta$ and $\mu_{r}$, are tuned according to the initial state $\ket{\psiin}$ as
\begin{align}
    \tr{\rhogge H} &= \bra{\psiin}H\ket{\psiin} =\Bar{E},\label{tuning-energy} \\
    \tr{\rhogge(U_1)^{q_2}(U_2)^{-q_1}} &= \bra{\psiin}(U_1)^{q_2}(U_2)^{-q_1}\ket{\psiin} \qquad \forall(q_1,q_2)\in\bbZ_N\times\bbZ_N. \label{tuning-conserved}
\end{align}

Now, the expectation value of a general charged operator $\cO^{q_1, q_2}$ measured with $\rhogge$ \eqref{def-GGE} can be computed as
\begin{align}
    \tr{\rhogge\cO^{q_1, q_2}} 
    &= \sum_{i,\alpha} \bbra{E_i}(U_2)^{-\alpha} \rhogge \cO^{q_1, q_2} (U_2)^\alpha \kket{E_i}
\end{align}
Inserting the completeness relation $\sum_{j,\alpha'}(U_2)^{\alpha'} \kket{E_j} \bbra{E_j} (U_2)^{-\alpha'} = 1$ into this expression and using the selection rule as in Eq.~\eqref{selection via projevtive phase}, we have
\begin{align}
    \tr{\rhogge\cO^{q_1, q_2}}
    &= \sum_{i,\alpha'} \bbra{E_i}(U_2)^{-\alpha'-q_1} \rhogge (U_2)^{\alpha'} \kket{E_i} \bbra{E_i}(U_2)^{-\alpha'} \cO^{q_1, q_2} (U_2)^{\alpha'+q_1} \kket{E_i} \no \\
    &= \sum_{i} \tr{[\rhogge(U_1)^{q_2}(U_2)^{-q_1}P_i]} \bbra{E_i} \cO^{q_1, q_2}(U_2)^{q_1}(U_1)^{-q_2} \kket{E_i}.\label{exp-GGE}
\end{align}
From the equality $\tr{\rhogge(H-\Bar{E})^2}=\tr{\rhoge(H-\Bar{E})^2}=\order{V}$,\footnote{Since the representations of $U_1$ and $U_2$ are identical within each degenerate subspace $P_i$, the trace of  $\exp\left(- \sum_{r=1}^{N^2-1} \mu_{r} Q^r\right)$ in the subspace~$P_i$ is independent of the choice of $i$. That is,
\begin{align}
    \forall\, i,j \quad 
    \tr{P_i \exp\left(- \sum_{r=1}^{N-1} \mu_{r} Q^r\right)}
    = \tr{P_j \exp\left(- \sum_{r=1}^{N-1} \mu_{r} Q^r\right)}.
    \label{energy sector independent}
\end{align}
Therefore, we obtain the following relation between 
$\tr{\rhogge(H-\Bar{E})^2}$ and $\tr{\rhoge(H-\Bar{E})^2}$:
\begin{align}
    \tr{\rhogge(H-\Bar{E})^2} 
    &= \frac{\tr{\rhoge\exp\left(-\sum_{r=1}^{N-1} \mu_{r} Q^r\right) (H-\Bar{E})^2}}
    {\tr{\rhoge\exp\left(-\sum_{r=1}^{N-1} \mu_{r} Q^r\right)}} \no \\
    &= \frac{\sum_i e^{-\beta E_i}(E_i-\Bar{E})^2 
    \tr{P_i \exp\left(- \sum_{r=1}^{N-1} \mu_{r} Q^r\right)}}
    {\sum_i e^{-\beta E_i} \tr{P_i \exp\left(-\sum_{r=1}^{N-1} \mu_{r} Q^r\right)}} \no \\
    &= \frac{\sum_i e^{-\beta E_i}(E_i-\Bar{E})^2}{\sum_i e^{-\beta E_i}}
    = \tr{\rhoge(H-\Bar{E})^2}.
\end{align}
Here, in going from the first to the second line, we have inserted $\sum_i P_i = 1$ inside the trace, and in going from the second to the third line, we have used Eq.~\eqref{energy sector independent}.}
and the prETH in Eq.~\eqref{projective-ETH} and the conjecture~\eqref{Conjecture-matrix-element-zero}, we obtain
\begin{align}
    &\tr{\rhogge \cO^{q_1, q_2}}\no\\
    &= 
    \begin{cases}
        \scrO^{0,0}(\tr\{\rhogge\,\varepsilon\}) + \order{V^{-1}} & \text{for neutral operators}, \\
        \lorder{V^0} & \text{for Type I charged operators},\\
         \tr{\left[\rhogge(U_1)^{q_2}(U_2)^{-q_1}\right]}\scrO^{q_1, q_2}(\tr\{\rhogge\,\varepsilon\})+\order{V^{-\frac{1}{2}}} & \text{for Type II charged operators}.
    \end{cases}\label{GGE-ETH-summary}
\end{align}
Furthermore, noting that Eqs.~\eqref{tuning-energy},~\eqref{tuning-conserved} and \eqref{time-average-charged-op-scaling}, in the thermodynamic limit $V \to \infty$,
\begin{align}
    \lim_{V\to\infty}\tr{\rhogge \cO^{q_1, q_2}} = \lim_{V\to\infty}\overline{\bra{\psiin} e^{\rmi Ht} \cO^{q_1, q_2} e^{-\rmi Ht} \ket{\psiin}}\label{Time is GGE}
\end{align}
is justified. Even for the case of Type II charged operators,  stationary values can be described correctly by the GGE, although the standard Gibbs ensemble fails in general since $\tr(\rhoge\cO^{q_1,q_2})=0$ for every charged operator.

\magenta{
Before ending this section, we stress that our use of nonlocal conserved quantities does \textit{not} mean that we allow arbitrary operators $Q^r$ in the exponent of Eq.~\eqref{def-GGE}. If one were to allow an arbitrary set of nonlocal operators, any density matrix could indeed be written in a ``generalized Gibbs'' form, rendering the construction vacuous.
In contrast, in our setup the charges $\{Q^r\}$ are \textit{fixed by symmetry} and form a finite-dimensional set generated by the projective symmetry operators $\{(U_1)^{r_1}(U_2)^{r_2}\}$ with $r_1$, $r_2\in\{0,1,\dots,N-1\}$, leading to $N^2-1$ independent Hermitian operators (excluding the identity).
Importantly, the number of chemical potentials is therefore $O(1)$ (independent of system size), while the space of all density matrices scales exponentially with system size.
Hence, our GGE is a highly constrained ensemble and cannot represent an arbitrary state.
Physically, these symmetry-determined charges encode the ``projective'' information stored in the symmetry-enforced degeneracy, and are required to reproduce the stationary values of observables with nontrivial overlap with this structure (Type~II).}

\section{Examples}\label{sec:examples}
In this section, we perform numerical analyses to verify the prETH \eqref{projective-ETH} and the GGE \eqref{Time is GGE} in systems where projective representations realize symmetries.
Time evolutions of observables in such systems are also computed numerically, and the resulting stationary values of observables are shown to be well-described by the GGE \eqref{def-GGE}.
To this end, we employ the $(1+1)$-dimensional $\bbZ_N\times\bbZ_N$-symmetric spin chains and the $(2+1)$-dimensional $\bbZ_2$ lattice gauge theory as concrete examples.

We first introduce the notions of the $d$-dimensional space~$T^d$ and its hypercubic discretization~$\Lambda_d$ as
\begin{align}
 T^d &:= \left\{ X \in \mathbb{R}^d \mid 0 \leq X_\mu < L_\mu\ \text{for all $\mu = 1$, \dots, $d$} \right\}, \\
 \Lambda_d &:= \left\{ v \in \bbZ^d \mid 0 \leq v_\mu < L_\mu\ \text{for all $\mu = 1$, \dots, $d$} \right\},
\end{align}
where $L_\mu$ is the lattice size along each direction, and the lattice spacing is set to unity.
The hypercubic lattice $\Lambda_d$ decomposes $T^d$ into $d$-dimensional hypercubes (cells)~$c(v)$ with $v\in\bbZ^d$, defined by
\begin{align}\label{cv}
 c(v) := \left\{ X \in \mathbb{R}^d \mid 0 \leq X_\mu - v_\mu \leq 1\ \text{for all $\mu$} \right\}.
\end{align}
We further define the fundamental lattice elements as follows:
\begin{itemize}
    \item \textbf{Vertices:} $v \in \Lambda_d$, which are equivalent to lattice sites. We also denote $v$ by a tuple of its coordinate as~($x$, $y$, \dots).
    \item \textbf{Links:} 
    \begin{align}
     \ell(v,\mu) := \left\{ X \in c(v) \mid 0 \leq X_\mu - v_\mu \leq 1;\  X_\nu = v_\nu\ \text{for $\nu \neq \mu$} \right\}.
    \end{align}
    These can be interpreted as oriented 1-cells connecting $v$ and $v + \Hat{\mu}$,
    where $\Hat{\mu}$ is a unit vector in the positive $\mu$ direction. 
    In what follows, we also use the notation such that $(v,n\Hat{\mu}) := \ell(v,\mu)\cup\ell(v+\Hat{\mu},\mu)\cup\cdots\cup\ell(v+(n-1)\Hat{\mu},\mu)$.
    \item \textbf{Plaquettes:} 
    \begin{align}
     p(v;\mu,\nu) := \left\{ X \in c(v) \mid 0 \leq X_\mu - v_\mu \leq 1,\ 0 \leq X_\nu - v_\nu \leq 1;\ X_\rho = v_\rho\ \text{for $\rho \neq \mu$, $\nu$} \right\},
    \end{align}
    which correspond to oriented 2-cells (elementary squares) spanning the $\mu$-$\nu$ square with $\mu\neq\nu$.
    \item \textbf{Higher-dimensional cells:} In general, a $k$-cell ($1 \leq k \leq d$) is specified by a tuple of $k$ orthogonal directions, $\mu_1$, $\mu_2$, \dots, and $\mu_k$ ($\mu_i\neq\mu_j$ for $\forall i\neq j$), and a point $v \in \Lambda$, denoted as:
    \begin{align}
     c^{(k)}(v;\mu_1,\dots,\mu_k).
    \end{align}
\end{itemize}

Each type of cells (vertices, links, plaquettes, etc.) serves as the support for corresponding degrees of freedom in lattice theory. In theories with higher-form symmetries, degrees of freedom naturally reside on higher-dimensional cells. For example, a variable, say $\sigma$, lives on a link; then, $\sigma_{(v,-\Hat{\mu})}$ denotes a physical degree of freedom supported on the link that extends from vertex $v$ in the negative $\mu$ direction.

\subsection{$\bbZ_N\times\bbZ_N$ spin chains}\label{ZN ZN spin chains}

We first consider $(1+1)$-dimensional spin chains exhibiting $\bbZ_N \times \bbZ_N$ symmetry.
A projective representation of the $\bbZ_N \times \bbZ_N$ symmetry can be realized on an $N$-dimensional local Hilbert space $\scrH_{\mathrm{loc}}$ spanned by the orthonormal basis states $\{ \ket{g} \}_{g=0}^{N-1}$.
The symmetry generators corresponding to each $\bbZ_N$ factor are represented by the ``clock'' and ``shift'' operators, denoted $Z$ and $X$, respectively.
These operators satisfy the algebraic relation $XZ = \omega ZX$, where $\omega = e^{2\pi \rmi / N}$ is the primitive $N$th root of unity, and both $X^N = Z^N = \bm{1}$~\cite{Alavirad:2019iea}.
They can be explicitly written in matrix form as
\begin{align}
    Z &= \begin{pmatrix}
    1 & 0 & 0 & \cdots & 0 \\
    0 & e^{2\pi \rmi / N} & 0 & \cdots & 0 \\
    0 & 0 & e^{4\pi \rmi / N} & \cdots & 0 \\
    \vdots & \vdots & \vdots & \ddots & \vdots \\
    0 & 0 & 0 & \cdots & e^{2\pi \rmi (N-1) / N}
    \end{pmatrix}, 
    &
    X &= \begin{pmatrix}
    0 & 0 & \cdots & 0 & 1 \\
    1 & 0 & \cdots & 0 & 0 \\
    0 & 1 & \cdots & 0 & 0 \\
    \vdots & \vdots & \ddots & \vdots & \vdots \\
    0 & 0 & \cdots & 1 & 0
    \end{pmatrix},
\end{align}
which act on the local Hilbert space as
\begin{align}
    Z \ket{g} &= e^{2\pi \rmi g / N} \ket{g}, \\
    X \ket{g} &= \ket{(g + 1) \bmod N}.
\end{align}
Note that for $N=2$, the operators $Z$ and $X$ reduce to the standard Pauli matrices.

In order to accommodate a spin chain with $L_1$ sites ($L_1 > N$), we introduce the tensor product Hilbert space 
$\scrH:=(\scrH_{\mathrm{loc}})^{\otimes L_1}$,
along with operators that act only on the $j$th site: $Z_j := \bm{1} \otimes \dots \otimes Z \otimes \dots \otimes \bm{1}$ and $X_j := \bm{1} \otimes \dots \otimes X \otimes \dots \otimes \bm{1}$.
We can then interpret the physical observables as residing on the vertices of the $1$-dimensional lattice.
The symmetry operators are given by
\begin{align}
    U_1 &:= \prod_{j=1}^{L_1} Z_j, &
    U_2 &:= \prod_{j=1}^{L_1} X_j.
\end{align}
We note that the projective phase between $U_1$ and $U_2$ becomes trivial when $\gcd(N, L_1) = N$ (i.e., $L_1$ is divided by $N$). Therefore, we assume $\gcd(N, L_1) \neq N$ so that the discussion in the previous section remains applicable. This projective phase can be interpreted as the $\bbZ_N$ action of $U_1$ on the other symmetry operators $U_2$, namely,\footnote{By considering a central extension of $\bbZ_N \times \bbZ_N$ by $\bbZ_{\lcm(L_1, N)/L_1}$, we can realize this relation as a linear representation, where the center $\bbZ_{\lcm(L_1, N)/L_1}$ is generated by
\begin{align}
    \prod_{j=1}^{L_1} e^{\frac{2\pi \rmi}{N}} = e^{\frac{2\pi \rmi L_1}{N}}.
\end{align}}
\begin{align}
    U_1 U_2 U_1^{-1} = \exp\left(2\pi \rmi\frac{L_1}{N}\right) U_2.
\end{align}
The minimal projective phase \eqref{Projective-representation} can be realized when the lattice size $L_1$ satisfies
\begin{align}
    L_1 \equiv 1 \pmod{N} \qquad (\Rightarrow\quad \lcm(L_1, N)/L_1 = N),
\end{align}
and we focus on this case in the following numerical calculations.

\paragraph{$\bbZ_2\times\bbZ_2$-symmetric spin chain}
For $N = 2$, a $(1+1)$-dimensional spin chain with $\bbZ_2 \times \bbZ_2$ symmetry is described by
\begin{align}
    H_{N=2} = \sum_{j=1}^{L_1} \left(
    J^x_j X_j X_{j+1} + J^y_j Y_j Y_{j+1} + J^z_j Z_j Z_{j+1}
    \right)
    +
    \alpha \sum_{j=1}^{L_1} X_j Y_{j+1} Z_{j+2},
    \label{2-Ham}
\end{align}
where $Y_j := i X_j Z_j$. We impose periodic boundary conditions by identifying $j \sim j + L_1$, which corresponds to a spatial topology of $S^1$. To eliminate unwanted spacetime symmetries, we introduce weak randomness into the couplings $J^x_j$, $J^y_j$, and $J^z_j$. The second term in Eq.~\eqref{2-Ham} is added to break an additional discrete symmetry that flips the sign of certain operators, such as $Y_j \mapsto -Y_j$ for all $j$. When $\alpha=0$, this Hamiltonian reduces to the well-known XYZ Heisenberg spin chain.

For the system given by Eq.~\eqref{2-Ham}, the GGE \eqref{def-GGE} reads
\begin{align}
    \rhogge&=
     \frac{1}{\cZ}
    \exp\left(-\beta H_{N=2} -\mu_{1}\cR^{0,1}-\mu_{2}\cR^{1,0}-\mu_{3}\cI^{1,1}\right),
    \\
    \cR^{0,1}&= U_2, \quad
    \cR^{1,0}=U_1, \quad
    \cI^{1,1}= U_1U_2/i.
\end{align}
Note here that the operators $\cR^{0,1}$, $\cR^{1,0}$, and $\cI^{1,1}$ are independent nonlocal conserved quantities.
As discussed in Eqs.~\eqref{tuning-energy} and \eqref{tuning-conserved}, the chemical potentials $\beta$ and $\mu_{r}$ should be set for a given initial state $\ket{\psiin}$ so that the density matrix $\rhogge$ satisfies
\begin{align}\begin{split}
    \tr\rhogge H&=\bra{\psiin}H\ket{\psiin},
    \\
    \tr\rhogge U_1&=\bra{\psiin}U_1\ket{\psiin},
    \qquad
    \tr\rhogge U_2=\bra{\psiin}U_2\ket{\psiin},
    \\
    \tr\rhogge U_1U_2&=\bra{\psiin}U_1U_2\ket{\psiin},
\end{split}
\end{align}

\begin{figure}[t]
\centering
\begin{minipage}{0.328\columnwidth}
\centering
\includegraphics[width=\columnwidth]{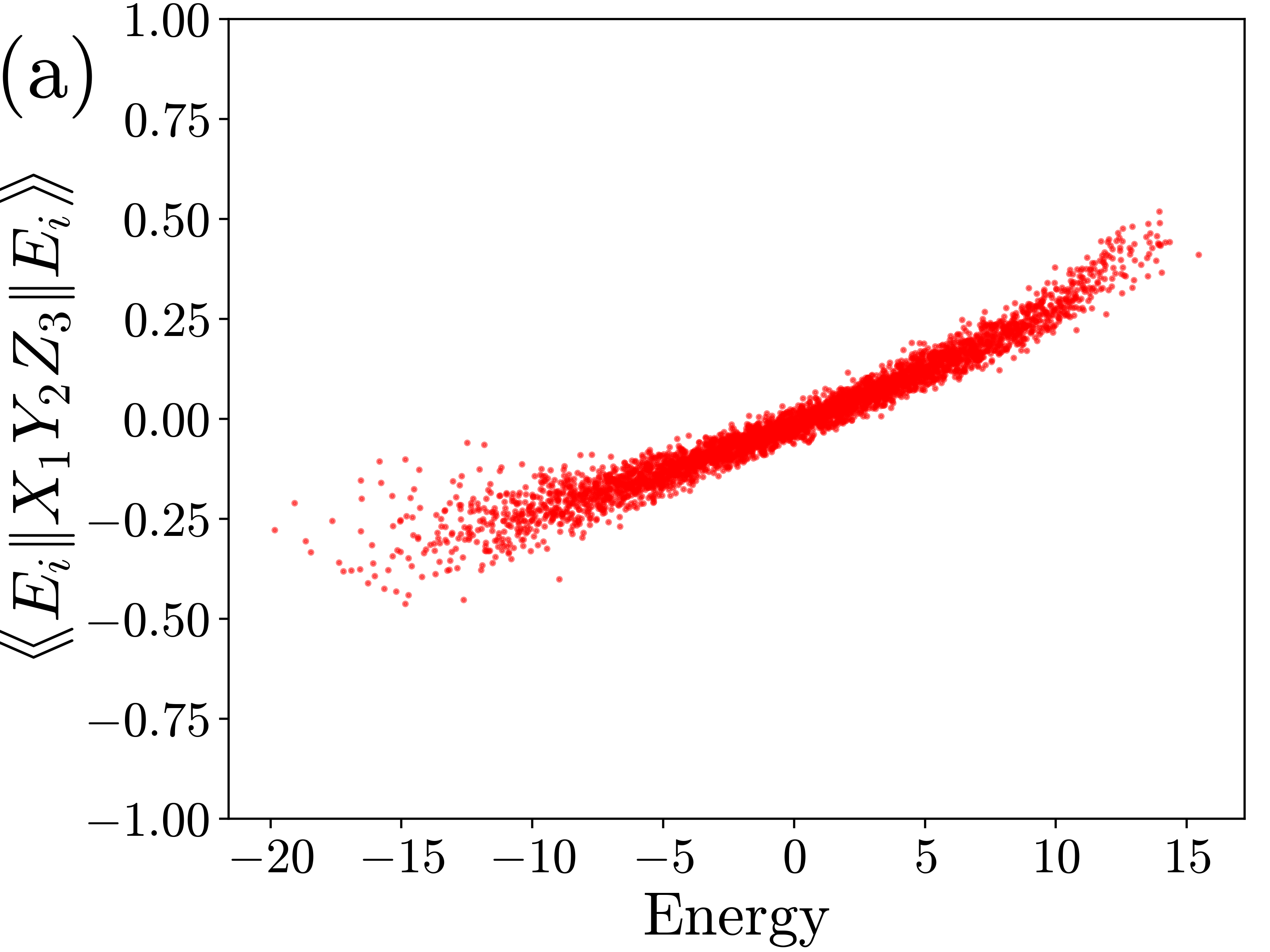}
\end{minipage} 
\begin{minipage}{0.328\columnwidth}
\centering
\includegraphics[width=\columnwidth]{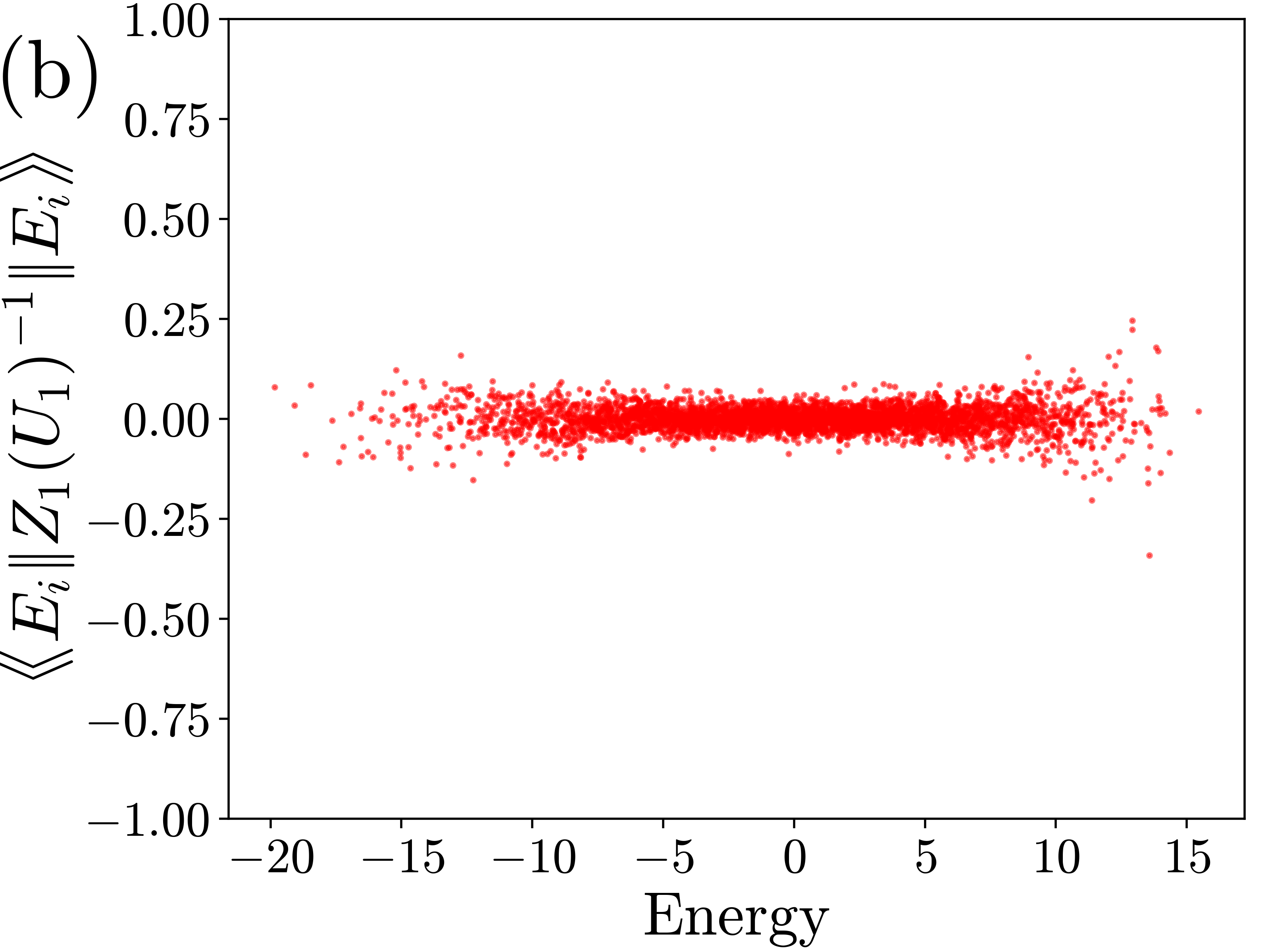}
\end{minipage}
\begin{minipage}{0.328\columnwidth}
\centering
\includegraphics[width=\columnwidth]{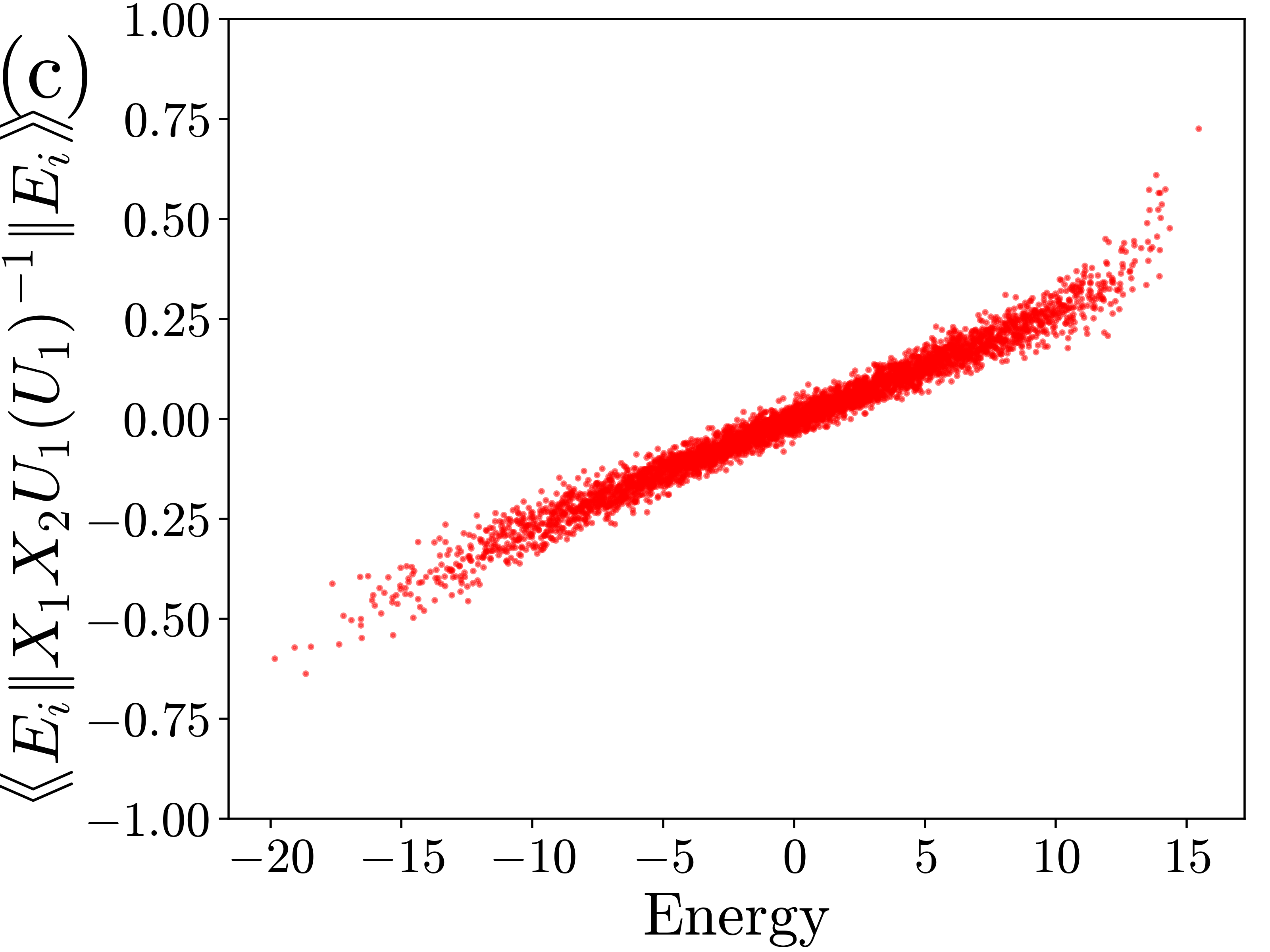}
\end{minipage}
\begin{minipage}{0.45\columnwidth}
\centering
\includegraphics[width=\columnwidth]{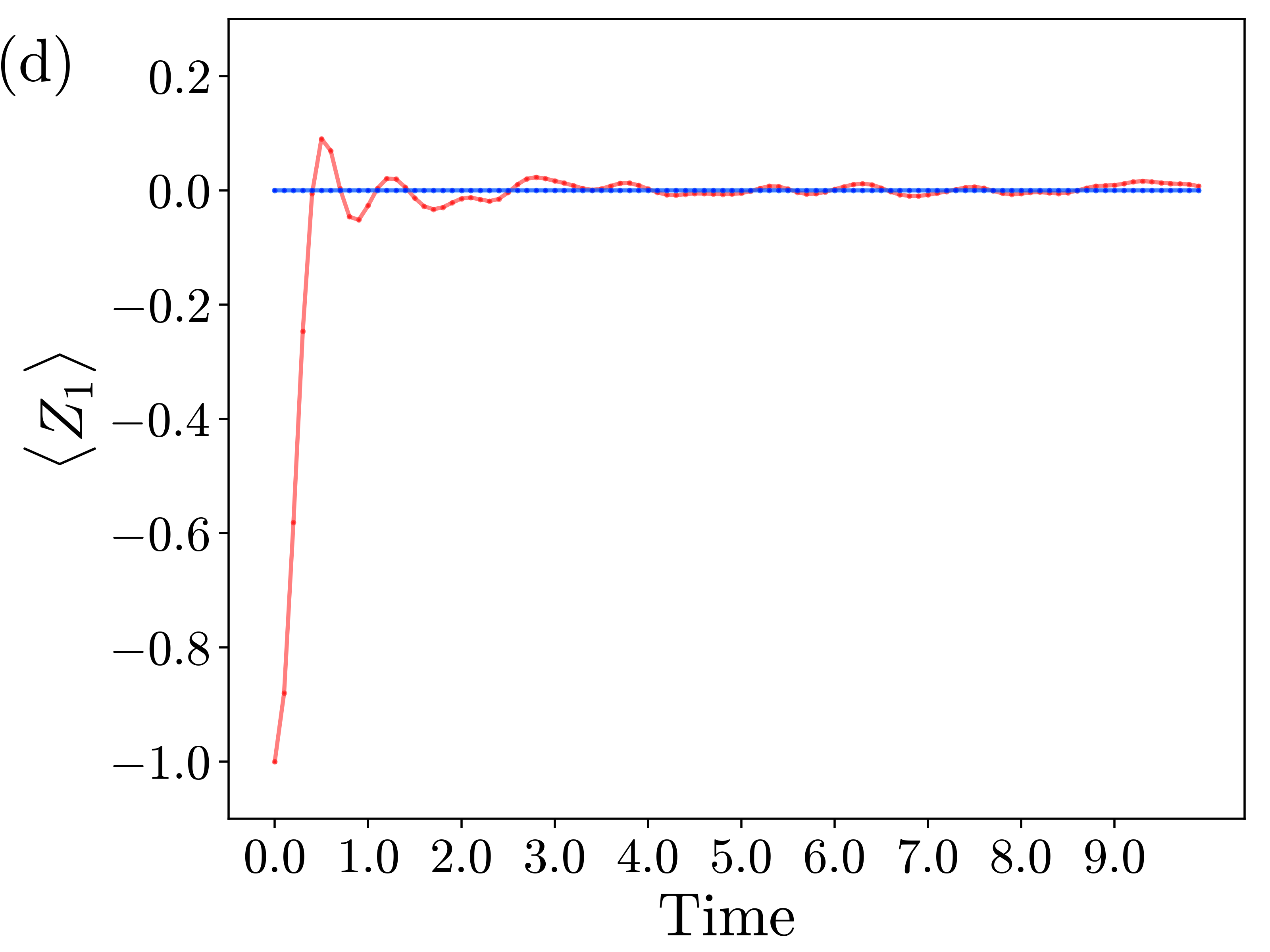}
\end{minipage} 
\begin{minipage}{0.45\columnwidth}
\centering
\includegraphics[width=\columnwidth]{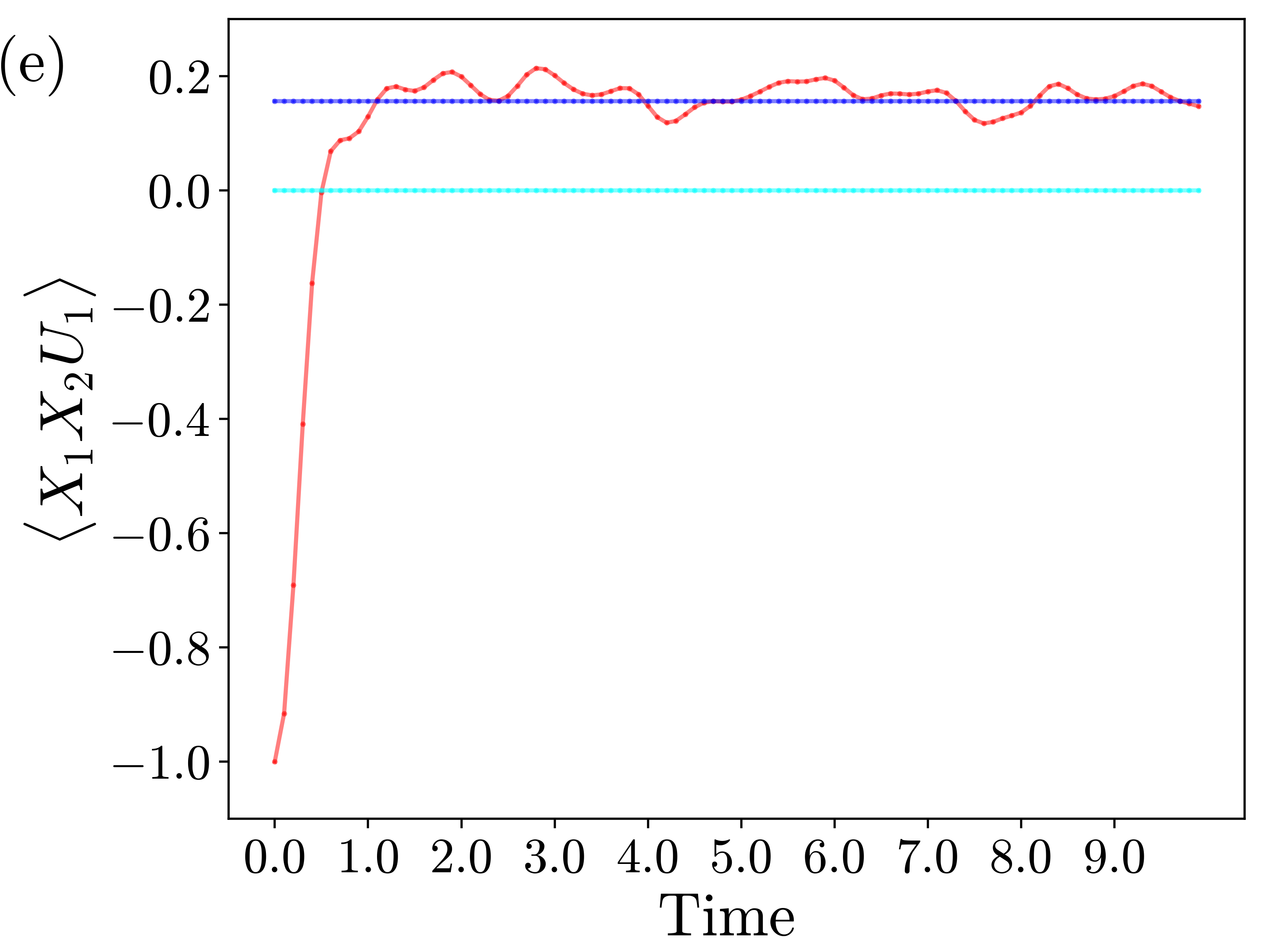}
\end{minipage} 
\hspace{1em}
\caption{(a)(b)(c) Diagonal matrix elements for 
the $\bbZ_2\times\bbZ_2$-symmetric spin chain for $L_1=13$.
The coupling constants are uniformly sampled from $J_j^x\in[0.9,1.0]$, $J_j^y\in[0.7,0.8]$, $J_j^z\in[0.6,0.7]$, and the symmetry-breaking parameter $\alpha$ is given by $\alpha=0.9$.
(a) The expectation values of a neutral operator $\cO^{0,0}=X_1Y_2Z_3$ are well approximated by a smooth function $\scrO^{0,0}(E/V)$.
(b) The matrix elements $\bbra{E_i}Z_1U_1^{-1}\kket{E_i}$ corresponding to  a Type~I charged operator $\cO_{\rm I}^{0,1}=Z_1$ become vanishing, $\scrO_{\rm I}^{0,1}(E/V)\simeq 0$.
(c) The matrix elements 
$\bbra{E_i}X_1X_2U_1U_1^{-1}\kket{E_i}$
corresponding to a Type II operator $\cO_{\rm II}^{0,1}=X_1X_2U_1$ can be described by a smooth function $\scrO_{\rm II}^{0,1}(E/V)$ in the same way as neutral operators.
(d)(e) Time-evolution of the operators (d) $\cO_{\rm I}^{0,1} = Z_1$ and (e) $\cO_{\rm II}^{0,1}=X_1X_2U_1$  for system size $L_1=13$.
The initial state is a random superposition of the eigenstates of $\cO_{\rm I/II}^{0,1}$ with the eigenvalue $-1$,
whose energy expectation values lie within an energy window $E\in[-6.0,5.5]$ and $E\in[-5.0,4.5]$, respectively.
The prediction from the GGE and the Gibbs ensemble are shown by the blue and cyan lines, respectively.
The stationary value of $\braket{\cO_{\rm I/II}^{0,1}}$ is accurately described by the GGE.
For $\cO_{\rm I}^{0,1}$, the predictions from the GGE and the standard Gibbs ensemble give almost the same value.
}\label{fig:Z2Z2-ETH}
\end{figure}

Numerical demonstration of the prETH for diagonal matrix elements is shown in Fig.~\ref{fig:Z2Z2-ETH}.
We compute the matrix elements $\bbra{E_i} \cO^{q_1,q_2}(U_2)^{q_1} (U_1)^{-q_2}\kket{E_i}$.
In particular, we consider a neutral operator $\cO^{0,0}=X_1Y_2Z_3$, a Type I charged operator $\cO_{\rm I}^{0,1}=Z_1$, and a Type II charged operator $\cO_{\rm II}^{0,1}=X_1X_2U_1$.
As expected from Eqs.~\eqref{projective-ETH} and \eqref{Conjecture-matrix-element-zero}, the matrix elements are well-described by a smooth function of energy, and they tend to vanish for Type I charged operators. This result for the Type I charged operator exemplifies the validity of our conjecture that local charged operators belong to Type I.
We also numerically calculate the time-evolution of operators $\cO_{\rm I}^{0,1}=Z_1$ and $\cO_{\rm II}^{0,1}=X_1X_2U_1$.
Since the operator $X_1X_2U_1$ belongs to the class of Type II charged operators of the form~\eqref{Type II neutral times symmetry}, as in Eq.~\eqref{GGE-ETH-summary}, the resultant thermal equilibrium should exhibit a discrepancy between the Gibbs ensemble and the GGE \eqref{def-GGE}.
We indeed observe that the expectation value of the Type II operator relaxes to the GGE prediction instead of the Gibbs one, whereas the Gibbs ensemble suffices for the Type I operator.

\paragraph{$\bbZ_3\times\bbZ_3$-symmetric spin chain}
For $N = 3$, we can consider a  Hamiltonian of a $(1+1)$-dimensional spin chain with $\bbZ_3 \times \bbZ_3$ symmetry that  takes the form
\begin{align}
    H_{N=3} := \sum_{j=1}^{L_1}
    \left(
    J^w_j W_j W_{j+1}^\dagger + J^x_j X_j X_{j+1}^\dagger + J^y_j Y_j Y_{j+1}^\dagger + J^z_j Z_j Z_{j+1}^\dagger
    \right)
    + (\mathrm{h.c.}),
    \label{3-Ham}
\end{align}
where $W_j := Z^\dagger_j X_j$ and $Y_j := Z_j X_j$. As in the $N=2$ case, we impose periodic boundary conditions by identifying $j \sim j + L_1$.
When the couplings $J^w_j$, $J^x_j$, $J^y_j$, and $J^z_j$ are weakly random and complex (i.e., not real), the system exhibits no symmetries other than the $\bbZ_3 \times \bbZ_3$ symmetry generated by $U_1$ and $U_2$. Under the action of these symmetry operators, the local operators transform as
\begin{align}
    U_1^\dagger W_j U_1 &= e^{\frac{4\pi \rmi}{3}} W_j,&
    U_1^\dagger X_j U_1 &= e^{\frac{4\pi \rmi}{3}} X_j,&
    U_1^\dagger Y_j U_1 &= e^{\frac{4\pi \rmi}{3}} Y_j,&
    U_1^\dagger Z_j U_1 &= Z_j, 
    \\
    U_2^\dagger W_j U_2 &= e^{\frac{4\pi \rmi}{3}} W_j,&
    U_2^\dagger X_j U_2 &= X_j,&
    U_2^\dagger Y_j U_2 &= e^{\frac{2\pi \rmi}{3}} Y_j,&
    U_2^\dagger Z_j U_2 &= e^{\frac{2\pi \rmi}{3}} Z_j.&
\end{align}
To consider the GGE for the $\bbZ_3\times\bbZ_3$-symmetric chain \eqref{3-Ham}, we need to introduce $3^2=9$ parameters as
\begin{align}
    \rhogge=&\, \frac{1}{\cZ}\exp\left(-\beta H_{N=3}-\sum_{r=1}^{8} \mu_{r}Q^r
    \right),
    \\
    \{Q^r\}_{r=1,\dots,8} =&\, \big\{\cR^{r_1,r_2},\cI^{r_1,r_2}\big\}_{(r_1,r_2)=(0,1),(1,0)(1,1),(1,2)}
\end{align}
so that the conditions \eqref{tuning-energy} and \eqref{tuning-conserved} are satisfied.

\begin{figure}[t]
\centering
\begin{minipage}{0.328\columnwidth}
\centering
\includegraphics[width=\columnwidth]{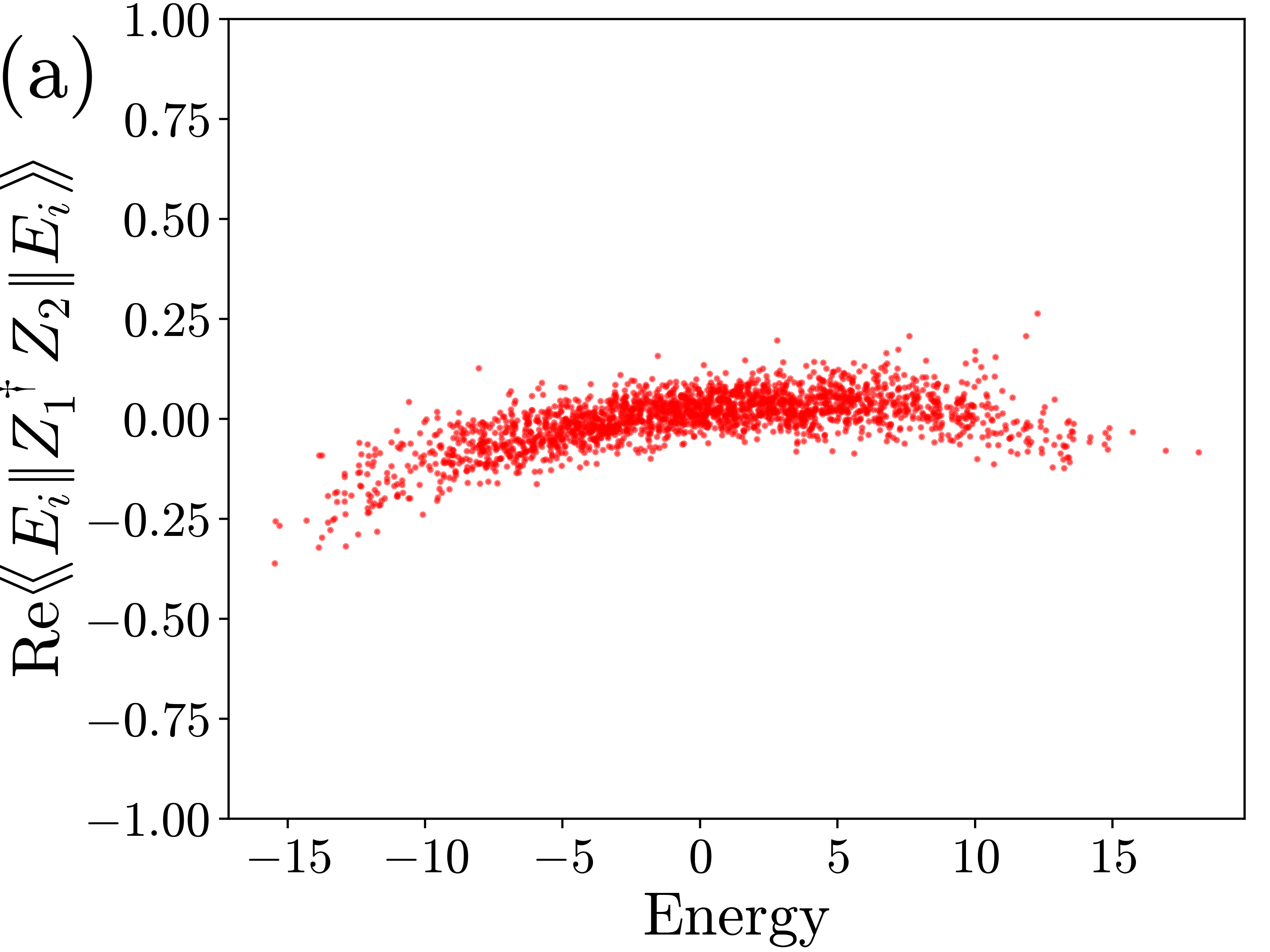}
\end{minipage} 
\begin{minipage}{0.328\columnwidth}
\centering
\includegraphics[width=\columnwidth]{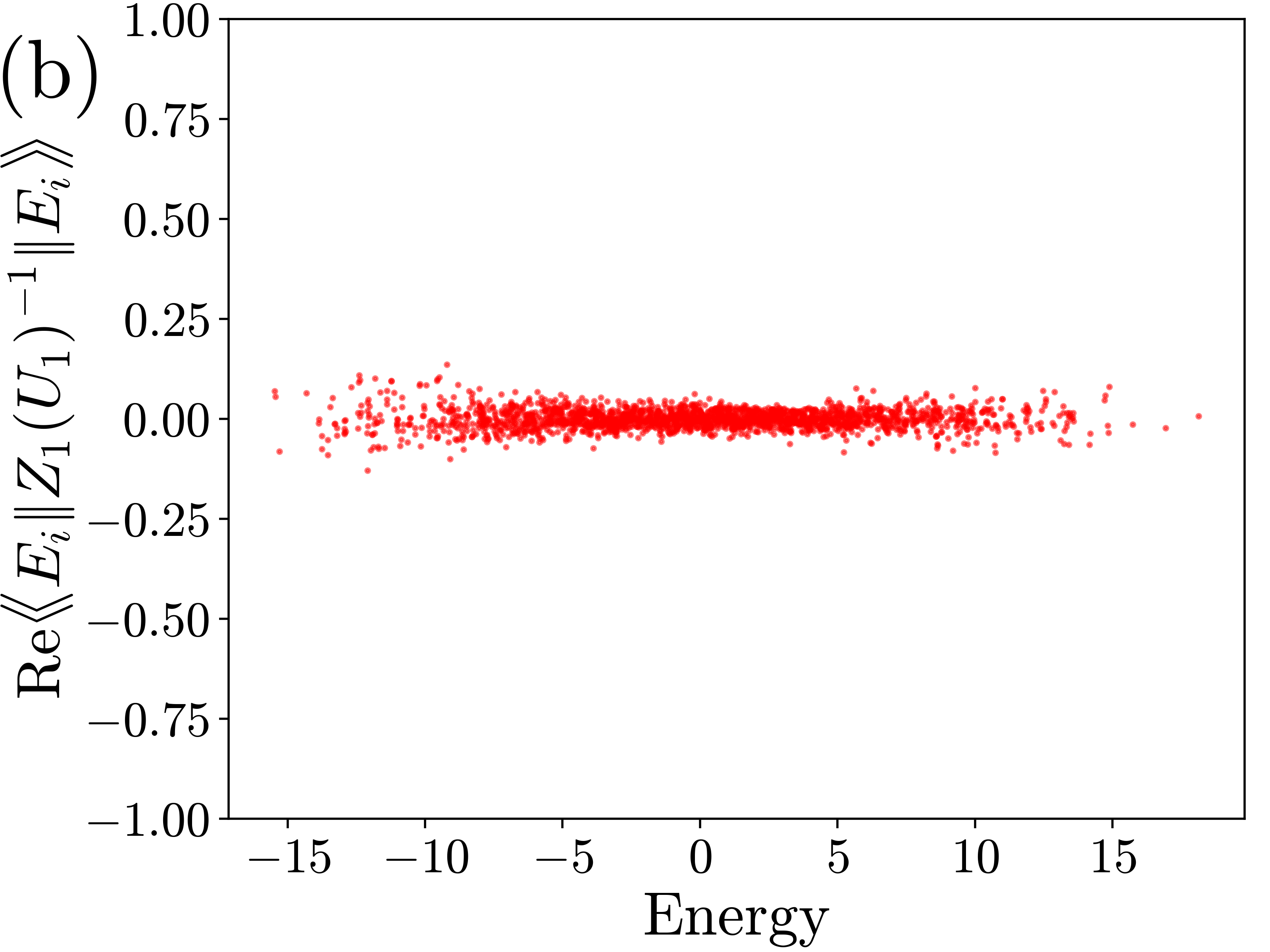}
\end{minipage}
\begin{minipage}{0.328\columnwidth}
\centering
\includegraphics[width=\columnwidth]{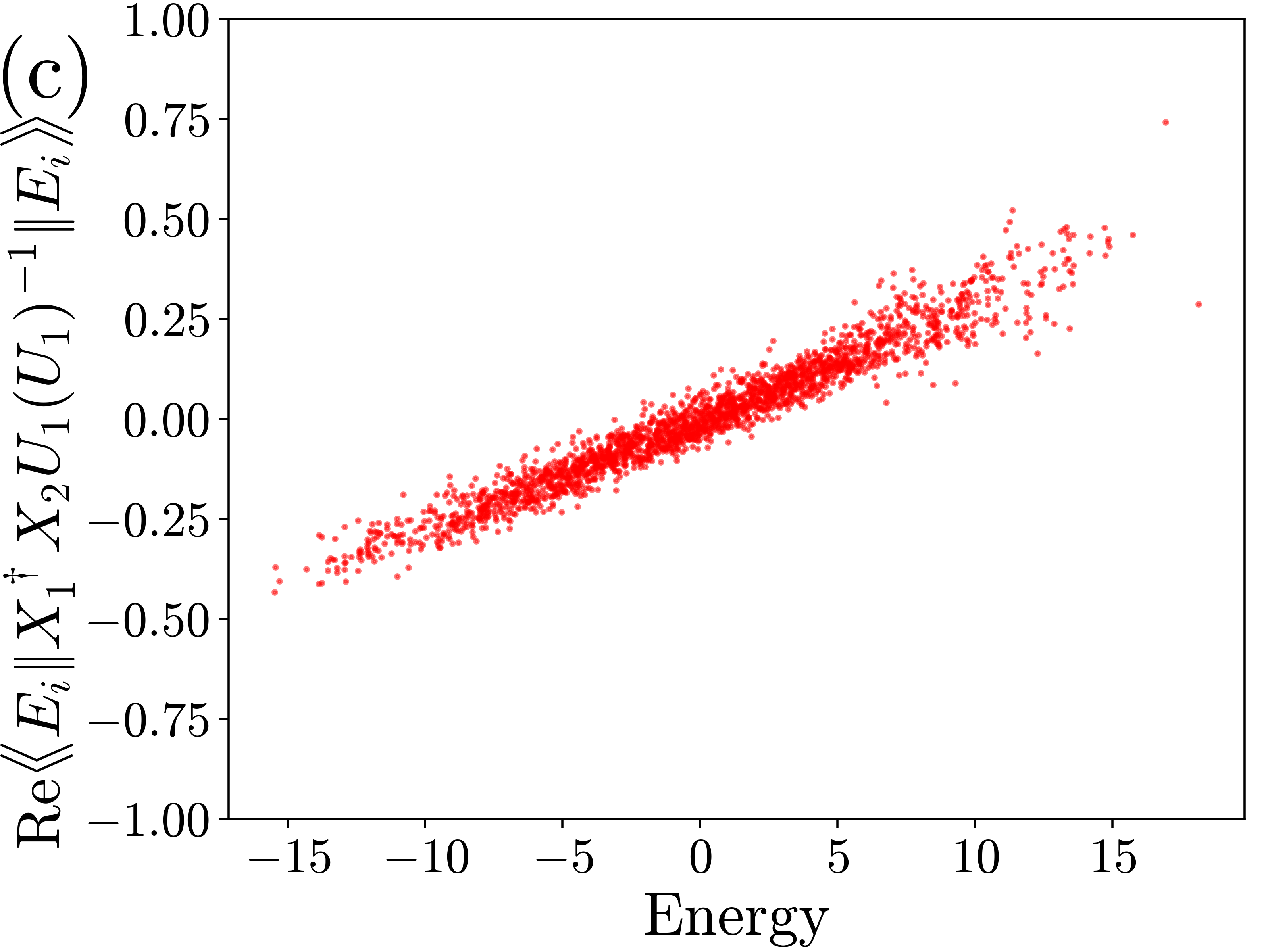}
\end{minipage} 
\begin{minipage}{0.45\columnwidth}
\centering
\includegraphics[width=\columnwidth]{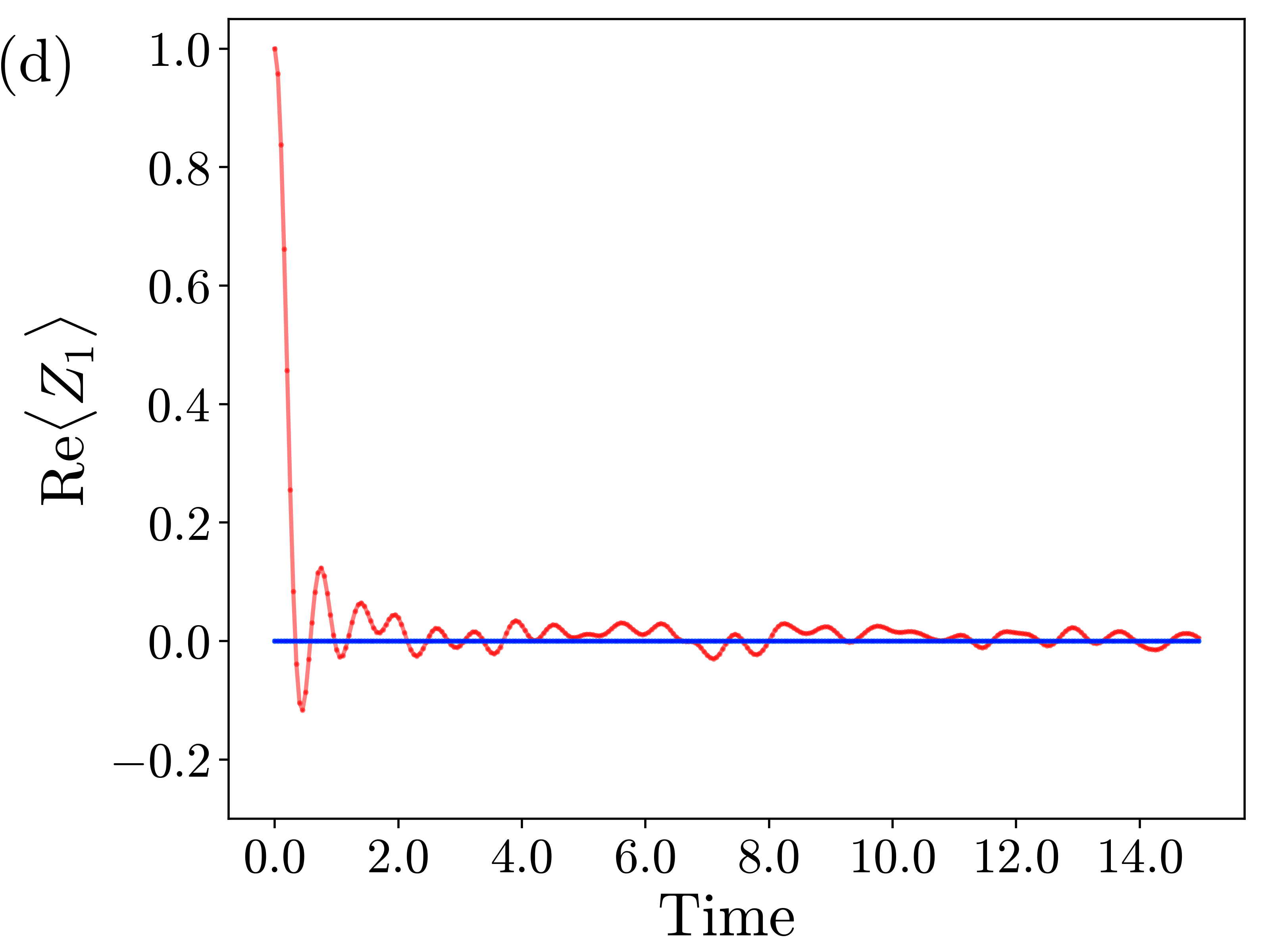}
\end{minipage} 
\hspace{1em}
\begin{minipage}{0.45\columnwidth}
\centering
\includegraphics[width=\columnwidth]{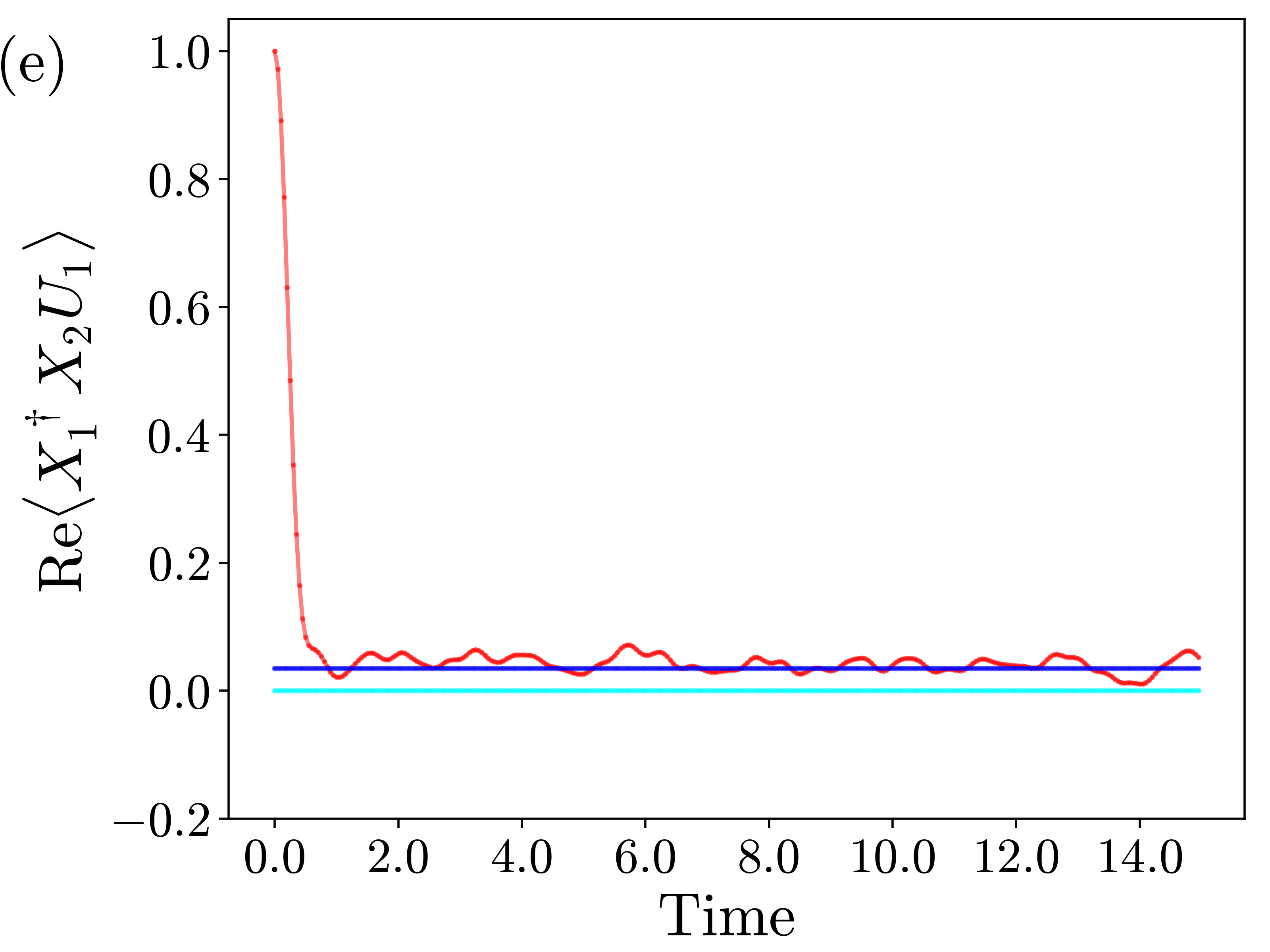}
\end{minipage} 
\caption{(a)(b)(c) Real part of the diagonal matrix elements for 
the $\bbZ_3\times\bbZ_3$-symmetric spin chain for $L_1=8$.
The coupling constants are uniformly distributed in $J_j^w\in[1.0,1.1]$, $J_j^x\in[0.9,1.0]+0.2\rmi$, $J_j^y\in[0.1,0.2]$, $J_j^z\in[0.2,0.3]$.
(a) The expectation values of a neutral operator $\cO^{0,0}=Z_1^\dagger Z_2$ are well-approximated by a smooth function $\scrO^{0,0}(E/V)$.
(b) The matrix elements 
$\re[\bbra{E_i}Z_1U_1^{-1}\kket{E_i}]$
corresponding to  a Type I charged operator $\cO_{\rm I}^{0,1}=Z_1$ become vanishing, i.e., $\re\scrO_{\rm I}^{0,1}(E/V)\simeq 0$.
(c) The matrix elements $\re[\bbra{E_i}X_1^\dagger X_2U_1U_1^{-1}\kket{E_i}]$ corresponding to
 an operator $\cO_{\rm II}^{0,1}=X_1^\dagger X_2U_1$ can be described by a smooth function $\re\scrO_{\rm II}^{0,1}(E/V)$ in the same way as neutral operators.
(d)(e) Time-evolution of the operators (d) $\cO_{\rm I}^{0,1} = Z_1$ and (e) $\cO_{\rm II}^{0,1}=X_1X_2U_1$ for system size $L_1=8$.
The initial state is a random superposition of the eigenstates of $\cO_{\rm I/II}^{0,1}$ with the eigenvalue $-1$,
whose energy expectation values lie within an energy window $E\in[-2.0,1.5]$.
The predictions from the GGE and the Gibbs ensemble are shown by the blue and cyan lines, respectively.
The stationary value of $\braket{\cO_{\rm I/II}^{0,1}}$ is accurately described by the GGE.
For $\cO_{\rm I}^{0,1}$, the predictions from the GGE and the standard Gibbs ensemble give almost the same value.
}\label{fig:Z3Z3-ETH}
\end{figure}

Figure~\ref{fig:Z3Z3-ETH} presents a numerical test of the prETH matrix elements in the same way as the $\bbZ_2\times\bbZ_2$-symmetric case.
We evaluate
$\operatorname{Re}\bbra{E_i}\,\cO^{q_1,q_2}(U_2)^{q_1}(U_1)^{-q_2}\,\kket{E_i}$
for the following operators;
a neutral operator $\cO^{0,0}=Z_1^\dagger Z_2$,
a Type I charged operator $\cO_{\rm II}^{0,1} =Z_1$, 
and
a Type II charged operator $\cO_{\rm I}^{0,1}=X_1^\dagger X_2U_1$.
 The results are consistent with the prediction of Eqs.~\eqref{projective-ETH} and \eqref{Conjecture-matrix-element-zero}.
We also compute the time evolution of $\cO_{\rm I}^{0,1}$ and $\cO_{\rm II}^{0,1}$.
The stationary values of the expectation values again agree with the prediction based on the GGE \eqref{def-GGE}, thereby confirming the prediction following from the prETH.

\subsection{$\bbZ_2$ lattice gauge theory}\label{Z2 lattice gauge theory}

In this section, we consider a $(2+1)$-dimensional $\bbZ_2$ lattice gauge theory\footnote{For a review, see Ref.~\cite{Kogut:1979wt}.} that possesses both an electric $1$-form $\bbZ_2$ symmetry and a $0$-form $\bbZ_2$ symmetry. The spatial manifold is taken to be a square lattice system $\cM_2$ of size $L_x \times L_y$ with periodic boundary conditions. The Hamiltonian is given by
\begin{align}
    H:=-\sum_{p\in\cM_2}f_p\prod_{\ell\in p}\sigma^z_\ell-\sum_{p\in\cM_2}g_p\prod_{\ell\in p}\sigma^x_\ell-\sum_{v\in\cM_2}h_v\,\sigma^x_{(v, \Hat{1})}\sigma^x_{(v, \Hat{2})}-\sum_{v\in\cM_2}h_v'\,\sigma^x_{(v, \Hat{1})}\sigma^x_{(v+\Hat{1}, \Hat{2})},
    \label{Hamiltonian Z2 gauge}
\end{align}
where $p,l$ and $v$ have been introduced after Eq.~\eqref{cv}.
The parameters $f_p$, $g_p$, $h_v$, and $h_v'$ are supposed to be nonzero coupling constants for all $p$ and $v$.\footnote{We do not consider cases where any of $f_p$, $g_p$, $h_v$, or $h_v'$ vanish in this subsection. A more general discussion including nontrivial symmetry structures with projective representations is given in Appendix~\ref{Symmetry structure of Hamiltonian}.}
The $\bbZ_2$ gauge transformations are defined by
\begin{align}
    G_v:=\prod_{\ell\ni v}\sigma^x_\ell=\sigma^x_{(v, \Hat{1})}\sigma^x_{(v, -\Hat{1})}\sigma^x_{(v, \Hat{2})}\sigma^x_{(v, -\Hat{2})},\qquad G_v^2=1.\label{gauge transf.}
\end{align}
We define the physical Hilbert space $\cH_{\text{phys}}$ such that it  consists of states $\ket{\psi}$ satisfying the Gauss law
\begin{align}
   \forall v\qquad G_v\ket{\psi}=+\ket{\psi}.\label{Gauss law}
\end{align}
As a result, the expectation value of any gauge non-invariant operator vanishes when evaluated on a physical state.

\begin{figure}[t]
\centering
\begin{tikzpicture}[scale=1.35]
  \draw[->] (-1.5,0) -- (-0.5,0) node[right] {$x$};
  \draw[->] (-1.5,0) -- (-1.5,1) node[above] {$y$};

  \draw[step=0.5, gray, thick, dashed] (0,0) grid (6,4);
  \draw[very thick] (0,0) grid (6,4);

  \fill[teal, opacity=0.5] (2,3) circle(4pt);
  \draw[teal, line width=6pt, opacity=0.5] (2,2) -- (2,4);
  \draw[teal, line width=6pt, opacity=0.5] (1,3) -- (3,3);
  \node[teal, below right] at (2,3) {$v$};
  \node[teal] at (1.4,3.7) {\large $G_v$};


  \foreach \x in {0,1,2,3,4,5,6} {
    \draw[blue, line width=6pt, opacity=0.4] (\x,0) -- (\x,1);
  }
  \node[blue,right] at (6,1.3) {\large $U^{(1)}_x$ on $\textcolor{orange}{C_x^*}$};

  \draw[orange, line width=4pt, opacity=0.5] (0,0.5) -- (6,0.5);

  \draw[red, line width=2pt, opacity=0.5] (5,0) -- (5,4);
  \draw[magenta, line width=6pt, opacity=0.2] (5,0) -- (5,4);
  \node[red,above right] at (5,4) {\large $W_y$ on $\textcolor{magenta}{C_y}$};
\end{tikzpicture}
\caption{Symmetry and loop-operator structures on a two-dimensional periodic square lattice. The black solid lines indicate the lattice points and links of the original lattice, while the dashed lines represent their duals. At each site $v$ (marked by a green point), the $\bbZ_2$ gauge symmetry operator $G_v$ defined in Eq.~\eqref{gauge transf.} acts on the four adjacent links $\ell \ni v$, depicted by green segments. This enforces the Gauss law condition given in Eq.~\eqref{Gauss law}.
A non-contractible loop on the dual lattice is illustrated by the orange path wrapping around the lattice; when oriented along the $x$- or $y$-direction, it is denoted by $C_x^*$ or $C_y^*$, respectively. The blue lines indicate links $\ell \in C_x^*$ that intersect this loop. The electric $1$-form symmetry operators $U_x^{(1)}$ and $U_y^{(1)}$ are then constructed as in Eq.~\eqref{eq:z2lattice_el1_z2_sym}. In contrast, the magenta line shows a non-contractible loop on the original lattice, say $C_{x,y}$, along which the Wilson loop (shown by the red line) $W_{x,y}$ is defined.}
\label{fig:operators in Z2 gauge}
\end{figure}
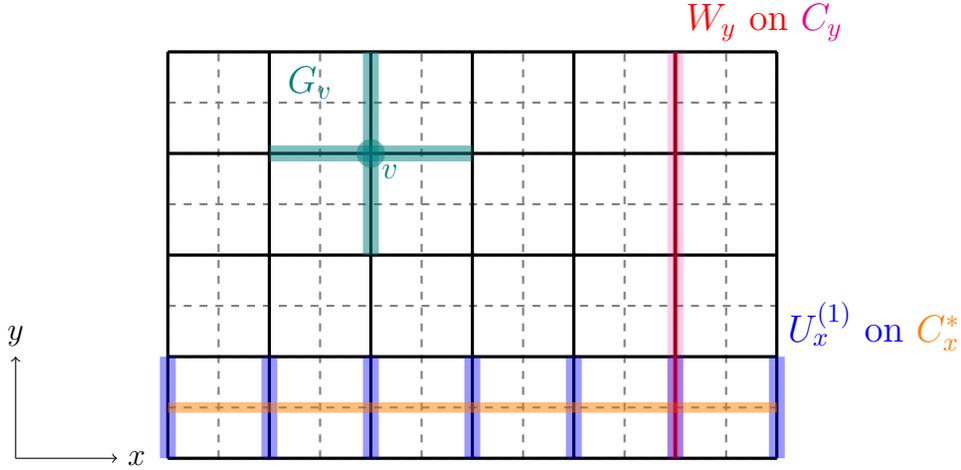

The Hamiltonian~\eqref{Hamiltonian Z2 gauge} possesses the electric $1$-form $\bbZ_2$ symmetry,
\begin{align}
    U_x^{(1)}&:=\prod_{\ell\in C^*_x}\sigma_\ell^x,& U_y^{(1)}&:=\prod_{\ell\in C^*_y}\sigma_\ell^x,& [U_{x,y}^{(1)},H]&=0, \label{eq:z2lattice_el1_z2_sym}
\end{align}
where $C^*_x$ and $C^*_y$ are non-contractible loops along the $x$ and $y$ directions, respectively, defined on the dual lattice~(see Fig.~\ref{fig:operators in Z2 gauge}). Because the Gauss law~\eqref{Gauss law} holds in the physical Hilbert space, the operators $U^{(1)}_{x,y}$ are invariant under topological deformations along a spatial direction of the paths $C^*$. This defines a genuine $1$-form symmetry. These $1$-form symmetry operators act nontrivially on the following Wilson loop operators:
\begin{align}
    W_x&:=\prod_{\ell\in C_x}\sigma_\ell^z,& (U_y^{(1)})^\dagger W_xU_y^{(1)}&=-W_x\label{Wilson line x}\\
    W_y&:=\prod_{\ell\in C_y}\sigma_\ell^z,& (U_x^{(1)})^\dagger W_yU_x^{(1)}&=-W_y\label{Wilson line y}
\end{align}
where $C_x$ and $C_y$ are non-contractible loops along the $x$ and $y$ directions, respectively, on the original lattice~(see Fig.~\ref{fig:operators in Z2 gauge}). Hence, $W_x$ and $W_y$ are the charged operators associated with the electric $1$-form symmetry.

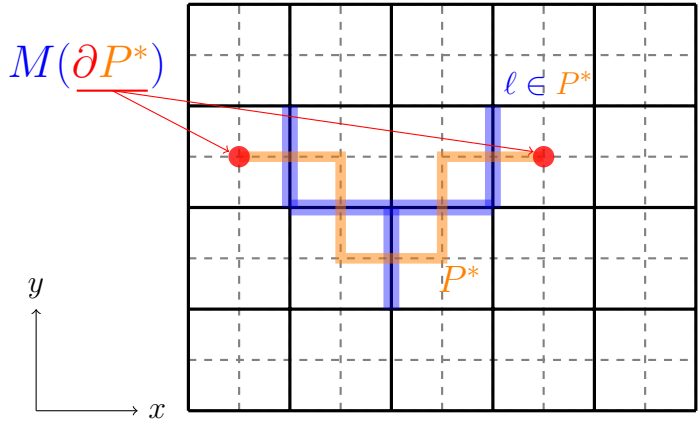
\begin{figure}[t]
\centering
\begin{tikzpicture}[scale=1.35]
  \draw[->] (-1.5,-1) -- (-0.5,-1) node[right] {$x$};
  \draw[->] (-1.5,-1) -- (-1.5,0) node[above] {$y$};

  \draw[step=0.5, gray, thick, dashed] (0,-1) grid (5,3);
  \draw[very thick] (0,-1) grid (5,3);

  \draw[blue, line width=6pt, opacity=0.4] (1,1) -- (1,2);
  \draw[blue, line width=6pt, opacity=0.4] (1,1) -- (2,1);
  \draw[blue, line width=6pt, opacity=0.4] (2,1) -- (2,0);
  \draw[blue, line width=6pt, opacity=0.4] (2,1) -- (3,1);
  \draw[blue, line width=6pt, opacity=0.4] (3,1) -- (3,2);

  \node[blue, above right] at (3,2) {$\ell\in\textcolor{orange}{P^*}$};

  \draw[orange, line width=4pt, opacity=0.5] (0.5,1.5) -- (1.5,1.5) -- (1.5,0.5) -- (2.5,0.5) -- (2.5,1.5) -- (3.5,1.5);
  \node[orange,right] at (2.35,0.3) {\large $P^*$};

  \fill [red, opacity=0.79] (0.5,1.5) circle [radius=3pt] ;
  \fill [red, opacity=0.79] (3.5,1.5) circle [radius=3pt] ;

  \node[blue] at (-1,2.4) {\Large $M(\textcolor{red}{\partial}\textcolor{orange}{P^*})$};
  \draw[red,thick] (-1.1,2.15) -- (-0.4,2.15);
  \draw[red,->] (-0.75,2.15) -- (0.4,1.55);
  \draw[red,->] (-0.75,2.15) -- (3.4,1.55);
\end{tikzpicture}
\caption{Illustration of the magnetic operator $M(\partial P^*)$ on the dual lattice. The orange path indicates the segments of $P^*$ on the dual lattice, while the blue links are those contributing to the product $\prod_{\ell\in P^*} \sigma_\ell^x$. The operator depends only on the endpoints $\partial P^*$ (marked by the two red points), i.e., it is invariant under topological deformations of the path that fix these endpoints, due to the Gauss law~\eqref{Gauss law}.}
\label{fig:Magnetic operator}
\end{figure}

Additionally, the Hamiltonian~\eqref{Hamiltonian Z2 gauge} possesses the $0$-form $\bbZ_2$ symmetry
\begin{align}
U^{(0)}&:=\prod_{\ell\in\text{all links}}\sigma_\ell^z,&[U^{(0)},H]&=0.
\end{align}
This $0$-form symmetry operator $U^{(0)}$ acts nontrivially on magnetic operators defined along segments $P^*$ on the dual lattice. Now, let us define a magnetic operator~$M(\partial P^*)$ as
\begin{align}
    M(\partial P^*)&:=\prod_{\ell\in P^*}\sigma_\ell^x ,
    &
    \cZ(P^*)&:=\prod_{\ell\in P^*}(-1), 
\end{align}
then we can find
\begin{align}
    (U^{(0)})^\dagger M(\partial P^*)U^{(0)}=\cZ(P^*)M(\partial P^*) .
\end{align}
Figure~\ref{fig:Magnetic operator} illustrates the construction and key features of $M(\partial P^*)$.
Here, $M(\partial P^*)$ may be regarded as a part of an electric $1$-form symmetry operator. In the physical Hilbert space, due to Gauss law \eqref{Gauss law}, $M(\partial P^*)$ is topologically invariant under deformations of $P^*$ that fix its endpoints $\partial P^*$.
Namely, $M$ is determined only by~$\partial P^*$, which is why we write it as $M(\partial P^*)$.

Importantly, the $1$-form and $0$-form symmetries exhibit a projective representation~\cite{Fukushima:2023gmj} in the following:
\begin{align}
    U^{(0)}\,U_{x,y}^{(1)}=\cZ(C^*_{x,y})U_{x,y}^{(1)}\,U^{(0)}=(-1)^{L_{x,y}}U_{x,y}^{(1)}\,U^{(0)}.
    \label{eq:proj-rep_u0_uxy}
\end{align}
This identity uses the fact that the number of links composing $C^*_{x,y}$ is $|C^*_{x,y}| = L_{x,y} \bmod 2$. Therefore, when either $L_x$ or $L_y$ is odd, the $1$-form and $0$-form symmetry operators obey a nontrivial projective representation. In contrast, if both $L_x$ and $L_y$ are even, such a projective relation does not arise, and so we do not consider this case in this work.

\begin{figure}[t]
\centering
\begin{minipage}{0.328\columnwidth}
\centering
\includegraphics[width=\columnwidth]{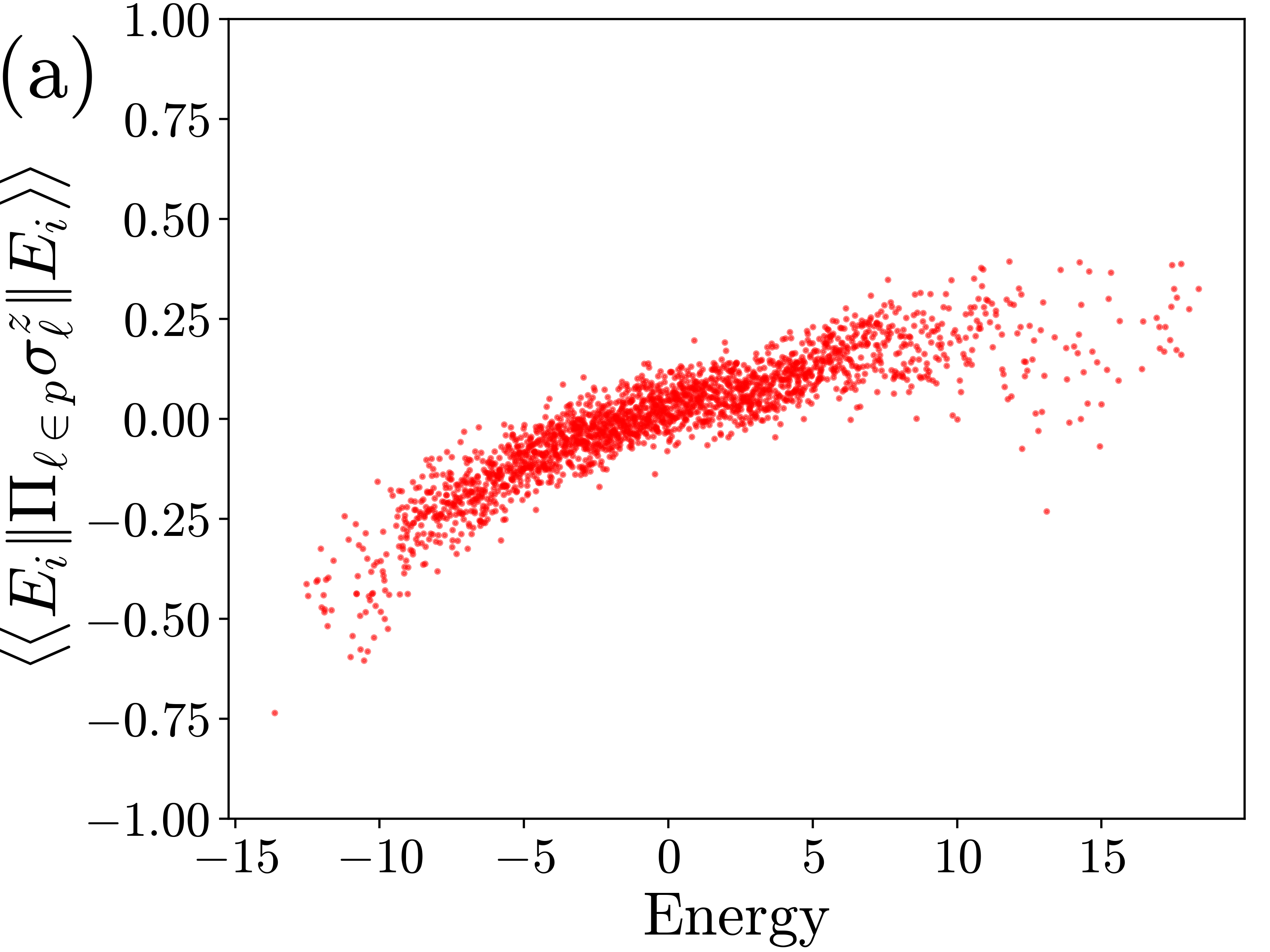}
\end{minipage} 
\begin{minipage}{0.328\columnwidth}
\centering
\includegraphics[width=\columnwidth]{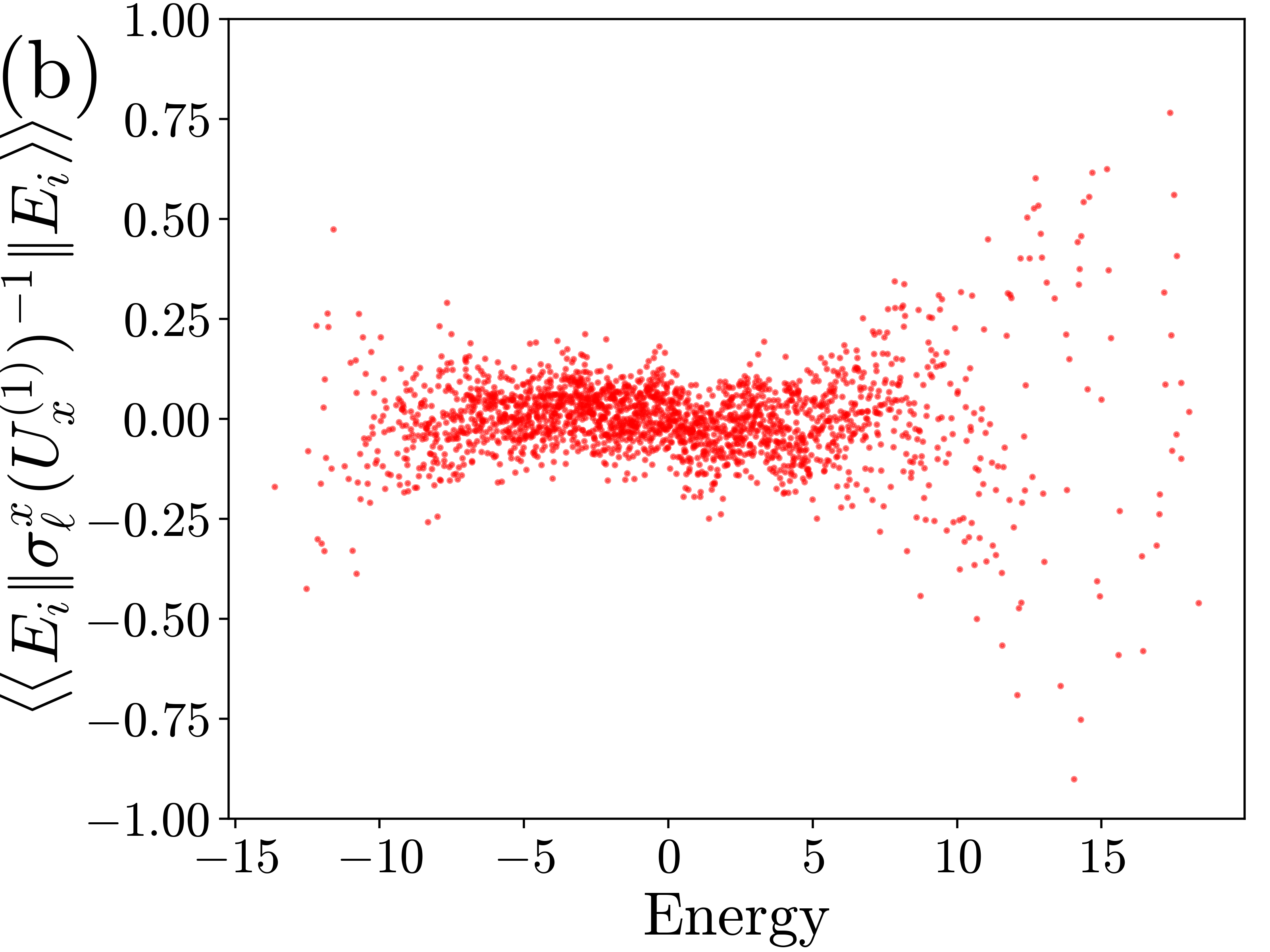}
\end{minipage} 
\begin{minipage}{0.328\columnwidth}
\centering
\includegraphics[width=\columnwidth]{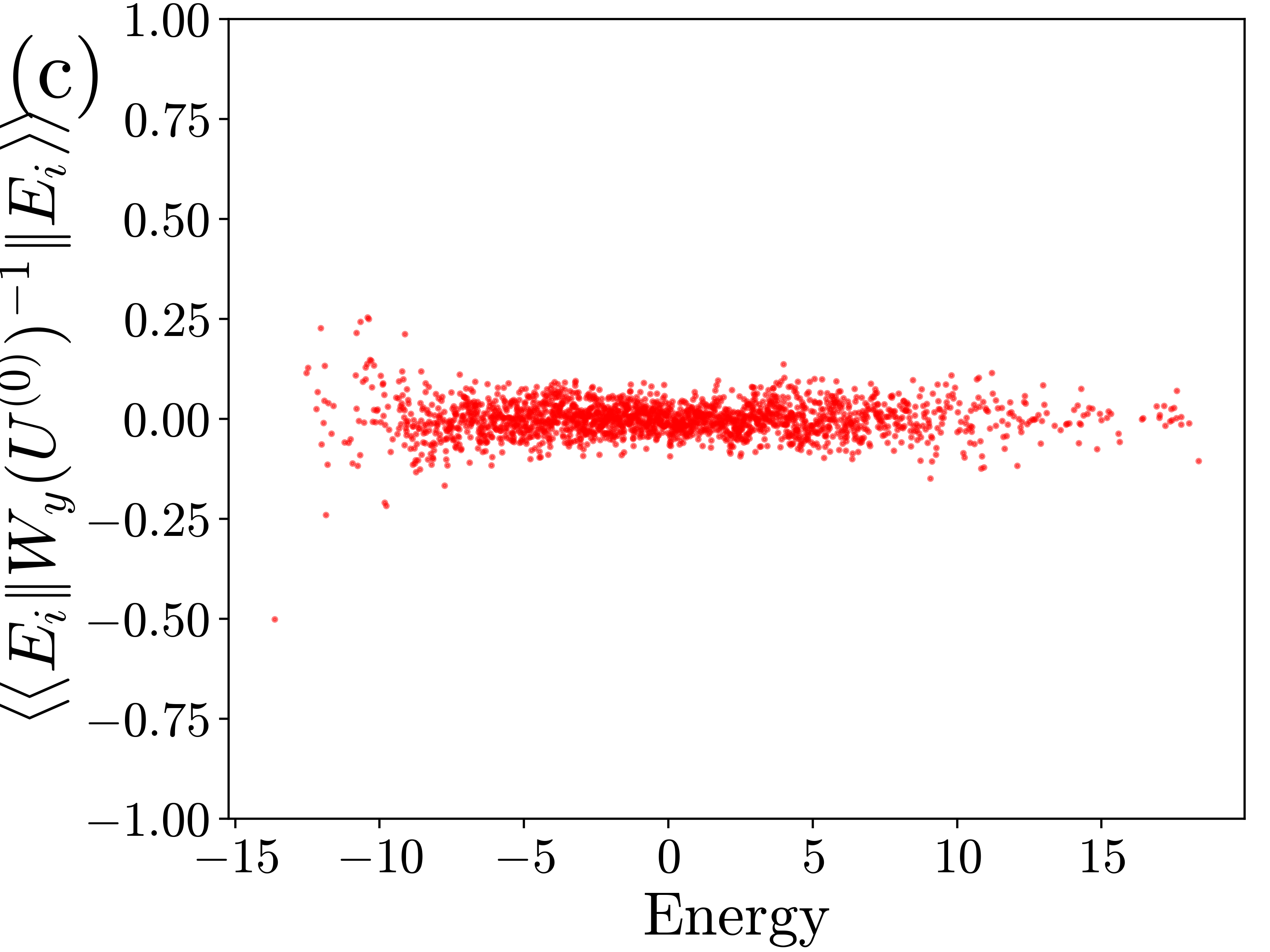}
\end{minipage}
\begin{minipage}{0.45\columnwidth}
\centering
\includegraphics[width=\columnwidth]{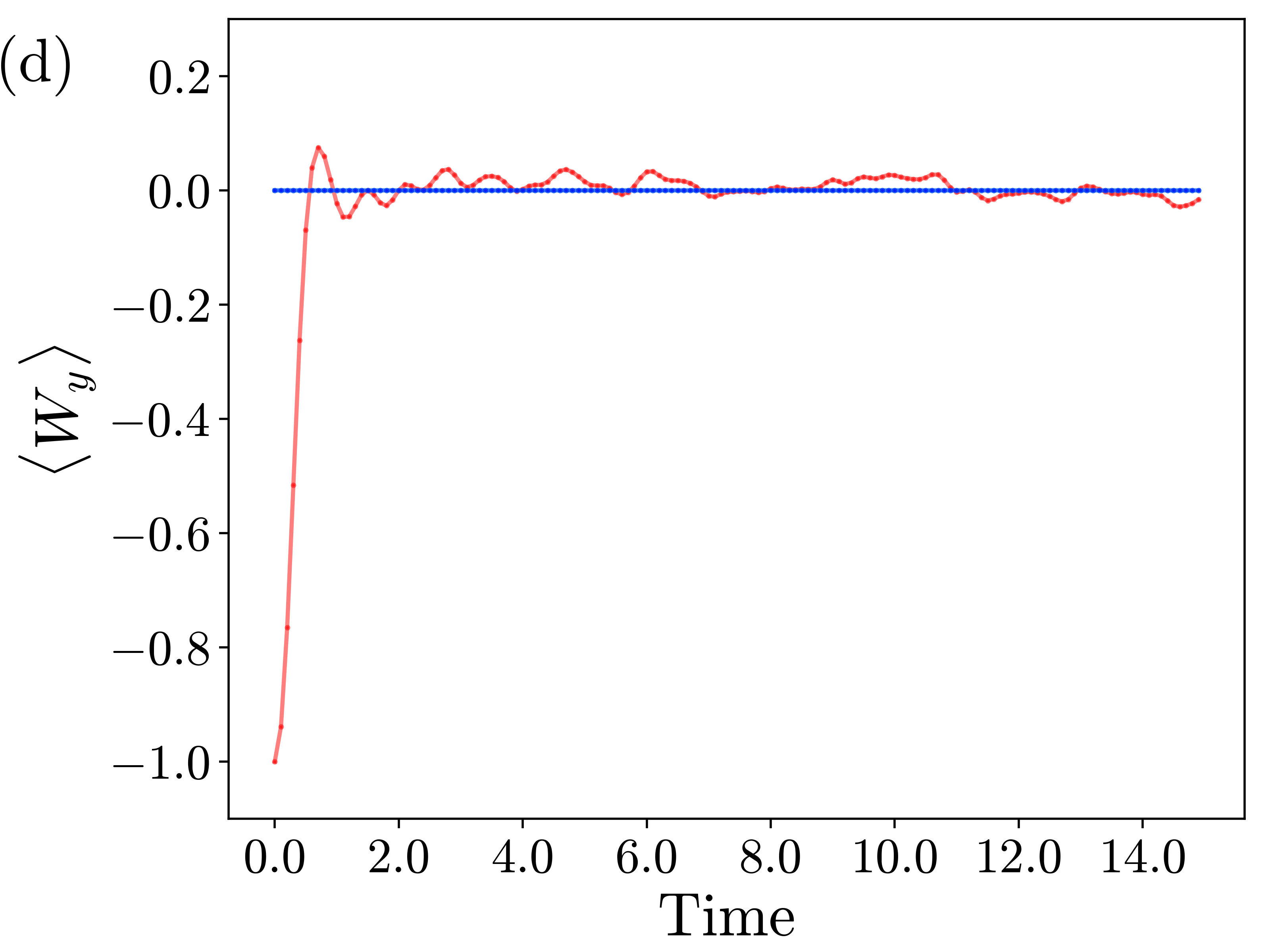}
\end{minipage} 
\begin{minipage}{0.45\columnwidth}
\centering
\includegraphics[width=\columnwidth]{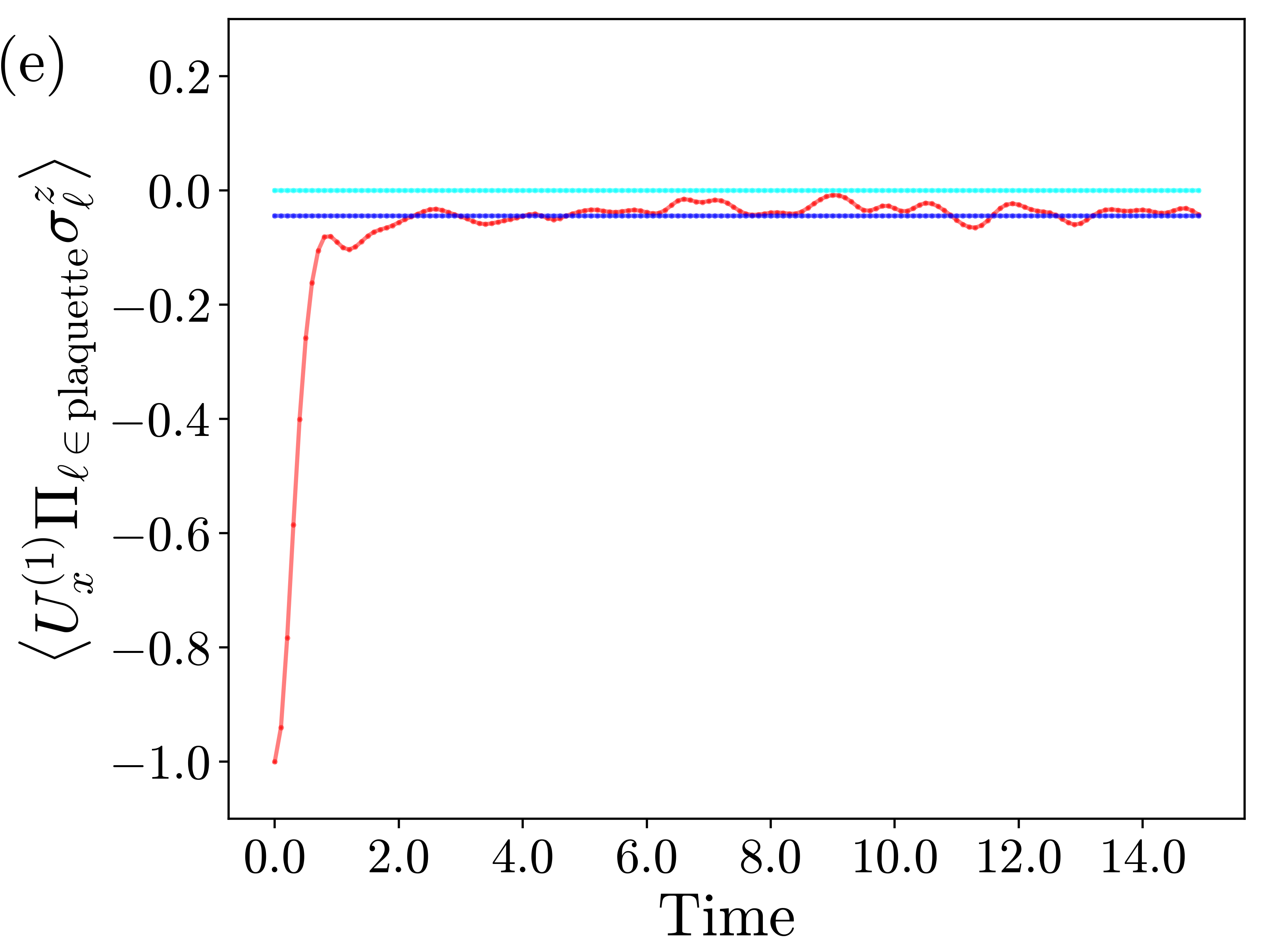}
\end{minipage} 
\hspace{1em}
\caption{(a)(b)(c) Diagonal matrix elements for 
the $\bbZ_2$ gauge theory for $(L_x,L_y)=(3,4)$.
The coupling constants are uniformly distributed in $f_p\in[0.7,0.9]$, $g_p\in[0.7,0.9]$, $h_v\in[0.7,0.9]$, $h_v'\in[0.7,0.9]$.
(a) The expectation values of a neutral operator $\cO^{0,0}=\prod_{\ell\in p}\sigma^z_\ell$ are well-approximated by a smooth function $\scrO^{0,0}(E/V)$.
%
These values are equivalent to the matrix elements $\bbra{E_i}\cO_{\rm II}^{1,0}(U_x^{(1)})^{-1}\kket{E_i}$ corresponding to the Type II operator $\cO_{\rm II}^{1,0}=\cO^{0,0} U_x^{(1)}$;
they are also described by a smooth function $\scrO^{0,0}(E/V)$ in the same way as the neutral operator $\cO^{0,0}$.
(b) Matrix elements $\bbra{E_i}\sigma_\ell^x(U_x^{(1)})^{-1}\kket{E_i}$ corresponding to a Type I local charged operator $\cO_{\rm I,local}^{1,0}=\sigma_\ell^x$, which become almost vanishing, i.e., $\scrO_{\rm I, local}^{0,1}(E/V)\simeq 0$.
(c)  Matrix elements $\bbra{E_i}W_y(U^{(0)})^{-1}\kket{E_i}$ corresponding to a Type I nonlocal charged operator $\cO_{\rm I,nonlocal}^{0,1}=W_y$, which become almost vanishing, i.e., $\scrO_{\rm I, nonlocal}^{0,1}(E/V)\simeq 0$.
(d)(e) Time evolution of the operators (d) $\cO_{\rm I,\mathrm{nonlocal}}^{0,1} = W_y$ and (e) $\cO_{\rm II}^{1,0}=U_x^{(1)}\prod_{\ell\in p}\sigma^z_\ell$ for system size $(L_x,L_y)=(3,4)$.
The initial states are random superpositions of the eigenstates of $\cO_{\rm I,\mathrm{nonlocal}}^{0,1}$ and $\cO_{\rm II}^{1,0}$ with the eigenvalue $-1$,
whose energy expectation values lie within an energy window $E\in[-6.0,-5.0]$.
The predictions from the GGE and the Gibbs ensemble are shown by the blue and cyan lines, respectively.
The stationary values of these charged operators are accurately described by the GGE.
For $\cO_{\rm I,nonlocal}^{0,1}$, the predictions from the GGE and the standard Gibbs ensemble give almost the same value.
}\label{fig:Z2gauge-ETH}
\end{figure}

Figure \ref{fig:Z2gauge-ETH} shows numerical results for the $\bbZ_2$ gauge theory \eqref{Hamiltonian Z2 gauge}, where we take $L_x=3$ and $L_y=4$.
Since $L_y$ is even, $U^{(0)}$ and $U_x^{(1)}$ become the crucial symmetries exhibiting the nontrivial projective representation; $U_y^{(1)}$ is trivial in Eq.~\eqref{eq:proj-rep_u0_uxy}.
We consider that $U^{(0)}$ and $(U^{(1)}_x)^{-1}$ correspond to the symmetries $\mathbb{Z}_{N_1}$ and $\mathbb{Z}_{N_2}$, respectively.
We focus on  the following operators\footnote{Note that, by definition, the classifications Type I/II are determined after we numerically calculate matrix elements of observables.}
\begin{alignat}{2}
\cO^{0,0} &= \prod_{\ell\in \mathrm{plaquette}}\sigma^z_\ell,
&\qquad &\text{neutral operator},\\
\cO_{\rm I,local}^{1,0} &= \sigma_\ell^x,
&\qquad &\text{Type I charged local operator}, \\
\cO_{\rm I,nonlocal}^{0,1} &= W_y,
&\qquad &\text{Type I charged nonlocal operator},\\
\cO_{\rm II}^{1,0} = \cO^{0,0} U_x^{(1)}&=
U_x^{(1)}\prod_{\ell\in\mathrm{plaquette}}\sigma^z_\ell,&\qquad &\text{Type II charged operator}.
\end{alignat}
The matrix elements are well-described by a smooth function $\scrO$ of energy.
We also numerically calculate the time-evolution of the Type I charged operator and the Type II charged operator, $\cO_{\rm I,nonlocal}^{0,1}$ and $\cO_{\rm II}^{1,0}$.
We observe that the expectation values of the operators relax to the GGE prediction in \eqref{def-GGE}.
As predicted by the conjectured scaling for Type I charged operators, the stationary value of $\cO_{\rm I,nonlocal}^{1,0}$ is approximated by zero.

\clearpage

\section{Anomalous scaling}\label{Anomalous scaling}
By comparing between Eq.~\eqref{time-average-neutral-op-scaling} and Eq.~\eqref{time-average-charged-op-scaling}, we observe that
\begin{align}
    &\overline{\bra{\psiin} e^{\rmi Ht}\cO^{q_1,q_2}e^{-\rmi Ht}\ket{\psiin}}\no\\
    &=
    \begin{cases}
        \scrO^{0,0}(\Bar{\varepsilon}) + \order{V^{-1}}
        &\text{for neutral operators  $\cO^{q_1,q_2}=\cO^{0,0}$},\\
        \bra{\psiin}(U_1)^{q_2}(U_2)^{-q_1}\ket{\psiin} \scrO^{q_1,q_2}(\Bar{\varepsilon}) + \order{V^{-1/2}}
        &\text{for Type II charged operators.}
    \end{cases}
\end{align}
Note that the scaling behaviors of the correction terms are of order $\order{V^{-1}}$ (standard scaling) and $\order{V^{-1/2}}$ (anomalous scaling), respectively. Such an {anomalous} discrepancy in finite-size scaling has also been reported in systems with $SU(2)$ symmetry~\cite{Murthy:2022dao}, originating from the symmetry operators' non-commutativity~\cite{Yunger_Halpern_2016}. This non-commutativity of symmetry operators also plays a central role in giving rise to anomalous scaling in the case of projective representations. 

In contrast with the conventional scaling~$\order{V^{-1}}$, the anomalous finite-size scaling of order $\order{V^{-1/2}}$ originates from the first-order term in the expansion around~$\Bar\varepsilon$, specifically the quantity $\sum_i \bra{\psiin} (U_1)^{q_2} (U_2)^{-q_1} P_i \ket{\psiin} (\varepsilon_i - \Bar{\varepsilon})$ [see Eq.~\eqref{Time-average-charged-op-with-symmetry-op}]. This term satisfies the inequality $|\sum_i \bra{\psiin} (U_1)^{q_2} (U_2)^{-q_1} P_i \ket{\psiin} (\varepsilon_i - \Bar{\varepsilon})| \leq \order{V^{-1/2}}$, where the larger finite-size effect may remain compared with the standard scaling ${O}(V^{-1})$. In particular, if the first-order coefficient~$(\scrO^{q_1,q_2})'(\Bar{\varepsilon})$ takes a non-zero value and the above inequality is almost saturated, this anomalous scaling behavior will become relevant. However, this bound merely provides an upper limit, and it is not obvious which initial states actually lead to the anomalous scaling.
In the following, we investigate this issue in more depth.

\subsection{Typicality of anomalous scaling}\label{Typicality of anomalous scaling}

First, we discuss how typical initial states lead the anomalous scaling if we choose a random initial state from a certain measure.
To see this, we start from rewriting the time-averaged expectation value for Type II charged operators by using the prETH~\eqref{projective-ETH} as
\begin{align}
    &\overline{\bra{\psiin} e^{\rmi Ht}\cO^{q_1,q_2}e^{-\rmi Ht}\ket{\psiin}} \no\\
    &= \sum_i \bra{\psiin}(U_1)^{q_2}(U_2)^{-q_1}P_i\ket{\psiin}\left[\scrO^{q_1,q_2}(\varepsilon_i) + e^{-\order{V}/2}\right] \no\\
    &= \sum_{i}  f(\varepsilon_i)\scrO^{q_1,q_2}(\varepsilon_i)
    + \bra{\psiin} (U_1)^{q_2}(U_2)^{-q_1} \ket{\psiin} \sum_{i} \bra{\psiin} P_i \ket{\psiin} \scrO^{q_1,q_2}(\varepsilon_i) + e^{-\order{V}/2}, \label{Rewrite-Time-average-charged-op-with-symmetry-op}
\end{align}
where we have introduced $f(\varepsilon)$ by
\begin{align}
f(\varepsilon_i) :=\bra{\psiin}(U_1)^{q_2}(U_2)^{-q_1}P_i\ket{\psiin} - \bra{\psiin} P_i \ket{\psiin}\bra{\psiin} (U_1)^{q_2}(U_2)^{-q_1} \ket{\psiin}. \label{def-differrence-f(epsilon)}
\end{align}
The second term on the right-hand side in Eq.~\eqref{Rewrite-Time-average-charged-op-with-symmetry-op} becomes
\begin{align}
    &\bra{\psiin} (U_1)^{q_2}(U_2)^{-q_1} \ket{\psiin} \sum_{i} \bra{\psiin} P_i \ket{\psiin} \scrO^{q_1,q_2}(\varepsilon_i)\no\\
    &=\bra{\psiin} (U_1)^{q_2}(U_2)^{-q_1}\ket{\psiin} \scrO^{q_1,q_2}(\Bar{\varepsilon})+\order{V^{-1}},
\end{align}
contributing to the stationary value with the finite-size deviation that follows the  conventional scaling~$\order{V^{-1}}$. Therefore, the first term
depending on~$f(\varepsilon_i)$
is responsible for the anomalous scaling.

For simplicity, we consider the case that~$\gcd(\gcd(q_1, q_2), N) = 1$. In this case, we define the $N$ eigenstates $\kket{E_i, (q_1, q_2), \alpha}$ of the operator $(U_1)^{q_2}(U_2)^{-q_1}$ by
\begin{align}
    (U_1)^{q_2}(U_2)^{-q_1}\kket{E_i,(q_1,q_2),\alpha}=F_N^{(q_2, -q_1)}e^{\frac{2\pi \rmi}{N}\alpha}\kket{E_i,(q_1,q_2),\alpha},
\end{align}
with $\alpha=0$, $1$, \dots, $N-1$. The $\alpha$-independent phase~$F_N^{(q_2, -q_1)}$ is defined in Eq.~\eqref{factor FN}. Noting that, since $(U_1)^{q_2}(U_2)^{-q_1}$ is a unitary operator, $|F_N^{(q_2, -q_1)}|=1$ holds. Using this basis, the initial state can be written as
\begin{align}
    &\ket{\psiin} = \sum_{i} \sum_{\alpha=0}^{N-1} c_{i,\alpha} \kket{E_i, (q_1, q_2), \alpha},\label{non trivial based initial state}
\end{align}
which behaves as
\begin{align}
    &\bra{\psiin}(U_1)^{q_2}(U_2)^{-q_1}P_i\ket{\psiin} = F_N^{(q_2, -q_1)}\sum_{\alpha=0}^{N-1} e^{\frac{2\pi \rmi}{N}\alpha} |c_{i,\alpha}|^2,\label{non trivial based conserved}\\
    &\bra{\psiin} P_i \ket{\psiin} = \sum_{\alpha=0}^{N-1} |c_{i,\alpha}|^2.\label{non trivial based distribution}
\end{align}

We aim to assess how \textit{typical} it is for randomly chosen initial states $\ket{\psiin}$ to satisfy $\sum_i  f(\varepsilon_i)\scrO^{q_1,q_2}(\varepsilon_i) = \order{V^{-\gamma}}$, where the scaling exponent $\gamma$ lies in the range $0 \leq \gamma \leq 1/2$.
To this end, we consider a probability distribution over initial states $\ket{\psiin}$ under the constraint that the weight in each energy subspace is fixed as Eq.~\eqref{non trivial based distribution}. That is, the distribution, $\Pop$, is defined such that the probability ${|c_{i,\alpha}|^2}$ for each energy sector sum is set to a prescribed value $p_i$;
\begin{align}
    \Pop[\{c_{i,\alpha}\}] \prod_{i,\alpha} \rmd\re c_{i,\alpha} \rmd\im c_{i,\alpha} \propto \prod_{i,\alpha} \rmd\re c_{i,\alpha} \rmd\im c_{i,\alpha} \prod_i \delta\left( \sum_{\alpha=0}^{N-1} |c_{i,\alpha}|^2 - p_i \right). \label{randomly-choose}
\end{align}
Here, $p_i$ is a fixed probability distribution satisfying $\sum_ip_i(E_i-\Bar{E})^2=\order{V}$, which is similar to Eq.~\eqref{Var-energy}. The expectation value and variance under this probability distribution are denoted by $\Eop\left[\cdot\right]$ and $\Vop\left[\cdot\right]$, respectively.
Under the probability distribution~\eqref{randomly-choose}, we find the vanishing expectation value of $f(\varepsilon_i)$ as
\begin{align}
    \Eop\left[f(\varepsilon_i)\right]
        &= \Eop\left[F_N^{(q_2, -q_1)}\sum_{\alpha=0}^{N-1}e^{\frac{2\pi \rmi}{N}\alpha}|c_{i,\alpha}|^2 -p_i F_N^{(q_2, -q_1)}\sum_i\sum_{\alpha=0}^{N-1}e^{\frac{2\pi \rmi}{N}\alpha}|c_{i,\alpha}|^2\right]\no\\
    &= e^{\frac{2\pi \rmi}{N}} \Eop[f(\varepsilon_i)]
    = 0.
\end{align}
Here, in the second equality, we have performed the shift of the variable $c_{i,\alpha}\to c_{i,\alpha+1}$ in the integration~\eqref{randomly-choose}.\footnote{When we have $m := N / \gcd(\gcd(q_1, q_2), N) \neq N$, i.e., $\gcd(\gcd(q_1, q_2), N)\neq1$,
\begin{align}
    &\bra{\psiin}P_i(U_1)^{q_2}(U_2)^{-q_1}P_i\ket{\psiin} = \Tilde{F}_N^{(q_2, -q_1)}\sum_{\alpha=0}^{\frac{N}{m}-1} e^{\frac{2\pi \rmi}{m}\alpha} \sum_{\beta=0}^{\frac{N}{m}-1} |c_{i,\alpha,\beta}|^2\no\\
    &\Tilde{F}_N^{(q_2, -q_1)}:=e^{-\frac{\pi i}{N}\Tilde{q}_1\Tilde{q}_2\gcd(q_1, q_2)\{\gcd(q_1, q_2)-1\}} \{F_N^{(\Tilde{q}_2, -\Tilde{q}_1)}\}^{\gcd(q_1, q_2)}
    \label{non-trivial degeneracy}
\end{align}
Therefore, the same result $\Eop\left[f(\varepsilon_i)\right] = 0$ can be obtained by using the shift~$c_{i,\alpha,\beta} \to c_{i,\alpha+1,\beta}$ in this representation~\eqref{non-trivial degeneracy}. To do this, in Appendix~\ref{sec:app_diag}, we show how to diagonalize $(U_1)^{q_2}(U_2)^{-q_1}$.}
Since the random variables $c_{i,\alpha}$ for different values of $i$ are mutually independent, we can compute $f(\varepsilon_i)^*f(\varepsilon_j)$ as
\begin{align}
    \Eop\left[f(\varepsilon_i)^* f(\varepsilon_j)\right]
    &=  \Eop\left[ \left| \bra{\psiin}(U_1)^{q_2}(U_2)^{-q_1}P_i\ket{\psiin} \right|^2 \right] \delta_{ij} \no \\&\qquad
    - p_i \Eop\left[ \left| \bra{\psiin}(U_1)^{q_2}(U_2)^{-q_1} P_j \ket{\psiin} \right|^2 \right] \no \\&\qquad
    - p_j \Eop\left[ \left| \bra{\psiin}(U_1)^{q_2}(U_2)^{-q_1}P_i\ket{\psiin} \right|^2 \right] \no \\&\qquad
    + \sum_k (p_k)^2 \Eop\left[ \left| \bra{\psiin} (U_1)^{q_2}(U_2)^{-q_1} P_k \ket{\psiin} \right|^2 \right].
\end{align}
After some calculations, we obtain,
\begin{align}
    &\Eop\left[ \sum_i f(\varepsilon_i) \scrO^{q_1,q_2}(\varepsilon_i) \right] = 0,
\end{align}
and
\begin{align}
    &\Vop\left[ \sum_i f(\varepsilon_i) \scrO^{q_1,q_2}(\varepsilon_i) \right]
    = \Eop\left[ \left| \sum_i f(\varepsilon_i) \scrO^{q_1,q_2}(\varepsilon_i) \right|^2 \right] \no \\
    &= \sum_i \Eop\left[ \left| \bra{\psiin}(U_1)^{q_2}(U_2)^{-q_1}P_i\ket{\psiin} \right|^2 \right] |\scrO^{q_1,q_2}(\varepsilon_i)|^2 \no \\
    &\qquad
    +\left| \sum_i p_i \scrO^{q_1,q_2}(\varepsilon_i) \right|^2 \sum_k \Eop\left[ \left| \bra{\psiin} (U_1)^{q_2}(U_2)^{-q_1} P_k \ket{\psiin} \right|^2 \right] \no \\
    &\qquad
    -2\re\left[ \left( \sum_i p_i \scrO^{q_1,q_2}(\varepsilon_i)^* \right) \sum_j \Eop\left[ \left| \bra{\psiin}(U_1)^{q_2}(U_2)^{-q_1} P_j \ket{\psiin} \right|^2 \right] \scrO^{q_1,q_2}(\varepsilon_j) \right].\label{long calculation V}
\end{align}
Using the Cauchy--Schwartz inequality for an absolute value of Eq.~\eqref{non trivial based conserved}, we obtain
\begin{align}
    \Eop\left[ \left| \bra{\psiin}(U_1)^{q_2}(U_2)^{-q_1}P_i\ket{\psiin} \right|^2 \right] \leq p_i^2.
    \label{expectation value of pi2}
\end{align}
Substituting Eq.~\eqref{expectation value of pi2} into Eq.~\eqref{long calculation V} and the assumption that $||\cO^{q_1,q_2}||=\order{V^0}$, and applying the Chebyshev's inequality, we obtain
\begin{align}
    \Pop \left[ \left| \sum_i f(\varepsilon_i) \scrO^{q_1,q_2}(\varepsilon_i) \right| \geq \delta \right]
    \leq \frac{\Vop\left[ \sum_i f(\varepsilon_i) \scrO^{q_1,q_2}(\varepsilon_i) \right]}{\delta^2}
    \leq\order{V^0}\times\frac{\sum_i(p_i)^2}{\delta^2}.
\end{align}
By substituting $\delta = \order{V^{-\gamma}}$, which corresponds to anomalous scaling, we arrive at
\begin{align}
    \Pop \left[ \left| \sum_i f(\varepsilon_i) \scrO^{q_1,q_2}(\varepsilon_i) \right| \geq \order{V^{-\gamma}} \right]
    \leq \order{V^{2\gamma}}\sum_i(p_i)^2.\label{bound Chebyshev}
\end{align}
Here, $\left(\sum_i (p_i)^2\right)^{-1}$ is called the \textit{effective dimension}, which estimates how many energy eigenstates are superposed in the initial state. In many situations,  one finds
$\left(\sum_i(p_i)^2\right)^{-1}\approx e^{\order{V}}$ (for example, if we take a randomly chosen state in the microcanonical energy shell, we find this scaling~\cite{reimann2008foundation}). In this case, we obtain
\begin{align}
    \Pop \left[ \left| \sum_i f(\varepsilon_i)\, \scrO^{q_1,q_2}(\varepsilon_i) \right| \geq \order{V^{-\gamma}} \right]
    \leq e^{-\order{V}} \times \order{V^{2\gamma}} \approx e^{-\order{V}}.\label{bound Chebyshev typicality}
\end{align}
Therefore, under the large effective dimension, initial states that possess an anomalous scaling~$\order{V^{-\gamma}}>\order{V^{-1}}$ are \textit{exponentially atypical} with respect to $\mathbb{P}$.

However, noting that the bound in the inequality~\eqref{bound Chebyshev} is determined by the effective dimension, one finds that if the effective dimension is sufficiently small, the exponentially small upper bound in Eq.~\eqref{bound Chebyshev typicality} can be avoided. 
In other words, when an initial state is constructed from a superposition of a small number of energy eigenstates, anomalous scaling may occur. 
A detailed discussion of this point will be provided in the next subsection.

\subsection{When does anomalous scaling occur?}

In what follows, we examine what kinds of initial states can lead to anomalous scaling behavior.
We define the standard deviation of the energy as~$\Delta\epsilon := \left[\bra{\psiin}\left(\varepsilon - \Bar{\varepsilon}\right)^2\ket{\psiin}\right]^{\frac{1}{2}}\leq\order{V^{-\frac{1}{2}}}$.
We consider a scenario in which the function $f(\varepsilon_i)$ is localized around two distinct regions:
(i) near $\Bar{\varepsilon} + \Delta\epsilon$, and
(ii) in a region where $|\varepsilon_i - \Bar{\varepsilon}| \ll \Delta\epsilon$.
In regions where $|\varepsilon_i - \Bar{\varepsilon}| \gg \Delta\epsilon$, we can assume that  the spectral weight $\bra{\psiin} P_i \ket{\psiin}$ becomes negligible, i.e., $0\approx|\bra{\psiin} P_i \ket{\psiin}|\geq|\bra{\psiin}(U_1)^{q_2}(U_2)^{-q_1}P_i\ket{\psiin}|$, which lead to~$\sum_{\Delta\epsilon \ll |\varepsilon_i - \Bar{\varepsilon}|}f(\varepsilon_i)(\varepsilon_i - \Bar{\varepsilon}) \approx 0$. This implies that contributions from states far from the mean energy are suppressed, and therefore, large deviations from $\Bar{\varepsilon}$ do not affect the scaling behavior.

In this case, the sum can be approximately decomposed as
\begin{align}
    \sum_{i}f(\varepsilon_i)
    \approx\sum_{\varepsilon_i \approx \Bar{\varepsilon}+\Delta\epsilon}f(\varepsilon_i)
    +\sum_{|\varepsilon_i - \Bar{\varepsilon}|\ll\Delta\epsilon}f(\varepsilon_i).
\end{align}
Noting that $\sum_{i}f(\varepsilon_i)=0$, we see
\begin{align}
\left|\sum_{\varepsilon_i\approx\Bar{\varepsilon}+\Delta\epsilon}f(\varepsilon_i)\right|
    \approx
    \left|\sum_{|\varepsilon_i - \Bar{\varepsilon}|\ll\Delta\epsilon}f(\varepsilon_i)\right|\gg
\left|\sum_{i\in\text{otherwise}}f(\varepsilon_i)\right|.
\end{align}
Then, we find
\begin{align}
    \sum_{i}f(\varepsilon_i)(\varepsilon_i - \Bar{\varepsilon})
    &\approx \sum_{\varepsilon_i \approx \Bar{\varepsilon} + \Delta\epsilon}f(\varepsilon_i)(\varepsilon_i - \Bar{\varepsilon})
    + \sum_{|\varepsilon_i - \Bar{\varepsilon}|\ll\Delta\epsilon}f(\varepsilon_i)(\varepsilon_i - \Bar{\varepsilon}) \no\\
    &\approx \sum_{\varepsilon_i \approx \Bar{\varepsilon} + \Delta\epsilon}f(\varepsilon_i)\Delta\epsilon \approx \order{V^{-\frac{1}{2}}}.
\end{align}
Finally, we have
\begin{align}
    \sum_{i}f(\varepsilon_i)\scrO^{q_1,q_2}(\varepsilon_i) = \sum_{i}f(\varepsilon_i)(\varepsilon_i - \Bar{\varepsilon})(\scrO^{q_1,q_2})'(\bar{\varepsilon}) + \order{V^{-1}} \approx \order{V^{-\frac{1}{2}}},
\end{align}
where we have assumed $(\scrO^{q_1,q_2})'(\Bar{\varepsilon})=\order{V^0}$.
A similar analysis applies when the distribution of $f(\varepsilon_i)$ is localized around $\Bar{\varepsilon} - \Delta\epsilon$ instead of $\Bar{\varepsilon} + \Delta\epsilon$. More generally, when the anomalous scaling is of order $\order{V^{-\gamma}}$ with $0 < \gamma \leq 1/2$, the relevant localization region shifts to $\varepsilon_i \approx \Bar{\varepsilon} + \order{V^{-\gamma}}$, rather than being tied to the standard deviation $\Delta\epsilon$.

In summary, our analysis indicates that anomalous thermalization emerges when the profile of $f(\varepsilon_i)$ is effectively localized around specific energy windows.
To see this, we provide an explicit example of an initial state that leads to anomalous scaling.  
For instance, let $E=\order{V}$ and $\Delta E=\order{V^{1/2}}$, and prepare the following state:
\begin{align}
    \ket{\psi_{\text{an}}}:=\frac{1}{2}\left(\kket{E-2\Delta E}+\kket{E}+(1+U_2)\kket{E+\Delta E}\right).
\end{align}
Here we can see the profile of~$f$ with vanishing $q_2$ explicitly
\begin{align}
    f(\varepsilon_i) =
    \begin{cases}
        - \frac{1}{16}(\delta_{q_1,1}+\delta_{q_1,-1})
        & \text{for $E_i = E-2\Delta E$, $E$} ,\\
        \frac{1}{8}(\delta_{q_1,1}+\delta_{q_1,-1})
        & \text{for $E_i = E+\Delta E$} .
    \end{cases}
\end{align}
By construction, this state satisfies 
\begin{align}
    \bra{\psi_{\text{an}}}H\ket{\psi_{\text{an}}}=E=\order{V},\qquad 
    \bra{\psi_{\text{an}}}(H-E)^2\ket{\psi_{\text{an}}}=\Delta E^2/2=\order{V},
\end{align}
and thus belongs to the same class of states considered so far.  
Taking this state as the initial state, we evaluate the long-time average of the expectation value of the Type II operator~$\cO^{1,0}$ using prETH:
\begin{align}
    \overline{\bra{\psi_\text{an}} e^{\rmi Ht}\cO^{1,0}e^{-\rmi Ht}\ket{\psi_\text{an}}}
    &=\frac{1}{4}\bbra{E+\Delta E}\cO^{1,0}U_2\kket{E+\Delta E}\no\\
    &=\frac{1}{4}\scrO^{1,0}(E/V+\Delta E/V)\no\\
    &=\frac{1}{4}\scrO^{1,0}(E/V)+\order{V^{-\frac{1}{2}}}.
\end{align}
In going from the second to the third line, we use the Taylor expansion of $\scrO^{1,0}$ together with the scaling $\Delta E=\order{V^{1/2}}$.  
Noting that the prefactor $1/4$ is consistent with $\bra{\psi_{\text{an}}}(U_2)^\dagger\ket{\psi_{\text{an}}}=1/4$.

\section{Conclusion}\label{sec:conclusion}
In this work, we have examined the validity of the Eigenstate Thermalization Hypothesis (ETH) in quantum many-body systems exhibiting nontrivial projective representations of Abelian symmetries. While conventional ETH mainly assumes non-degenerate energy eigenstates, we have pointed out that 't~Hooft anomalies can naturally induce degeneracies even in excited states through projective symmetry structures. 
We have formulated the projective-representation ETH (prETH) under such degeneracies due to such projective representations.

Notably, when the operator under consideration carries projective charges sourced from the symmetry operators themselves, our prETH predicts that the standard Gibbs ensemble fails to describe the stationary value of the operator.
Instead, we argue that the  stationary value is correctly captured by an appropriate generalized Gibbs ensemble. These results suggest that projective representations provide a new avenue for exploring the interplay between symmetry, degeneracy, and thermalization in isolated quantum systems.
Our findings not only shed light on the generalization of ETH to anomalous symmetry settings but also open the door to studying more exotic thermalization dynamics in gauge theories, systems with higher-form symmetries, and beyond. 

\section*{Acknowledgment}
R.H.\ was supported by JST ERATO Grant Number JPMJER2302, Japan.
This work was partially supported by Japan Society for the Promotion of Science (JSPS)
Grant-in-Aid for Scientific Research Grant Numbers
JP24K22890 (O.F.), JP24K16982 (R.H.), JP25K17402 (O.M.), JP25KJ1954 (S.O.).
O.M.\ acknowledges the RIKEN Special Postdoctoral Researcher Program
and RIKEN FY2025 Incentive Research Projects.

\appendix

\section{Projective representation on more general case: $G = \bbZ_{N_1} \times \bbZ_{N_2}$}\label{N_1, N_2}

Here, we consider the case where the Hamiltonian $H$ possesses a $\bbZ_{N_1} \times \bbZ_{N_2}$ symmetry with $N_1\neq N_2$. In the operator formalism, we denote the operators corresponding to the subgroup $\bbZ_{N_1}$ and $\bbZ_{N_2}$ symmetries by $U_1$ and $U_2$, respectively. That is, the representation of $\bbZ_{N_i}$ on the Hilbert space is given by $(U_i)^\alpha$ for $\alpha = 0$, $1$, \dots, $N_i-1$ with $i=1$ or $2$.
They satisfy
\begin{align}
    (U_1)^{N_1} &= 1, & (U_2)^{N_2} &= 1, \label{rep-of-Z_{1,2}}\\
    [(U_1)^\alpha, (U_1)^\beta] &= 0, & [(U_2)^\alpha, (U_2)^\beta] &= 0, \label{anomaly-free-general}\\
    U_2 U_1 &= e^{-\frac{2\pi \rmi}{n}} U_1 U_2 \label{Projective-representation-general}
\end{align}
for $\forall \alpha$, $\beta \in \bbZ$. Equation~\eqref{rep-of-Z_{1,2}} indicates that $U_{1,2}$ is a representation of $\bbZ_{N_{1,2}}$ symmetry; Eq.~\eqref{anomaly-free-general} shows that the subgroups $\bbZ_{N_{1,2}}$ are anomaly-free; and the third line expresses a projective representation corresponding to a mixed 't~Hooft anomaly between $\bbZ_{N_1}$ and $\bbZ_{N_2}$. Here, $n$ is taken to be any common divisor of $N_1$ and $N_2$.

For two integers,
\begin{align}
    m_1 := \frac{N_1}{n}, \qquad m_2 := \frac{N_2}{n}, \qquad \text{$m_1$, $m_2 \in \bbZ$}.
\end{align}
the elements
\begin{align}
    (U_1)^{k_1 n},\quad \text{$k_1 = 0$, $1$, \dots, $m_1 - 1$}; \quad (U_2)^{k_2 n},\quad \text{$k_2 = 0$, $1$, \dots, $m_2 - 1$}
\end{align}
commute with all $(U_1)^\alpha$ and $(U_2)^\beta$, and thus the subgroup $\bbZ_{m_1} \times \bbZ_{m_2}$ can be interpreted as the center. Let us now simultaneously diagonalize these with~$H$. That is, for each energy eigenvalue $E_i$, we define an eigenstate $\kket{E_i, (x_i^1, x_i^2)}$ such that
\begin{align}
    H \kket{E_i, (x_i^1, x_i^2)} &= E_i \kket{E_i, (x_i^1, x_i^2)},\label{energy eigenstate}\\
    (U_1)^n \kket{E_i, (x_i^1, x_i^2)} &= e^{\frac{2\pi \rmi x_i^1}{m_1}} \kket{E_i, (x_i^1, x_i^2)},\label{diagonal center N1}\\
    (U_2)^n \kket{E_i, (x_i^1, x_i^2)} &= e^{\frac{2\pi \rmi x_i^2}{m_2}} \kket{E_i, (x_i^1, x_i^2)},\label{diagonal center N2}\\
    U_1 \kket{E_i, (x_i^1, x_i^2)} &= e^{\frac{2\pi \rmi x_i^1}{N_1}} \kket{E_i, (x_i^1, x_i^2)}.
\end{align}
for $x_i^1 = 0$, $1$, \dots, $m_1 - 1$, and $x_i^2 = 0$, $1$, \dots, $m_2 - 1$. We assume that no degeneracy arises from the anomaly-free subgroup $\bbZ_{m_1} \times \bbZ_{m_2}$. In other words, each label $(x_i^1, x_i^2)$ is uniquely associated with a distinct state indexed by $i$. On the other hand, there exist $n$ distinct eigenvalues labeled by the projective representation in Eq.~\eqref{Projective-representation-general}; eigenvalues of $U_1$ (or $U_2$) can be classified into $n$ distinct types as
\begin{align}
    \exp\left(\frac{2\pi \rmi x_i^1}{n m_1} + \frac{2\pi \rmi \alpha}{n}\right) = \exp\left(\frac{2\pi \rmi x_i^1}{N_1} + \frac{2\pi \rmi \alpha}{n}\right),\qquad \text{$\alpha = 0$, \dots, $n - 1$}.
\end{align}
The eigenstate is not neutral under the action of $U_1$ in contrast to the main text.
Then, we find that, for $\alpha = 0$, \dots, $n - 1$,
\begin{align}
    H (U_2)^\alpha \kket{E_i, (x_i^1, x_i^2)} &= E_i (U_2)^\alpha \kket{E_i, (x_i^1, x_i^2)},\\
    U_1 (U_2)^\alpha \kket{E_i, (x_i^1, x_i^2)} &= \exp\left(\frac{2\pi \rmi x_i^1}{N_1} + \frac{2\pi \rmi \alpha}{n}\right) (U_2)^\alpha \kket{E_i, (x_i^1, x_i^2)}.
\end{align}
Here, the index $\alpha$ can be interpreted as labeling eigenstates of $U_1$ in the degeneracy subspace arising from the projective representation.
Since the eigenstates corresponding to all possible eigenvalues of $U_1$ are generated by $(U_2)^\alpha \kket{E_i, (x_i^1, x_i^2)}$, the Hilbert space is spanned by these states. A general state can be expressed as
\begin{align}
    \ket{\psi} = \sum_i \sum_{\alpha = 0}^{n - 1} z_{i, \alpha} (U_2)^\alpha \kket{E_i, (x_i^1, x_i^2)},
\end{align}
where the coefficients $z_{i,\alpha} \in \mathbb{C}$ satisfy the normalization condition $\sum_i \sum_{\alpha=0}^{n-1} |z_{i,\alpha}|^2 = 1$.

\subsection{Time-averaged expectation value}

We define the initial state as
\begin{align}
    \ket{\psiin} := \sum_{i}\sum_{\alpha=0}^{n-1}c_{i,\alpha}(U_2)^\alpha\kket{E_i,(x_i^1,x_i^2)},
\end{align}
with the conditions in Eqs.~\eqref{Energy} and \eqref{Var-energy}. In this case, the time-averaged expectation value of the charged operator $\cO^{q_1,q_2}$ can be written as
\begin{align}
    &\lim_{T\to\infty}\frac{1}{T}\int_0^{T}\rmd t\bra{\psiin} e^{\rmi Ht}\cO^{q_1, q_2}e^{-\rmi Ht}\ket{\psiin} \no\\
    &= \lim_{T\to\infty}\frac{1}{T}\int_0^{T}\rmd t\sum_{i,j}\sum_{\alpha, \beta}c_{i,\alpha}^* c_{j,\beta}e^{\rmi(E_i-E_j)t}\bbra{E_i,(x_i^1,x_i^2)}(U_2)^{-\alpha} \cO^{q_1, q_2}(U_2)^\beta\kket{E_j,(x_j^1,x_j^2)} \no\\
    &= \sum_{i}\sum_{\alpha, \beta}c_{i,\alpha}^* c_{i,\beta}\bbra{E_i,(x_i^1,x_i^2)}(U_2)^{-\alpha} \cO^{q_1, q_2}(U_2)^\beta\kket{E_i,(x_i^1,x_i^2)}.
\end{align}
Here, we have used the identity $\lim_{T\to\infty}\frac{1}{T}\int_0^{T}\rmd t\,e^{i(E_i-E_j)t}=\delta_{ij}$, assuming that there are no degeneracies aside from the ones due to the projective representation. From Eqs.~\eqref{diagonal center N1} and~\eqref{diagonal center N2}, we obtain
\begin{align}
    &\bbra{E_i,(x_i^1,x_i^2)}(U_2)^{-\alpha} \cO^{q_1, q_2}(U_2)^\beta\kket{E_i,(x_i^1,x_i^2)} \no\\
    &= \bbra{E_i,(x_i^1,x_i^2)}(U_1)^{-n}(U_2)^{-n}(U_2)^{-\alpha} \cO^{q_1, q_2}(U_2)^\beta(U_2)^{n}(U_1)^{n}\kket{E_i,(x_i^1,x_i^2)} \no\\
    &= e^{\frac{2\pi i}{m_1}q_1}e^{\frac{2\pi i}{m_2}q_2}\bbra{E_i,(x_i^1,x_i^2)}(U_2)^{-\alpha} \cO^{q_1, q_2}(U_2)^\beta\kket{E_i,(x_i^1,x_i^2)}.
\end{align}
The selection rules become
\begin{align}
    &\bbra{E_i,(x_i^1,x_i^2)}(U_2)^{-\alpha} \cO^{q_1, q_2}(U_2)^\beta\kket{E_i,(x_i^1,x_i^2)} = 0&
    &\text{for $(q_1,q_2)\notin(m_1\bbZ,m_2\bbZ)$},
    \label{selection-rule-matrix-element-by-center}
\end{align}
and hence
\begin{align}
    \lim_{T\to\infty}\frac{1}{T}\int_0^{T}\rmd t\bra{\psiin} e^{\rmi Ht}\cO^{q_1, q_2}e^{-\rmi Ht}\ket{\psiin} = 0.
    \label{selection-rule-by-center}
\end{align}

On the other hand, for $(q_1,q_2)\in(m_1\bbZ,m_2\bbZ)$, we can obtain nonzero matrix elements. For $(q_1',q_2') := (q_1/m_1,q_2/m_2)$, we define $\fO^{q_1', q_2'}$ by
\begin{align}
    \fO^{q_1', q_2'}&:=\cO^{m_1q_1', m_2q_2'},&
    U_1^\dagger \fO^{q_1', q_2'}U_1 &= e^{\frac{2\pi i}{n}q_1'}\fO^{q_1', q_2'},& 
    U_2^\dagger \fO^{q_1', q_2'}U_2 &= e^{\frac{2\pi i}{n}q_2'}\fO^{q_1', q_2'}.
    \label{dashed-Charged-operator}
\end{align}
Using this, the time-averaged expectation value can be computed in the same manner as in Section~\eqref{subsection:Time-averaged expectation value}:
\begin{align}
    &\lim_{T\to\infty}\frac{1}{T}\int_0^{T}\rmd t\bra{\psiin} e^{\rmi Ht}\fO^{q_1', q_2'}e^{-\rmi Ht}\ket{\psiin} \no\\
    &= \sum_i\bra{\psiin}(U_1)^{q_2'}(U_2)^{-q_1'}P_i\ket{\psiin}\bbra{E_i,(x_i^1,x_i^2)} \fO^{q_1', q_2'}(U_2)^{q_1'}(U_1)^{-q_2'}\kket{E_i,(x_i^1,x_i^2)}.
    \label{Time-averaged-expectation-value-general}
\end{align}

\subsection{Violation of diagonal prETH}
Here, we explain that the diagonal prETH can be violated due to the presence of symmetry operators of the form $(U_1)^{k_1 n}(U_2)^{k_2 n}$ that do not generate degeneracies.  
Let us consider a charged operator defined by $\tilde{\fO}^{q_1', q_2'}=\fO^{q_1', q_2'}(U_1)^{n}(U_2)^{n}$. Then one finds
\begin{align}
    &\bbra{E_i,(x_i^1,x_i^2)} \tilde{\fO}^{q_1', q_2'}(U_2)^{q_1'}(U_1)^{-q_2'}\kket{E_i,(x_i^1,x_i^2)}\no\\
    &=e^{\frac{2\pi\rmi x_i^1}{m_1}}\,e^{\frac{2\pi\rmi x_i^2}{m_2}}
    \bbra{E_i,(x_i^1,x_i^2)}\fO^{q_1', q_2'}(U_2)^{q_1'}(U_1)^{-q_2'}\kket{E_i,(x_i^1,x_i^2)}.
\end{align}
Both sides in the above equation can be nontrivial since there exist Type~II-like operators defined in the main text, which take nonzero expectation values.
Since the exponent $e^{\frac{2\pi\rmi x_i^1}{m_1}}e^{\frac{2\pi\rmi x_i^2}{m_2}}$ varies discretely under changes in $(x_i^1,x_i^2)$, at least one of diagonal terms
\[\bbra{E_i,(x_i^1,x_i^2)} \tilde{\fO}^{q_1', q_2'}(U_2)^{q_1'}(U_1)^{-q_2'}\kket{E_i,(x_i^1,x_i^2)},\quad\bbra{E_i,(x_i^1,x_i^2)} \fO^{q_1', q_2'}(U_2)^{q_1'}(U_1)^{-q_2'}\kket{E_i,(x_i^1,x_i^2)}\]
cannot be a smooth function of energy density~$\varepsilon_i$ as in the first term of Eq.~\eqref{projective-ETH}.  
In other words, either $\tilde{\fO}^{q_1', q_2'}$ or $\fO^{q_1', q_2'}$ necessarily violates the diagonal prETH.

In the case of $0$-form symmetry, it is natural that prETH is violated because $(U_1)^{n}(U_2)^{n}$ is generally a non-local operator. However, for higher-form symmetries, in the infinite volume limit, the support of $(U_1)^{n}(U_2)^{n}$ can become infinitesimally small compared to the thermal bath (system). Therefore, due to the symmetry structure, sufficiently local operators exist that violate prETH~(similar to Ref.~\cite{Fukushima:2023svf}). 

Nevertheless, even when an operator explicitly violates prETH due to the symmetry structure,  prETH can hold within a symmetry sector characterized by fixed $(x_i^1,x_i^2)$. That is, for $i$, $j$ that satisfy $(x_i^1,x_i^2)=(x_j^1,x_j^2)=(x^1,x^2)$,
\begin{align}
    &\bbra{E_i,(x^1,x^2)} \fO^{q_1', q_2'}(U_2)^{q_1'}(U_1)^{-q_2'}\kket{E_j,(x^1,x^2)}\no\\
    &=
    \scrO^{(q_1',q_2')}_{(x^1,x^2)}(\cE/V)\delta_{ij}+
    e^{-S_{(x^1,x^2)}(\cE)/2}f^{(q_1',q_2')}_{(x^1,x^2)}(\cE/V,\omega)R_{ij}^{(x^1,x^2)}.\label{projective-ETH fixed symmetry sector}
\end{align}
Here, the variables defined in Eq.~\eqref{projective-ETH fixed symmetry sector} are the same as those in Eq.~\eqref{projective-ETH}, except that the dependence on the symmetry sector~$(x^1,x^2)$ is written explicitly. We note that $\scrO^{(q_1',q_2')}_{(x^1,x^2)}(\cE/V)$ is a smooth function of $\cE/V$. Using Eq.~\eqref{projective-ETH fixed symmetry sector} and a Taylor expansion around $\Bar{\varepsilon}$, the time-averaged expectation value~\eqref{Time-averaged-expectation-value-general} can be evaluated as
\begin{align}
    &\lim_{T\to\infty}\frac{1}{T}\int_0^{T}\rmd t\bra{\psiin} e^{\rmi Ht}\fO^{q_1', q_2'}e^{-\rmi Ht}\ket{\psiin} \no\\
    &= \sum_i\bra{\psiin}(U_1)^{q_2'}(U_2)^{-q_1'} P_i\ket{\psiin}\bbra{E_i,(x_i^1,x_i^2)} \fO^{q_1', q_2'}(U_2)^{q_1'}(U_1)^{-q_2'}\kket{E_i,(x_i^1,x_i^2)}\no\\
    &=\sum_{x_1, x_2}\sum_{i:(x_i^1,x_i^2)=(x^1,x^2)}\bra{\psiin}(U_1)^{q_2'}(U_2)^{-q_1'}P_i\ket{\psiin}\left(\scrO^{(q_1',q_2')}_{(x^1,x^2)}(\varepsilon_i)+e^{-\order{V}/2}\right)\no\\
    &=\sum_{x_1, x_2}\bra{\psiin}(U_1)^{q_2'}(U_2)^{-q_1'}P_{x^1}P_{x^2}\ket{\psiin}\scrO^{(q_1',q_2')}_{(x^1,x^2)}(\Bar{\varepsilon})\no\\
    &\quad+\sum_{x_1, x_2}\sum_{n\geq1}\left(\scrO^{(q_1',q_2')}_{(x^1,x^2)}\right)^{(n)}(\Bar{\varepsilon})\sum_{i:(x_i^1,x_i^2)=(x^1,x^2)}\bra{\psiin}(U_1)^{q_2'}(U_2)^{-q_1'}P_i\ket{\psiin}(\varepsilon_i-\Bar{\varepsilon})^n+e^{-\order{V}/2}.
\end{align}
 Here, we have introduced projection operators onto the symmetry sectors defined as
 \begin{align}
     P_{x^1}&:=\frac{1}{m_1}\sum_{\alpha=0}^{m_1-1}(U_1)^{n\alpha},\no\\
     P_{x^2}&:=\frac{1}{m_2}\sum_{\alpha=0}^{m_2-1}(U_2)^{n\alpha},
 \end{align}
which satisfy the following property:
 \begin{align}
     \sum_{i:(x_i^1,x_i^2)=(x^1,x^2)}P_i=P_{x^1}P_{x^2}.
 \end{align}
 Since $n\geq1$, the higher order terms can be bounded in the same manner as in Eq.~\eqref{n-geq-2-inequality-with-charged-op} under certain assumptions (see the discussion there),
 \begin{align}
     &\left|\sum_{i:(x_i^1,x_i^2)=(x^1,x^2)}\bra{\psiin}(U_1)^{q_2'}(U_2)^{-q_1'}P_i\ket{\psiin}(\varepsilon_i-\Bar{\varepsilon})^n\right|\no\\
     &\qquad\leq
     \sum_{i:(x_i^1,x_i^2)=(x^1,x^2)}\bra{\psiin}P_i\ket{\psiin}|\varepsilon_i-\Bar{\varepsilon}|^n\no\\
     &\qquad\leq
\sum_{i}\bra{\psiin}P_i\ket{\psiin}|\varepsilon_i-\Bar{\varepsilon}|^n=\order{V^{-\frac{1}{2}}}.
 \end{align}
 Hence, we finally obtain
 \begin{align}
     &\lim_{T\to\infty}\frac{1}{T}\int_0^{T}\rmd t\bra{\psiin} e^{\rmi Ht}\fO^{q_1', q_2'}e^{-\rmi Ht}\ket{\psiin} \no\\
    &=\sum_{x_1, x_2}\bra{\psiin}(U_1)^{q_2'}(U_2)^{-q_1'}P_{x^1}P_{x^2}\ket{\psiin}\scrO^{(q_1',q_2')}_{(x^1,x^2)}(\Bar{\varepsilon})+\order{V^{-\frac{1}{2}}}.
 \end{align}


\subsection{Generalized Gibbs Ensemble}

Here, we give a GGE similar to that in Section~\ref{Generalized Gibbs Ensemble}. The $N_1N_2$-parameters $\mu_{r_1,r_2}$ and $\beta$ are tuned so that the following properties are satisfied:
\begin{align}
    \rhogge&:=\frac{1}{\cZ}\exp\left(-\beta H-\sum_{r=1}^{N_1N_2-1}\mu_{r} Q^r\right),\label{def-GGE-general}\\
    \cZ&:=\tr\left[\exp\left(-\beta H-\sum_{r=1}^{N_1N_2-1}\mu_{r} Q^r\right)\right],\label{Partition-function-GGE-general}\\
    \tr{\rhogge H}&=\bra{\psiin}H\ket{\psiin}=\Bar{E},\label{tuning-energy-general}\\
    \tr{\rhogge(U_1)^{q_2}(U_2)^{-q_1}}&=\bra{\psiin}(U_1)^{q_2}(U_2)^{-q_1}\ket{\psiin}\quad\forall(q_1,q_2)\in\bbZ_{N_1}\times\bbZ_{N_2}\label{tuning-conserved-general}.
\end{align}
where $Q^r$ are linearly independent Hermitian operators constructed as linear combinations of the conserved quantities $(U_1)^{r_1}(U_2)^{r_2}$ for $r_i=0$, $1$, \dots, $N_i-1$~($i=1,2$). 
When the condition~\eqref{tuning-conserved-general} is satisfied, one can also obtain the following relation:
\begin{align}
&\forall(q_1',q_2')\in\bbZ_{n}\times\bbZ_{n},\quad\forall(x_1,x_2)\in\bbZ_{m_1}\times\bbZ_{m_2},\no\\
     &\tr{\rhogge(U_1)^{q_2'}(U_2)^{-q_1'}P_{x^1}P_{x^2}}=\bra{\psiin}(U_1)^{q_2'}(U_2)^{-q_1'}P_{x^1}P_{x^2}\ket{\psiin}.\label{tuning-conserved-general projection}
\end{align}

At this point, since $(U_1)^n$ and $(U_2)^n$ commute with $\rhogge$, by using Eq.~\eqref{selection-rule-matrix-element-by-center}, we obtain
\begin{align}
    \tr{\rhogge\cO^{q_1, q_2}}=0
    =\lim_{T\to\infty}\frac{1}{T}\int_0^{T}\rmd t\bra{\psiin} e^{\rmi Ht}\cO^{q_1, q_2}e^{-\rmi Ht}\ket{\psiin}\quad\text{for $(q_1,q_2)\notin(m_1\bbZ,m_2\bbZ)$}\label{selection-rule-by-center-GGE}
\end{align}
 This indicates that, when $(q_1,q_2)\notin(m_1\bbZ,m_2\bbZ)$, the expectation value of the charged operator $\cO^{q_1, q_2}$ in the stationary state trivially equals that measured in  $\rhogge$ due to the selection rule~\eqref{selection-rule-by-center-GGE}.

In the case $(q_1,q_2)\in(m_1\bbZ,m_2\bbZ)$, by applying the arguments of Subsection~\eqref{Generalized Gibbs Ensemble} within each symmetry sector, we obtain
\begin{align}
    \tr{\rhogge\fO^{q_1, q_2}}=\sum_{x_1, x_2}\tr{\left[\rhogge (U_1)^{q_2'}(U_2)^{-q_1'}P_{x^1}P_{x^2}\right]}\scrO^{(q_1',q_2')}_{(x^1,x^2)}(\tr\{\rhogge\,\varepsilon\})+\order{V^{-\frac{1}{2}}}.
\end{align}
Furthermore, by utilizing the requirements~\eqref{tuning-energy-general} and~\eqref{tuning-conserved-general projection}, the following relation can be justified:
\begin{align}
    \tr{\rhogge\fO^{q_1, q_2}}=
    \lim_{T\to\infty}\frac{1}{T}\int_0^{T}\rmd t\bra{\psiin} e^{\rmi Ht}\fO^{q_1', q_2'}e^{-\rmi Ht}\ket{\psiin}+\order{V^{-\frac{1}{2}}}.
\end{align}

\section{Justification of small temporal fluctuations}\label{Justification of Stationarization}
In this appendix, we show that when the charged operator~$\cO^{q_1,q_2}$ satisfies the prETH~\eqref{Time-averaged-exp-ETH}
and the non-resonance condition for eigenvalues
\begin{align}
    E_i-E_j=E_k-E_l\neq0\Longrightarrow i=k,\,j=l,
\end{align}
the time fluctuation of the charged operator~$\cO^{q_1,q_2}$ 
vanishes in the thermodynamic limit~$V\to\infty$. Namely,
\begin{align}
    \sigma(\cO^{q_1, q_2})&:= \left( \overline{\left| \bra{\psiin}e^{\rmi Ht}\cO^{q_1, q_2}e^{-\rmi Ht}\ket{\psiin}-\overline{\bra{\psiin} e^{\rmi Ht}\cO^{q_1, q_2}e^{-\rmi Ht} \ket{\psiin}} \right|^2} \right)^{1/2}=\lorder{V^0}
\end{align}
holds.

First, using the completeness of the energy eigenstates~$\sum_iP_i=1$, we obtain
\begin{align}
    \bra{\psiin} e^{\rmi Ht} \cO^{q_1, q_2} e^{-\rmi Ht} \ket{\psiin}
    &=\sum_{i,j}\bra{\psiin}P_i e^{\rmi Ht} \cO^{q_1, q_2} e^{-\rmi Ht}P_j\ket{\psiin}\no\\
    &=\sum_{i,j}e^{\rmi(E_i-E_j)t}\bra{\psiin}P_i\cO^{q_1, q_2}P_j\ket{\psiin}.
\end{align}
Hence,
\begin{align}
    \sigma(\cO^{q_1,q_2})&= \left( \overline{   \left| \bra{\psiin}e^{\rmi Ht}\cO^{q_1,q_2} e^{-\rmi Ht}\ket{\psiin}-\overline{\bra{\psiin} e^{\rmi Ht} \cO^{q_1,q_2} e^{-\rmi Ht} \ket{\psiin}} \right|^2} \right)^{1/2}\no\\
    &=\left( \overline{\left|\sum_{i\neq j}e^{\rmi(E_i-E_j)t}\bra{\psiin}P_i\cO^{q_1, q_2}P_j\ket{\psiin}\right|^2} \right)^{1/2}\no\\
    &=\left( \sum_{i\neq j}\sum_{k\neq l}\overline{e^{\rmi\{E_i-E_j-(E_k-E_l)\}t}}\bra{\psiin}P_i\cO^{q_1, q_2}P_j\ket{\psiin}\times\bra{\psiin}P_l(\cO^{q_1, q_2})^\dagger P_k\ket{\psiin}\right)^{1/2}\no\\
    &=\left( \sum_{i\neq j}\left|\bra{\psiin}P_i\cO^{q_1, q_2}P_j\ket{\psiin}\right|^2\right)^{1/2},\label{fluctuation charged op}
\end{align}
where we have used the non-resonance condition in obtaining the final line.

Next, substituting
\begin{align}
    \ket{\psiin} := \sum_{i}\sum_{\alpha=0}^{N-1}c_{i,\alpha}(U_2)^\alpha\kket{E_i}
\end{align}
into $\bra{\psiin}P_i\cO^{q_1, q_2}P_j\ket{\psiin}$, we obtain
\begin{align}
    \bra{\psiin}P_i\cO^{q_1, q_2}P_j\ket{\psiin}
    &=\sum_{\alpha,\beta}c_{i,\alpha}^*c_{j,\beta}\bbra{E_i}(U_2)^{-\alpha}\cO^{q_1, q_2}(U_2)^\beta\kket{E_j},\quad(\alpha+q_1-\beta=0\bmod{N})\no\\
     &= \left(\sum_{\alpha=0}^{N-q_1-1}c_{i,\alpha}^* c_{j,\alpha+q_1} e^{\frac{2\pi \rmi}{N}q_2 \alpha}+\sum_{\alpha=N-q_1}^{N-1}c_{i,\alpha}^* c_{j,\alpha+q_1-N} e^{\frac{2\pi \rmi}{N}q_2 \alpha}\right)\no\\
   &\qquad\qquad
   \times\bbra{E_i} \cO^{q_1, q_2}(U_2)^{q_1}(U_1)^{-q_2} \kket{E_j}.\label{substituting initial into off doagonal}
\end{align}
Taking the absolute value of Eq.~\eqref{substituting initial into off doagonal} gives
\begin{align}
    \left|\bra{\psiin}P_i\cO^{q_1, q_2}P_j\ket{\psiin}\right|
   &\leq
   \left(\sum_{\alpha=0}^{N-q_1-1}\left|c_{i,\alpha}\right|\left| c_{j,\alpha+q_1}\right| +\sum_{\alpha=N-q_1}^{N-1}\left|c_{i,\alpha}\right| \left|c_{j,\alpha+q_1-N}\right| \right)\no\\
   &\qquad\qquad
   \times\left|\bbra{E_i} \cO^{q_1, q_2}(U_2)^{q_1}(U_1)^{-q_2} \kket{E_j}\right|\no\\
   &\leq
   \left(\frac{1}{2}\sum_{\alpha=0}^{N-1}\left(\left|c_{i,\alpha}\right|^2+\left| c_{j,\alpha+q_1}\right|^2\right)\right)\no\\
   &\qquad\qquad
   \times\left|\bbra{E_i} \cO^{q_1, q_2}(U_2)^{q_1}(U_1)^{-q_2} \kket{E_j}\right|\no\\
    &=
   \frac{1}{2}\left(\sum_{\alpha=0}^{N-1}\left|c_{i,\alpha}\right|^2+\sum_{\beta=0}^{N-1}\left|c_{j,\beta}\right|^2\right)
   \times\left|\bbra{E_i} \cO^{q_1, q_2}(U_2)^{q_1}(U_1)^{-q_2} \kket{E_j}\right|,\label{bound off diagonal part}
\end{align}
where we have used the arithmetic-geometric mean inequality.

Finally, by applying the estimate in Eq.~\eqref{bound off diagonal part} to Eq.~\eqref{fluctuation charged op}, we obtain
\begin{align}
    \sigma(\cO^{q_1, q_2})&=\left( \sum_{i\neq j}\left|\bra{\psiin}P_i\cO^{q_1, q_2}P_j\ket{\psiin}\right|^2\right)^{1/2}\no\\
    &\leq\left( \sum_{i\neq j}\frac{1}{4}\left(\sum_{\alpha=0}^{N-1}\left|c_{i,\alpha}\right|^2+\sum_{\beta=0}^{N-1}\left|c_{j,\beta}\right|^2\right)^2\times\left|\bbra{E_i} \cO^{q_1, q_2}(U_2)^{q_1}(U_1)^{-q_2} \kket{E_j}\right|^2\right)^{1/2}\no\\
    &\leq\max_{i,j(i\neq j)}\left|\bbra{E_i} \cO^{q_1, q_2}(U_2)^{q_1}(U_1)^{-q_2} \kket{E_j}\right|\leq e^{-\order{V}/2}.
\end{align}
Therefore,
\begin{align}
    \sigma(\cO^{q_1, q_2})&:= \left( \overline{\left| \bra{\psiin}e^{\rmi Ht}\cO^{q_1, q_2}e^{-\rmi Ht}\ket{\psiin}-\overline{\bra{\psiin} e^{\rmi Ht}\cO^{q_1, q_2}e^{-\rmi Ht} \ket{\psiin}} \right|^2} \right)^{1/2}=\lorder{V^0}\xrightarrow{V\to\infty}0,
\end{align}
which justifies that the fluctuations 
$\sigma(\cO^{q_1, q_2})$ vanish in the thermodynamic limit. In other words, the time-dependent expectation value 
$\bra{\psiin} e^{\rmi Ht} \cO^{q_1, q_2} e^{-\rmi Ht} \ket{\psiin}$ becomes stationary under the assumption of the prETH~\eqref{Time-averaged-exp-ETH}.

\section{Symmetry structure of Hamiltonian in Eq.~\eqref{Hamiltonian Z2 gauge}}
\label{Symmetry structure of Hamiltonian}

Let us discuss the symmetry structure of the $\bbZ_2$ gauge theory Hamiltonian introduced in Eq.~\eqref{Hamiltonian Z2 gauge}:
\begin{align}
    H := -\sum_{p\in\cM_2}f_p\prod_{\ell\in p}\sigma^z_\ell 
        - \sum_{p\in\cM_2}g_p\prod_{\ell\in p}\sigma^x_\ell 
        - \sum_{v\in\cM_2}h_v\,\sigma^x_{(v, \Hat{1})}\sigma^x_{(v, \Hat{2})} 
        - \sum_{v\in\cM_2}h_v'\,\sigma^x_{(v, \Hat{1})}\sigma^x_{(v+\Hat{1}, \Hat{2})} .
    \label{Re: Hamiltonian Z2 gauge}
\end{align}
The system described by this Hamiltonian exhibits interesting symmetry structures in certain regions of the parameter space, in addition to those discussed in Section~\ref{Z2 lattice gauge theory}. It remains a $\bbZ_2$ gauge theory with the Gauss' law constraint given by Eq.~\eqref{Gauss law} even when the parameters are varied, which possesses the electric $1$-form symmetry,
\begin{align}
    U_x^{(1)}&:=\prod_{\ell\in C^*_x}\sigma_\ell^x,& U_y^{(1)}&:= \prod_{\ell\in C^*_y}\sigma_\ell^x
\end{align}
However, in certain regions of the parameter space, the $0$-form $\bbZ_2$ symmetry
\begin{align}
    U^{(0)}&:=\prod_{\ell\in\text{all links}}\sigma_\ell^z,& [U^{(0)},H]&=0
\end{align}
is not the most fundamental one.


\subsection{Case of $h_v=h_v'=0$}\label{hv=hv'=0}

\begin{figure}[t]
\centering
\begin{tikzpicture}[scale=1.35]
  \draw[->] (-1.5,0) -- (-0.5,0) node[right] {$x$};
  \draw[->] (-1.5,0) -- (-1.5,1) node[above] {$y$};

  \draw[very thick] (0,0) grid (5,4);

  \foreach \x in {0,...,4} {
    \foreach \y in {0,...,3} {
      \draw[red, line width=3pt, opacity=0.5, ->] (\x,\y+0.1) -- (\x,\y+0.9);
    }
  }
  \foreach \y in {0,...,3} {
      \draw[red, line width=3pt, opacity=0.5, ->, dashed] (5,\y+0.1) -- (5,\y+0.9);
  }
  \node[left, red] at (2,2.5) {\large $(v,\Hat{y})$};

  \draw[red, ultra thick, opacity=0.5] (2,-0.3) -- (2,4.3);
  \node[below, red] at (2,-0.2) {winding around $y$-direction};

  \foreach \x in {0,...,4} {
    \foreach \y in {0,...,3} {
      \draw[blue, line width=3pt, opacity=0.5, ->] (\x+0.1,\y) -- (\x+0.9,\y);
    }
  }
  \foreach \x in {0,...,4} {
      \draw[blue, line width=3pt, opacity=0.5, ->, dashed] (\x+0.1,4) -- (\x+0.9,4);
  }
  \node[below, blue] at (2.5,2) {\large $(v,\Hat{x})$};

  \draw[blue, ultra thick, opacity=0.5] (5.2,2) -- (-0.7,2);
  \node[left, blue, rotate around={90:(-0.5,2)}] at (-1.95,2) {winding around $x$-direction};

  \fill (2,2) circle(3pt) node[above right] {\large $v$};
\end{tikzpicture}\hspace{1em}
\begin{tikzpicture}[scale=2]
  \pgfmathsetmacro{\R}{1.5} 
  \pgfmathsetmacro{\r}{0.5} 

  \foreach \v in {0,10,...,360} {
    \pgfmathsetmacro{\theta}{\v}
    \foreach \u in {0,10,...,360} {
      \pgfmathsetmacro{\phi}{\u}
      \pgfmathsetmacro{\x}{(\R + \r * cos(\phi)) * cos(\theta)}
      \pgfmathsetmacro{\y}{(\R + \r * cos(\phi)) * sin(\theta)}
      \pgfmathsetmacro{\z}{\r * sin(\phi)}
      \fill[blue] (\x/2,\y/2,\z/2) circle (0.01);
    }
  }

  \node[blue, above] at (0,1.3) {\large $\cW^x$};

  \node[blue, below] at (0,-1.2) {windings around $x$-direction}; 
  \draw[magenta, ->, very thick] (1,-0.45) arc(-30:60:1) node[above] {\large along $y$-axis};
\end{tikzpicture}
\caption{
Visualization of the Wilson loop operator $\cW^x$ ($\cW^y$), which winds around the $x$-direction ($y$-direction) along the $y$-axis ($x$-axis). 
\textbf{Left:} A schematic of the original $2$-dimensional periodic lattice. The operator $\cW^x$ consists of a product of $\sigma^z$ on all horizontal links $(v,\Hat{x})$ arranged along the $y$-axis. Each blue arrow denotes the contribution from $\sigma^z$ on a link in the $x$-direction; the  blue line indicates the winding structure across the entire $y$-axis. Also, the counterpart colored by red is for $\cW^y$.
\textbf{Right:} A conceptual $3$-dimensional illustration of the toroidal topology of the lattice. The torus represents periodicity in both $x$ and $y$ directions. The blue circle indicates a winding around the $x$-direction, while the magenta arc highlights the $y$-axis path along which $\cW^x$ is constructed to wrap around the $x$-cycle of the torus.}
\label{fig:mathcal Wxy}
\end{figure}
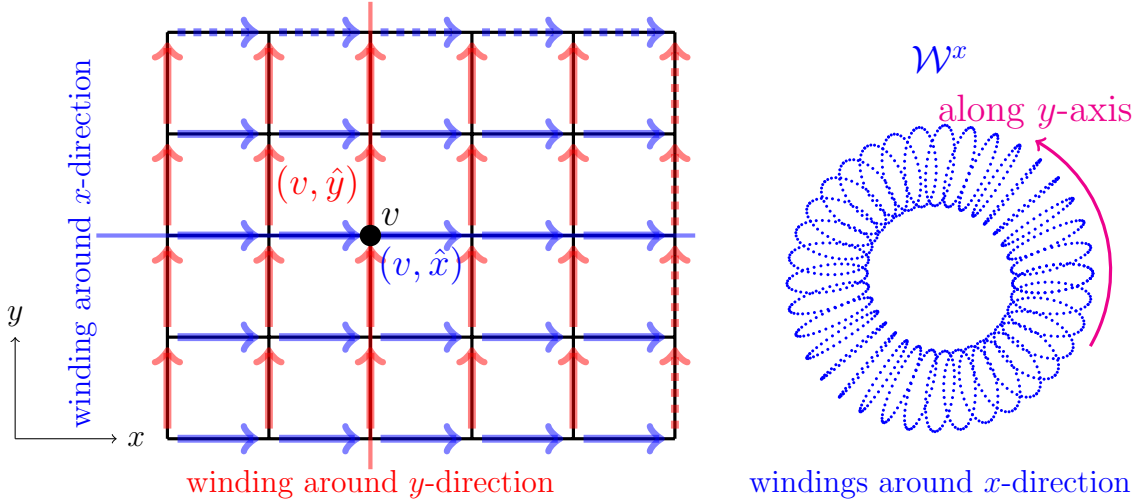

First, let $h_v=h_v'=0$. The Hamiltonian~\eqref{Re: Hamiltonian Z2 gauge} becomes
\begin{align}
    H_{h_v=h_v'=0} := -\sum_{p\in\cM_2} f_p \prod_{\ell\in p} \sigma^z_\ell - \sum_{p\in\cM_2} g_p \prod_{\ell\in p} \sigma^x_\ell.
\end{align}
This Hamiltonian commutes with an operator constructed by aligning Wilson lines that wind around the $x$-direction along the $y$-axis, defined as:
\begin{align}
    \cW^x &:=
    \left(\prod_{\substack{\ell=(v,\Hat{x})\\ v\in\{(x,0)\mid \forall x\}}} \sigma_\ell^z\right)
    \left(\prod_{\substack{\ell=(v,\Hat{x})\\ v\in\{(x,1)\mid \forall x\}}} \sigma_\ell^z\right)
    \cdots
    \left(\prod_{\substack{\ell=(v,\Hat{x})\\ v\in\{(x,L_y-1)\mid \forall x\}}} \sigma_\ell^z\right)
    \no\\
    &= \prod_{i=0}^{L_y-1} \left(\prod_{\substack{\ell=(v,\Hat{x})\\ v\in\{(x,i)\mid \forall x\}}} \sigma_\ell^z\right),\\
    (\cW^x)^2 &= 1.
\end{align}
 Similarly, the operator constructed by aligning Wilson lines winding around the $y$-direction along the $x$-axis also commutes with the Hamiltonian~\eqref{Hamiltonian two plaquettes}. That is given by
\begin{align}
    \cW^y &:= \prod_{i=0}^{L_x-1} \left(\prod_{\substack{\ell=(v,\Hat{y})\\ v\in\{(i,y)\mid \forall y\}}} \sigma_\ell^z\right),& (\cW^y)^2 &= 1,
\end{align}
Furthermore, since they satisfy
\begin{align}
    U^{(0)} = \cW^x \cW^y = \cW^y \cW^x, \label{decomposition U^0 by WxWy}
\end{align}
they can be regarded as a decomposition of the $0$-form $\bbZ_2$ symmetry operator $U^{(0)}$. The support of $\cW^{x,y}$ does not cover the entire space, but only ``half'' of it (see Fig.~\ref{fig:mathcal Wxy}).

The operators $\cW^{x,y}$ commute/anti-commute with the electric $1$-form symmetry operators $U_{x,y}^{(1)}$, depending on the even/odd values of $L_{x,y}$. We obtain the following projective representation:
\begin{align}
    &U_x^{(1)} \cW^y = (-1)^{L_y} \cW^y U_x^{(1)},\\
    &U_y^{(1)} \cW^x = (-1)^{L_x} \cW^x U_y^{(1)}.
\end{align}
As in Section~\eqref{Z2 lattice gauge theory}, when both $L_x$ and $L_y$ are even, there is no non-trivial projective representation.

\subsubsection{Case of $L_x \in 2\bbZ + 1$ and $L_y \in 2\bbZ$}

In the case that $L_x \in 2\bbZ + 1$ and $L_y \in 2\bbZ$, the only non-trivial projective representation is given by
\begin{align}
    U_{y}^{(1)}\cW^x = (-1)\cW^x U_{y}^{(1)}. \label{Lx odd Z2 projective rep}
\end{align}
Then, let us take a simultaneous eigenstate\footnote{Since $U_{y}^{(1)}$ commutes with both $U_{x}^{(1)}$ and $\mathcal{W}^y$, the state $\kket{E_i}_y$ can also be taken as an eigenstate of these operators.} of energy and $U_{y}^{(1)}$, denoted as $\kket{E_i}_y$, satisfying
\begin{align}
   H \kket{E_i}_y &= E_i \kket{E_i}_y,& U_{y}^{(1)} \kket{E_i}_y &= \kket{E_i}_y.
\end{align}
From Eq.~\eqref{Lx odd Z2 projective rep}, we obtain, for $\alpha=0$ and $1$,
\begin{align}
   H (\cW^x)^\alpha \kket{E_i}_y &= E_i (\cW^x)^\alpha \kket{E_i}_y, \\
   U_{y}^{(1)} (\cW^x)^\alpha \kket{E_i}_y &= (-1)^\alpha (\cW^x)^\alpha \kket{E_i}_y.
\end{align}
Thus, the energy eigenstates exhibit a two-fold degeneracy. A similar argument can apply for the case for $L_x \in 2\bbZ$ and $L_y \in 2\bbZ + 1$, which also leads to a two-fold degeneracy.

\subsubsection{Case of $L_x$, $L_y \in 2\bbZ + 1$}

Suppose that $L_x$ and $L_y$ are both odd.
In this case, the non-trivial projective representation is given by
\begin{align}
    U_{x}^{(1)} \cW^y &= (-1) \cW^y U_{x}^{(1)}, &
    U_{y}^{(1)} \cW^x &= (-1) \cW^x U_{y}^{(1)}. \label{Lx Ly odd Z2 projective rep}
\end{align}
Then, we obtain the following relations:
\begin{align}
H (\cW^x)^\alpha (\cW^y)^\beta \kket{E_i}_y &= E_i (\cW^x)^\alpha (\cW^y)^\beta \kket{E_i}_y, \\
U_{y}^{(1)} (\cW^x)^\alpha (\cW^y)^\beta \kket{E_i}_y &= (-1)^\alpha (\cW^x)^\alpha (\cW^y)^\beta \kket{E_i}_y, \\
U_{x}^{(1)} (\cW^x)^\alpha (\cW^y)^\beta \kket{E_i}_y &= (-1)^\beta (\cW^x)^\alpha (\cW^y)^\beta \kket{E_i}_y
\end{align}
with $\alpha$, $\beta = 0$, $1$.
Hence, in this case, the energy eigenstates exhibit a four-fold degeneracy.

\subsection{Case of $h_v'=0$}

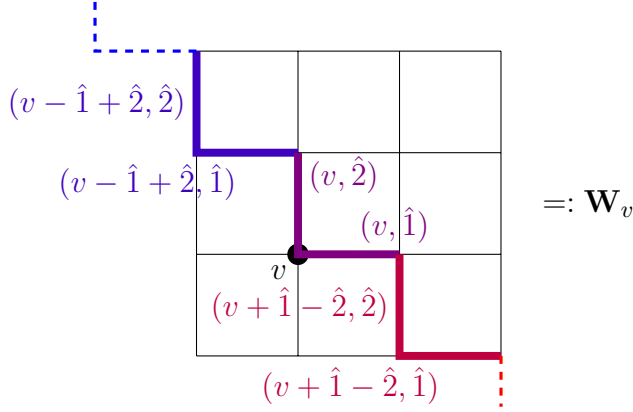
\begin{figure}[t]
\centering
\begin{tikzpicture}[scale=1.35]
  \draw (0,0) grid (3,3);
  \fill (1,1) circle(3pt) node[below left] {$v$};

  \draw[line width=3pt,blue!75!red] (0,3) -- (0,2) -- (1,2);
  \draw[line width=3pt,blue!50!red] (1,2) -- (1,1) -- (2,1);
  \draw[line width=3pt,blue!25!red] (2,1) -- (2,0) -- (3,0);
  \draw[very thick,dashed,blue] (0,3) -- (-1,3) -- (-1,3.5);
  \draw[very thick,dashed,red] (3,0) -- (3,-0.5);

  \node[blue!75!red,left] at (0,2.5) {$(v-\Hat{1}+\Hat{2},\Hat{2})$};
  \node[blue!75!red,below left] at (0.5,2) {$(v-\Hat{1}+\Hat{2},\Hat{1})$};
  \node[blue!50!red,above right] at (1,1.5) {$(v,\Hat{2})$};
  \node[blue!50!red,above right] at (1.5,1) {$(v,\Hat{1})$};
  \node[blue!25!red,left] at (2,0.5) {$(v+\Hat{1}-\Hat{2},\Hat{2})$};
  \node[blue!25!red,below left] at (2.5,0) {$(v+\Hat{1}-\Hat{2},\Hat{1})$};

  \node[right] at (3.3,1.5) {$=:\mathbf{W}_v$};
\end{tikzpicture}
\caption{Description of the definition of $\mathbf{W}_v$. Graphical representation of the Wilson operator $\mathbf{W}_v$ is shown locally around site~$v$ (black dot). The bold colored links indicate the support of $\mathbf{W}_v$ and are grouped into three pairs, shown in different color shades. Each link is labeled by its associated coordinate and direction, e.g., $(v, \Hat{1})$. In this simple figure, the operator $\mathbf{W}_v$ consists of six $\sigma^z_\ell$ operators acting on links adjacent to $v$ and its nearest diagonal neighbors. Dashed lines indicate the continuation of the zigzag path beyond the displayed region; so $\mathbf{W}_v$ is actually nonlocal. If extended periodically, the zigzag path eventually covers $\lcm(L_x, L_y)$ link variables. In particular, when $\gcd(L_x, L_y) = 1$, we have $\lcm(L_x, L_y) = L_x L_y$, and the entire lattice is covered by the zigzag structure (see Fig.~\ref{fig:W_zigzag_cover}).}
\label{fig:W_zigzag}
\end{figure}

\begin{figure}[t]
\centering
\begin{tikzpicture}[scale=1.1]
  \draw (0,0) grid (6,3);
  \fill (1,1) circle(3pt) node[below left] {$v$};

  \draw[line width=3pt,blue!75!red,rounded corners=10pt] (0,3) -- (0,2) -- (1,2);
  \draw[line width=3pt,blue!50!red,rounded corners=10pt] (1,2) -- (1,1) -- (2,1);
  \draw[line width=3pt,blue!25!red,rounded corners=10pt] (2,1) -- (2,0) -- (3,0);

  \draw[line width=3pt,blue!75!red,dashed] (6,2) -- (6,3);
  \draw[line width=3pt,teal,dashed] (6,3) -- (5,3);

  \draw[line width=3pt,teal,rounded corners=10pt] (6,0) -- (5,0) -- (5,1);
  \draw[line width=3pt,blue!25!red,rounded corners=10pt] (5,1) -- (4,1) -- (4,2);
  \draw[line width=3pt,blue!50!red,rounded corners=10pt] (4,2) -- (3,2) -- (3,3);
  \draw[line width=3pt,blue!75!red,dashed] (3,3) -- (2,3);

  \node[blue!25!red,below] at (2.5,0) {$\ell_1$};
  \node[blue!75!red,left] at (0,2.5) {$\ell_2$};
  \node[blue!75!red,right] at (6,2.5) {$\ell_2$};
  \node[teal,above] at (5.5,3) {$\ell_3$};
  \node[teal,below] at (5.5,0) {$\ell_3$};
  \node[blue!75!red,above] at (2.5,3) {$\ell_4=\ell_1$};

\end{tikzpicture}
\hspace{1em}
\begin{tikzpicture}[scale=1.1]
  \draw (0,0) grid (5,3);
  \fill (1,1) circle(3pt) node[below left] {$v$};

  \draw[line width=3pt,blue!75!red,rounded corners=10pt] (0,3) -- (0,2) -- (1,2);
  \draw[line width=3pt,blue!50!red,rounded corners=10pt] (1,2) -- (1,1) -- (2,1);
  \draw[line width=3pt,blue!25!red,rounded corners=10pt] (2,1) -- (2,0) -- (3,0);

  \draw[line width=3pt,blue!75!red,dashed] (5,2) -- (5,3);
  \draw[line width=3pt,teal,dashed] (5,3) -- (4,3);

  \draw[line width=3pt,teal,rounded corners=10pt] (5,0) -- (4,0) -- (4,1);
  \draw[line width=3pt,blue!25!red,rounded corners=10pt] (4,1) -- (3,1) -- (3,2);
  \draw[line width=3pt,blue!50!red,rounded corners=10pt] (3,2) -- (2,2) -- (2,3);
  \draw[line width=3pt,blue!75!red,dashed] (2,3) -- (1,3);

  \draw[line width=3pt,blue!75!red,rounded corners=10pt] (2,0) -- (1,0) -- (1,1);
  \draw[line width=3pt,teal,rounded corners=10pt] (1,1) -- (0,1) -- (0,2);

  \draw[line width=3pt,teal,dashed] (5,1) -- (5,2);
  \draw[line width=3pt,blue!25!red,rounded corners=10pt] (5,2) -- (4,2) -- (4,3);
  \draw[line width=3pt,blue!50!red,dashed] (4,3) -- (3,3);

  \draw[line width=3pt,blue!50!red,rounded corners=10pt] (4,0) -- (3,0) -- (3,1);
  \draw[line width=3pt,blue!75!red,rounded corners=10pt] (3,1) -- (2,1) -- (2,2);
  \draw[line width=3pt,teal,rounded corners=10pt] (2,2) -- (1,2) -- (1,3);
  \draw[line width=3pt,blue!25!red,dashed] (1,3) -- (0,3);

  \draw[line width=3pt,blue!25!red,rounded corners=10pt] (1,0) -- (0,0) -- (0,1);

  \draw[line width=3pt,blue!25!red,dashed] (5,0) -- (5,1);
  \draw[line width=3pt,blue!50!red,rounded corners=10pt] (5,1) -- (4,1) -- (4,2);
  \draw[line width=3pt,blue!75!red,rounded corners=10pt] (4,2) -- (3,2) -- (3,3);
  \draw[line width=3pt,teal,dashed] (3,3) -- (2,3);

  \node[blue!25!red,below] at (2.5,0) {$\ell_1$};
  \node[blue!75!red,left] at (0,2.5) {$\ell_2$};
  \node[blue!75!red,right] at (5,2.5) {$\ell_2$};
  \node[teal,above] at (4.5,3) {$\ell_3$};
  \node[teal,below] at (4.5,0) {$\ell_3$};
  \node[blue!75!red,above] at (1.5,3) {$\ell_4$};
  \node[blue!75!red,below] at (1.5,0) {$\ell_4$};
  \node[teal,left] at (0,1.5) {$\ell_5$};
  \node[teal,right] at (5,1.5) {$\ell_5$};
  \node[blue!50!red,above] at (3.5,3) {$\ell_6$};
  \node[blue!50!red,below] at (3.5,0) {$\ell_6$};
  \node[blue!25!red,above] at (0.5,3) {$\ell_7$};
  \node[blue!25!red,below] at (0.5,0) {$\ell_7$};
  \node[blue!25!red,left] at (0,0.5) {$\ell_8$};
  \node[blue!25!red,right] at (5,0.5) {$\ell_8$};
  \node[teal,above] at (2.5,3) {$\ell_9=\ell_1$};

\end{tikzpicture}
\caption{Local-to-global structure of the zigzag Wilson operator $\mathbf{W}_v$ for two different lattice sizes.
\textbf{Left:} $6\times 3$ lattice, where the zigzag path repeats after $\lcm(6,3)=6$ link variables.
\textbf{Right:} $5\times 3$ lattice, where $\gcd(5,3)=1$ implies $\lcm(5,3)=15$, so the path covers all $5\times 3=15$ link positions before repeating.
Colored thick links indicate the successive segments along the zigzag trajectory starting near site $v$ (black dot). Dashed segments, $\ell_1, \ell_2, \dots$, show the continuation across the periodic boundary. The color shading groups link into repeating triples, illustrating the nonlocal and periodic nature of $\mathbf{W}_v$ on the torus.}
\label{fig:W_zigzag_cover}
\end{figure}
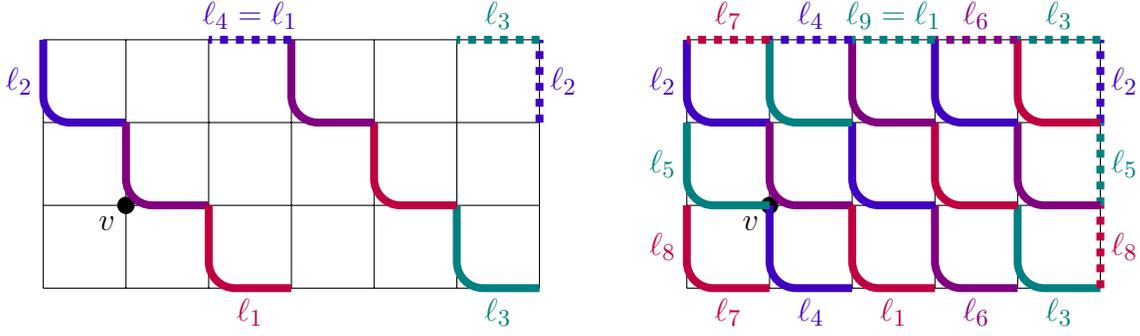

Here, we consider the Hamiltonian~\eqref{Re: Hamiltonian Z2 gauge} with $h_v' = 0$
\begin{align}
    H_{h_v'=0} := -\sum_{p\in\cM_2} f_p \prod_{\ell\in p} \sigma^z_\ell 
    - \sum_{p\in\cM_2} g_p \prod_{\ell\in p} \sigma^x_\ell 
    - \sum_{v\in\cM_2} h_v\, \sigma^x_{(v,\Hat{1})}\sigma^x_{(v,\Hat{2})}.
    \label{Hamiltonian two plaquettes}
\end{align}
In this case, due to the presence of the third term ~$\sum_{v\in\cM_2} h_v\, \sigma^x_{(v,\Hat{1})}\sigma^x_{(v,\Hat{2})}$, 
the operators $\mathcal{W}^{x,y}$ no longer commute with the Hamiltonian. However, a new symmetry emerges that respects the form of this term. The following Wilson line operator originated from $v$ commutes with the Hamiltonian~\eqref{Hamiltonian two plaquettes}:
\begin{align}
    \mathbf{W}_v &:= \cdots \sigma^z_{(v-\Hat{1}+\Hat{2},\Hat{1})}\sigma^z_{(v-\Hat{1}+\Hat{2},\Hat{2})}
    \sigma^z_{(v,\Hat{1})}\sigma^z_{(v,\Hat{2})}
    \sigma^z_{(v+\Hat{1}-\Hat{2},\Hat{1})}\sigma^z_{(v+\Hat{1}-\Hat{2},\Hat{2})} \cdots \no\\
    &= \prod_{i=0}^{\lcm(L_x,L_y)} \sigma^z_{(v - i\Hat{1} + i\Hat{2}, \Hat{1})} \sigma^z_{(v - i\Hat{1} + i\Hat{2}, \Hat{2})}.
\end{align}
Figure~\ref{fig:W_zigzag} shows a graphical illustration of these products of $\sigma_\ell^z$ in $\mathbf{W}_v$.
When $\gcd(L_x,L_y)=1$, we have $\lcm(L_x,L_y)=L_x L_y$. Since $L_x L_y$ is equal to the total number of lattice sites, it follows that
\begin{align}
    \forall v \qquad \mathbf{W}_v = U^{(0)}.
\end{align}
In Fig.~\ref{fig:W_zigzag_cover}, in both cases that $\gcd(L_x,L_y)\neq1$ and $\gcd(L_x,L_y)=1$, the paths in $\mathbf{W}_v$ on different lattices are depicted, respectively, and it is shown that a lattice with $\gcd(L_x,L_y)=1$ is covered by the path.
In other words, in this case, there is no symmetry that is more fundamental than $U^{(0)}$.

Next, consider the case for~$\gcd(L_x, L_y)\neq1$. Now, there exist $\gcd(L_x, L_y)$ independent Wilson lines (each commuting with the Hamiltonian):\footnote{Alternatively, one may express them as
\begin{align}
    \mathbf{W}_v, \quad \mathbf{W}_{v+\Hat{2}}, \quad \mathbf{W}_{v+2\Hat{2}}, \ldots,\quad \mathbf{W}_{v+(\gcd(L_x, L_y)-1)\Hat{2}}
\end{align}
by using the unit vector~$\Hat{2}$.
}
\begin{align}
    \mathbf{W}_v, \quad \mathbf{W}_{v+\Hat{1}}, \quad \mathbf{W}_{v+2\Hat{1}}, \ldots,\quad \mathbf{W}_{v+(\gcd(L_x, L_y)-1)\Hat{1}}.
\end{align}
Furthermore, $\mathbf{W}_v$ satisfies the following properties:\footnote{To see this, let us ask what is the integer $k$ which satisfies $\mathbf{W}_v = \mathbf{W}_{v+k\Hat{1}}$. From the definition of $\mathbf{W}_v$, the integer $k$ must be satisfied the relations
\begin{align}
    \exists i\in\bbZ\qquad -i &= k \bmod{L_x}, \no\\
    i &= 0 \bmod{L_y}.
\end{align}
From this relations, $k$ must be a multiple of $\gcd(L_x, L_y)$. In other words, when $k$ is not a multiple of $\gcd(L_x, L_y)$, then $\mathbf{W}_{v+k\Hat{1}}$ is another Wilson line from $\mathbf{W}_v$.}
\begin{align}
    &\prod_{k=0}^{\gcd(L_x, L_y)-1} \mathbf{W}_{v + k\Hat{1}} = U^{(0)}, \\
    &U_{x}^{(1)} \mathbf{W}_{v + k\Hat{1}} = (-1)^{\frac{L_x}{\gcd(L_x, L_y)}} \mathbf{W}_{v + k\Hat{1}} U_{x}^{(1)}, \\
    &U_{y}^{(1)} \mathbf{W}_{v + k\Hat{1}} = (-1)^{\frac{L_y}{\gcd(L_x, L_y)}} \mathbf{W}_{v + k\Hat{1}} U_{y}^{(1)}.
\end{align}
As in Section~\ref{hv=hv'=0}, depending on the values of $L_x$ and $L_y$, the energy eigenstates exhibit either two-fold or four-fold degeneracy. However, the classification of these degeneracies becomes complicated in this case. To avoid this complexity, we instead consider the Hamiltonian~\eqref{Hamiltonian Z2 gauge} in the main text. The residual term 
$\sum_{v\in\mathcal{M}_2} h_v' \, \sigma^x_{(v, \Hat{1})} \sigma^x_{(v+\Hat{1}, \Hat{2})}$ in the Hamiltonian~\eqref{Re: Hamiltonian Z2 gauge}
does not commute with $\mathbf{W}_v$, and thus the  symmetries~$\mathbf{W}_{v + k\Hat{1}}$ are explicitly broken. As a result, the $0$-form symmetry~$U^{(0)}$ becomes the most fundamental symmetry of the Hamiltonian~\eqref{Re: Hamiltonian Z2 gauge}.

\section{Diagonalization of $(U_1)^{q_2}(U_2)^{-q_1}$}\label{sec:app_diag}

In this Appendix, we consider the diagonalization of $(U_1)^{q_2}(U_2)^{-q_1}$ in the presence of a $\bbZ_N \times \bbZ_N$ symmetry. The following discussion can be straightforwardly extended to the case of $\bbZ_{N_1} \times \bbZ_{N_2}$ symmetry.

{In order to determine the possible eigenvalues of $(U_1)^{q_2}(U_2)^{-q_1}$, we consider taking its $N$th power:
\begin{align}
    \{(U_1)^{q_2}(U_2)^{-q_1}\}^N 
    &= (U_1)^{Nq_2}(U_2)^{-Nq_1} 
    e^{\frac{2\pi\rmi}{N}q_1q_2(1+2+\cdots+N-1)}\no\\
    &= e^{(N-1)q_1q_2\pi\rmi} \no\\
    &= \begin{cases}
        e^{-q_1q_2\pi\rmi} &\text{if } N \in 2\bbZ, \\
        1 &\text{if } N \in 2\bbZ + 1.
    \end{cases}
\end{align}
Therefore, when $N \in 2\bbZ + 1$, the eigenvalues of $(U_1)^{q_2}(U_2)^{-q_1}$ are $\bbZ_N$ phase. On the other hand, when $N \in 2\bbZ$, they are not necessarily $\bbZ_N$ phase. That is, in the case of $N \in 2\bbZ$, the eigenvalues of $(U_1)^{q_2}(U_2)^{-q_1}$ can generally be written as
\footnote{Note that the factor $e^{-\frac{\pi\rmi}{N}q_1q_2}$ is ambiguous up to multiplication by a $\bbZ_N$ phase. Accordingly, when $q_1q_2 \in 2\bbZ$, this factor can be absorbed into the definition of $\alpha$. Nevertheless, for notational simplicity, we choose to write $e^{-\frac{\pi\rmi}{N}q_1q_2}$ explicitly throughout this paper, regardless of the value of $q_1q_2$.}}
\begin{align}
    e^{-\frac{\pi\rmi}{N}q_1q_2} \times e^{\frac{2\pi\rmi}{N}\alpha}, \qquad\alpha = 0, \dots, N - 1.\label{eigenvalues N is even}
\end{align}

We begin with the case $\gcd(q_1, q_2) = 1$ and define a ``reference state'' as 
\begin{align}
    &\kket{E_i;(q_1,q_2)} := P^{q_2,-q_1} \kket{E_i}, \\
    &P^{q_2,-q_1}:=
    \begin{cases}
        \frac{1}{N} \sum_{\alpha=0}^{N-1} 
    {e^{-\frac{\pi\rmi}{N}q_1q_2\alpha}}
    \left[(U_1)^{q_2}(U_2)^{-q_2}\right]^\alpha &N\in2\bbZ,\\
    \frac{1}{N} \sum_{\alpha=0}^{N-1} 
    \left[(U_1)^{q_2}(U_2)^{-q_2}\right]^\alpha &N\in2\bbZ+1.
    \end{cases}, \\
    &(U_1)^{q_2}(U_2)^{-q_1} \kket{E_i;(q_1,q_2)} = F_N^{(q_2, -q_1)}\kket{E_i;(q_1,q_2)}.
\end{align}
Here, the factor~$F_N^{(q_2, -q_1)}$ is defined as
\begin{align}
    F_N^{(q_2, -q_1)}:=
    \begin{cases}
        e^{-\frac{\pi\rmi}{N}q_1q_2} &N\in2\bbZ,\\
    1 &N\in2\bbZ+1.
    \end{cases}\label{factor FN}
\end{align}
Starting from a ``reference state'' that is one of the eigenstates corresponding to the operator $(U_1)^{q_2}(U_2)^{-q_1}$, we can generate new eigenstates of this operator by acting with $U_1$ and $U_2$ as follows:
\begin{align}
    (U_1)^{q_2}(U_2)^{-q_1}(U_1)^{-\alpha}(U_2)^{\beta}\kket{E_i;(q_1,q_2)} = F_N^{(q_2, -q_1)}\,e^{\frac{2\pi\rmi}{N}(\alpha q_1 + \beta q_2)}(U_1)^{-\alpha}(U_2)^{\beta}\kket{E_i;(q_1,q_2)}.
\end{align}
Since $\gcd(q_1,q_2)=1$, the combination $\alpha q_1 + \beta q_2$ can take any integer modulo $N$. Therefore, an eigenvalue of $(U_1)^{q_2}(U_2)^{-q_1}$ is one of~$e^{\frac{2\pi\rmi}{N}}$, $e^{\frac{2\pi\rmi}{N}2}$, \dots, $e^{\frac{2\pi\rmi}{N}(N-1)}$ up to $F_N^{(q_2, -q_1)}$. The eigenvalues are consistent with the observation in Eq.~\eqref{eigenvalues N is even}.

We now show that there is no degeneracy in the eigenvalues of $(U_1)^{q_2}(U_2)^{-q_1}$ due to different choices of $(\alpha,\beta)$. First, we consider the case that $\alpha q_1 + \beta q_2 = \alpha' q_1 + \beta' q_2\bmod{N}$ holds for two distinct pairs $(\alpha,\beta)\neq(\alpha',\beta') \mod N$. Then, since $\gcd(q_1, q_2) = 1$, we have
\begin{align}
    (\alpha - \alpha')q_1 = -(\beta - \beta')q_2\quad \Longrightarrow\quad \exists \ell \in \bbZ\quad \alpha = \alpha' + \ell q_2,\quad \beta = \beta' - \ell q_1,
\end{align}
and hence,
\begin{align}
    (U_1)^{-\alpha'}(U_2)^{\beta'}\kket{E_i;(q_1,q_2)}
    &\propto (U_1)^{-\alpha}(U_2)^{\beta} [(U_1)^{q_2}(U_2)^{-q_1}]^{-\ell} \kket{E_i;(q_1,q_2)} \no \\
    &\propto (U_1)^{-\alpha}(U_2)^{\beta} \kket{E_i;(q_1,q_2)}.
\end{align}
$(U_1)^{-\alpha'}(U_2)^{\beta'}\kket{E_i\,;(q_1,q_2)}$ and $(U_1)^{-\alpha}(U_2)^{\beta}\kket{E_i\,;(q_1,q_2)}$ are linearly dependent. Therefore, we conclude that there is no degeneracy in the eigenvalues of $(U_1)^{q_2}(U_2)^{-q_1}$ due to different choices of $(\alpha,\beta)$. 

Next, let us consider the case $\gcd(q_1, q_2) \neq 1$. In this case, there exists a pair $(\Tilde{q}_1, \Tilde{q}_2)$ satisfying $(q_1,q_2) = (\gcd(q_1, q_2)\Tilde{q}_1, \gcd(q_1, q_2)\Tilde{q}_2)$ with $\gcd(\Tilde{q}_1, \Tilde{q}_2) = 1$.
Then, we can find
\begin{align}
    (U_1)^{q_2}(U_2)^{-q_1} =e^{-\frac{\pi\rmi}{N}\Tilde{q}_1\Tilde{q}_2\gcd(q_1, q_2)\{\gcd(q_1, q_2)-1\}} \left[(U_1)^{\Tilde{q}_2}(U_2)^{-\Tilde{q}_1}\right]^{\gcd(q_1, q_2)}.
\end{align}
Hence, the diagonalization of $(U_1)^{q_2}(U_2)^{-q_1}$ is equivalent to a diagonalization of $(U_1)^{\Tilde{q}_2}(U_2)^{-\Tilde{q}_1}$. That is, the states $(U_1)^{-\alpha}(U_2)^{\beta}\kket{E_i;(\Tilde{q}_1, \Tilde{q}_2)}$ are also eigenstates of $(U_1)^{q_2}(U_2)^{-q_1}$.
Furthermore, the corresponding eigenvalues can generally be written as
\begin{align}
    &e^{-\frac{\pi\rmi}{N}\Tilde{q}_1\Tilde{q}_2\gcd(q_1, q_2)\{\gcd(q_1, q_2)-1\}} \left[F_N^{(\Tilde{q}_2, -\Tilde{q}_1)}\,e^{\frac{2\pi\rmi}{N}(\alpha \Tilde{q}_1 + \beta \Tilde{q}_2)}\right]^{\gcd(q_1, q_2)}\no\\
    &=e^{-\frac{\pi\rmi}{N}\Tilde{q}_1\Tilde{q}_2\gcd(q_1, q_2)\{\gcd(q_1, q_2)-1\}} \left[F_N^{(\Tilde{q}_2, -\Tilde{q}_1)}\right]^{\gcd(q_1, q_2)}\times e^{\frac{2\pi\rmi}{N}\gcd(q_1, q_2)(\alpha \Tilde{q}_1 + \beta \Tilde{q}_2)}.
\end{align}
In the case of $\gcd(N,\gcd(q_1, q_2))\neq1$, the number of possible eigenvalues labeled by the factor~$e^{\frac{2\pi\rmi}{N}\gcd(q_1, q_2)(\alpha \Tilde{q}_1 + \beta \Tilde{q}_2)}$  is reduced to $N/\gcd(q_1, q_2) < N$. In this case, the eigenvalues of $(U_1)^{q_2}(U_2)^{-q_1}$ are $\gcd(q_1, q_2)$-fold degenerate. This degeneracy can be classified by diagonalizing $(U_1)^{\Tilde{q}_2}(U_2)^{-\Tilde{q}_1}$. Thus, when $\gcd(\gcd(q_1, q_2), N)\neq1$, by using the basis~$(U_1)^{-\alpha}(U_2)^{\beta}\kket{E_i;(\Tilde{q}_1, \Tilde{q}_2)}$, we can produce Eq.~\eqref{non-trivial degeneracy}.

\bibliographystyle{utphys}
\bibliography{ref}
\end{document}